\documentclass[11pt]{article}
\usepackage{amsmath,amsfonts,amsbsy}
\usepackage[dvips]{graphicx,epsfig}

 \csname
@addtoreset\endcsname{equation}{section}

\newcommand{\hoch}[1]{$^{#1}$}

\def\mso{\mathfrak{so}}
\def\msu{\mathfrak{su}}
\def\msl{\mathfrak{sl}}
\def\msp{\mathfrak{sp}}
\def\mosp{\mathfrak{osp}}
\def\psu{\mathfrak{psu}(2,2|4)}
\def\mhs{\mathfrak{hs}}
\def\mhso{\mathfrak{ho}}
\def\mg{\mathfrak{g}}
\def\mh{\mathfrak{h}}

\def\Real{{\mathbb R}}
\def\Comp{{\mathbb C}}
\def\integ{{\mathbb Z}}

\def\bec{\begin{center}}
\def\ec{\end{center}}
\def\a{\alpha} \def\ad{\dot{\a}} 

\def\b{\beta}   
\def\c{\gamma} 
\def\C{\Gamma}
\def\d{\delta} 
\def\D{\Delta}
\def\e{\epsilon}

\def\k{\kappa}
\def\l{\lambda}
\def\L{\Lambda}
\def\m{\mu}
\def\n{\nu}
\def\r{\rho}
\def\s{\sigma}
\def\S{\Sigma}
\def\t{\tau}
\def\th{\theta} 
\def\Th{\Theta}
\def\x{\xi}
\def\y{\eta}

\def\O{\Omega}
\def\o{\omega}

\def\sfD{{\mathfrak D}}

\def\cL{{\cal L}}

\def\cF{{\cal F}}

\def\cN{{\cal N}}

\def\cV{{\cal V}}
\def\cH{{\cal H}}

\def\yb{{\bar y}}
\def\zb{{\bar z}}

\let\la=\label

\def\nn{\nonumber}
\newcommand{\eq}[1]{(\ref{#1})}

\def\eqs#1#2{(\ref{#1}-\ref{#2})} \def\det{{\rm det\,}}
\def\be{\begin{equation}}
\def\ee{\end{equation}}
\def\bea{\begin{eqnarray}}
\def\eea{\end{eqnarray}}
\def\ba{\begin{array}}
\def\ea{\end{array}}
\def\se{\;\;=\;\;}

\def\mx#1#2#3#4{\left#1\begin{array}{#2} #3 \end{array}\right#4}

\def\ft#1#2{{\textstyle{{\scriptstyle #1}
\over {\scriptstyle #2}}}}

\def\ket#1{|#1\rangle}
\def\bra#1{\langle#1|}
\def\scs#1{\section{\sc #1}}
\def\scss#1{\subsection{\sc #1}}
\def\scsss#1{\subsubsection{\sc #1}}

\begin{document}

\hfill{UUITP-13/05}
\\[-20pt]

\hfill{hep-th/0508124}

\hfill{\today}

\vspace{20pt}

\begin{center}


{\Large\sc Brane Partons and Singleton Strings}


\vspace{30pt}
{\sc J. Engquist\hoch1 and P. Sundell\hoch2}\\[15pt]

{\it\small Department of Theoretical Physics\\ Uppsala
University\\ Box 803, SE-751 08 Uppsala, Sweden}


\vspace{30pt} {\sc\large Abstract}\end{center}

\noindent We examine $p$-branes in $AdS_D$ in two limits where
they exhibit partonic behavior: rotating branes with energy
concentrated to cusp-like solitons; tensionless branes with energy
distributed over singletonic partons on the Dirac hypercone.
Evidence for a smooth transition from cusps to partons is found.
First, each cusp yields $D-2$ normal-coordinate bound states with
protected frequencies (for $p>2$ there are additional bound
states); and can moreover be related to a short open $p$-brane
whose tension diverges at the AdS boundary leading to a decoupled
singular CFT at the ``brane at the end-of-the-universe''. Second,
discretizing the closed $p$-brane and keeping the number $N$ of
discrete partons finite yields an $\msp(2N)$-gauged phase-space
sigma model giving rise to symmetrized $N$-tupletons of the
minimal higher-spin algebra $\mhso_0(D-1,2)\supset \mso(D-1,2)$.
The continuum limit leads to a 2d chiral $\msp(2)$-gauged sigma
model which is critical in $D=7$; equivalent \emph{\'a la}
Bars-Vasiliev to an $\msu(2)$-gauged spinor string; and
furthermore dual to a WZW model in turn containing a topological
$\widehat{\mso}(6,2)_{-2}/(\widehat{\mso}(6)\oplus
\widehat\mso(2))_{-2}$ coset model with a chiral ring generated by
singleton-valued weight-$0$ spin fields. Moreover, the two-parton
truncation can be linked via a reformulation \emph{\'a la}
Cattaneo-Felder-Kontsevich to a topological open string on the
phase space of the $D$-dimensional Dirac hypercone. We present
evidence that a suitable deformation of the open string leads to
the Vasiliev equations based on vector oscillators and weak
$\msp(2)$-projection. Geometrically, the bi-locality reflects
broken boundary-singleton worldlines, while Vasiliev's intertwiner
$\kappa$ can be seen to relate T and R-ordered deformations of the
boundary and the bulk of the worldsheet, respectively.

\vspace{0.5cm}

\noindent{\it Based in part on the material presented by P.S. at
the ``First Solvay Conference of Higher-Spin Gauge Theories'',
Bruxelles, May 12-14, 2004.}

{\vfill\leftline{}\vfill
\vskip  10pt
\footnoterule
{\footnotesize
\hoch{1} E-mail: \texttt{johan.engquist@teorfys.uu.se}
\\
\hoch{2} E-mail: \texttt{per.sundell@teorfys.uu.se}
\vskip -12pt}}

\setcounter{page}{1}

\pagebreak

\tableofcontents


\scs{INTRODUCTION AND SUMMARY}\label{sec:Intro}


\scss{General Discussion}

The most intriguing property of String Theory is the absence of a
global minimum principle. This is essentially a manifestation of
general covariance. The main obstacle in formulating the theory
covariantly is the fact that the standard quantization takes place
inside spacetime, whereas the covariant formulation -- whether
perturbative or not -- ought to involve a dynamical reconstruction
of spacetime.

This motivates revisiting the principles of gauging in field
theory, focusing on stringy forms of covariance, with the aim of
developing tools for extracting background independent physical
information. Indeed, such a program based on higher-spin
symmetries -- initiated and pursued in its early days mainly by
Fradkin and Vasiliev in a fashion quite independent from parallel
trends in String Theory -- led early to the Vasiliev equations in
$D=3$ and $D=4$ \cite{hssymmetries1,Vaseqs,Vasiliev:1999ba}, whose
higher-dimensional generalizations have started to be understood
more recently
\cite{Sezgin:2001zs,Vasiliev:2001wa,Sezgin:2001yf,Sezgin:2001ij,
Alkalaev:2002rq,Vasiliev:2003ev,Bekaert:2005vh,Sagnotti:2005ns,MixedYT}.

A key development has been the unfolding principle
\cite{dfre,unfolding,misha2,Vasiliev:2005zu}, according to which
the geometry is determined by a in general deformed algebra of
differential forms arising from gauging. In this spirit, here we
examine a ``doubling'' proposal formulated on a fiberbundle
$E[{\cal M},{\cal Z}]$ (for recent, related work on geometric
quantization see \cite{Barnich:2005ru}), whereby a quantum string
living in a rigid phase space ${\cal Z}$ produces a zero-form
master field -- the phase-space analog of the string field -- in
turn determining a one-form master field -- that contains a
vielbein as well as higher-spin gauge fields -- describing a
classical spacetime ${\cal M}$ through the unfolding principle.

This also reflects the spirit of open-closed string duality
\cite{'tHooft:1973jz} (see \cite{openclosed,gopakumar} for recent developments),
with the open-string side corresponding to the fiber theory and
the closed-string side to the generally covariant classical
master-field equation on the base-manifold ${\cal M}$ resulting
from projection of the quantum master-field equation on $E$. We
emphasize, however, that the fiber theory is ultimately a
topological closed string/open membrane.

The resulting master-field equations harmonize with the strong
duality principle that String Field Theory, or M-theory, is
completely void of free parameters. Viewed from this angle, the
Landscape \cite{Susskind:2003kw} would unfold via expansions of a
strictly non-perturbative Master Equation around solutions
exhibiting various unbroken symmetries in turn gauged via
current-photon couplings and subsumed in a ``kitchen-sink''
fashion into a truly unified symmetry algebra.

Here we emphasize a crucial -- though not completely
uncontroversial -- standpoint, namely that the string tension is a
background \emph{dependent} quantity determined by the solution,
whereby tensionless limits
\cite{schild,nocritdim,lindstrom,Bakas:2004jq} open windows into
the Landscape which may ultimately turn out to be wider than those
offered by traditional tensile strings and supergravity. Moreover,
since excitations of strings and other extended objects carry
general angular momenta, the unifying algebra must incorporate
higher-spin symmetries
\cite{hssymmetries1,Bergshoeff:1988jm,Gunaydin:1989um,
Sezgin:1998gg,Sezgin:2001zs,Sezgin:2001yf,Sezgin:2001ij,misha}
(see also \cite{Eastwood:2002su}). The developments initiated by
\cite{holography} later led to the appreciation that these
symmetries play a natural role within holographic space-time
reconstruction
\cite{ferfrohol,Sezgin:1998gg,Sundborg,Sezgin:2001zs,Sezgin:2001yf,
wittenspeech,konstein,Mikhailov:2002bp,Sezgin:2002rt,On,
Girardello:2002pp,Bianchi:2003wx,Bianchi:2005ze}.

This paper treats aspects of tensionless limits, and how the
resulting brane dynamics is described by phase-space strings
subsuming higher-spin gauge theories. Our main working hypothesis
is that the standard target space is actually part of a rigid
fiber, whose unbroken structure group emerges in the tensionless limit,
whereupon the unfolding yields the true spacetime, where, for
example, references can be made to holographic duals.


\scss{Higher-Spin Gauge Theory and Singleton
Strings}\label{sec:hsgt}


The higher-spin gauge theories are generalizations of pure AdS
gravity, in which the metric is accompanied by an infinite tower
of higher-spin fields \emph{and} special sets of lower-spin
fields. In the minimal setting, the spectrum consists of massless
symmetric Lorentz tensors of rank $s=0,2,4,\dots$ making up an
irreducible massless higher-spin multiplet, which can be ``packed
away'' into a convenient master field introducing an oscillator in
turn playing a crucial role for the formulation of the full
equation.

The theories are thus based on associative oscillator algebras
intimately related to singletons -- the remarkable ultra-short
unitary $\mso(D-1,2)$ irreps without flat-space limit -- and come
in varieties equipped with additional matrix groups. This points
to a geometrical realization of singleton-valued Chan-Paton factors via open topological
strings\footnote{The idea of realizing ordinary finite-dimensional
Chan-Paton factors as a two-dimensional topological theory of
fermionic ``dipoles'' dates back to \cite{CPfactor} and has more
recently been examined in using the superembedding approach to
non-abelian Born-Infeld theory \cite{Howe:2005jz}.}.

In Section \ref{sec:tos} we shall see that the $D$-dimensional
Vasiliev equations with vector oscillators and weak $\msp(2)$
projection \cite{Vasiliev:2003ev,misha} originate from a
topological open string on the phase space ${\Gamma}$ of Dirac's
$D$-dimensional hypercone \cite{dirac}, described by the
sigma-model action
\be S\ =\ \frac12 \int_{\Sigma}\left( D Y^{Ai}\wedge \eta_{Ai}
+\frac12 \eta^{Ai}\wedge \eta_{Ai}+\xi^{ij} F_{ij}\right)\
,\label{intro:cfSP}\ee
where $Y^{Ai}$ parameterize a $2(D+1)$-dimensional ambient phase
space ${\cal Z}$. The classical model has a critical point with
unbroken $\msp(2)$ gauge symmetry, where the zero-modes $(x^A,p^A)$
of $Y^{iA}|_{\partial\S}$ are confined to
$\Gamma$, \emph{viz.}
\be x^A x_A\ =\ x^A p_A\ =\ p^A p_A\ =\ 0\ .\label{diraccone}\ee
In a slight abuse of terminology, this implies that $\partial\S$
carries non-compact Chan-Paton factors valued in the scalar
singleton and anti-singleton (see Appendix \ref{sec:rep} for conventions)
\be \sfD\ \equiv \ \sfD(\e_0,(0))\ ,\qquad \widetilde\sfD\ \equiv \ \widetilde\sfD(-\e_0,(0))\ ,\qquad \e_0\ =\ {D-3\over 2}\
.\label{sfD}\ee
Here we stress that, unlike the introduction of ordinary compact Chan-Paton factors --
which are typically introduced by hand -- the singleton-valued factors arise geometrically from the topological action \eq{intro:cfSP}.

Treating the singleton as a point-particle
\cite{Marnelius,Flato:1980we,Siegel:1988ru}, the phase-space
formulation concerns deformations generated by insertions of
vertex operators on closed singleton worldlines in turn
representing traces (see Section \ref{sec:sp4}). The vertices are functions on ${\cal Z}$
subject to $\msp(2)$ gauge conditions \cite{Bars:2001ma}
projecting them to operators mapping $\widetilde\sfD$ to $\sfD$, \emph{i.e.}
elements of $\sfD\otimes \widetilde\sfD^\star$, representing external
\emph{massless two-singleton composites}. As we shall see, these states are given by the
Flato-Fronsdal formula \cite{Flato:1978qz,misha}
\be \sfD\otimes \sfD\ = \ \bigoplus_{s=0}^\infty
\sfD(2\e_0+s,(s))\ .\label{specintro1}\ee
This germ of an extended object dates back to the highly
influential work of Flato and Fronsdal
\cite{Flato:1978qz,Fronsdal:1980aa,Fronsdal:1978vb} who introduced
the notion of \emph{local and bi-local master fields}, \emph{i.e.}
functions on ${\cal Z}$ and ${\cal Z}\times {\cal Z}$, and was
later turned by Fradkin and Vasiliev \cite{hssymmetries1,Vaseqs}
into an elegant algebraic machinery used to write the full
higher-spin equations \cite{Vaseqs}, while the geometric aspects
have been pursued more recently mainly by Bars et al
\cite{Bars,Barsreview,Bars:2001ma}.

Here we unify the approaches using Cattaneo-Felder-Kontsevich's
stringy reformulation of phase-space quantum mechanics
\cite{Kontsevich:1997vb,Cattaneo:1999fm} -- leading to
\eq{intro:cfSP} -- refined further by an algebraic treatment of
embedding-field \emph{branch points where the boundary-singleton
worldline breaks to form an asymptotic two-singleton composite in
turn described by a bi-local operator reducing to a local operator
only at the linearized level}. This provides a natural geometric
realization of Vasiliev's algebraic structures and indeed
facilitates the construction of a deformation potentially giving rise
to the full Vasiliev equation.

To specify our results, let us first recall the $D$-dimensional
Vasiliev equations \cite{Vasiliev:2003ev,Sagnotti:2005ns}. These
can be written in a \emph{strongly $\msp(2)$-projected} form as
\cite{Sagnotti:2005ns}
\be \widehat d\,\widehat A+\widehat A\star \widehat A+\widehat
\Phi\star \widehat J'\ =\ 0\ ,\qquad \widehat
d\,\widehat\Phi+[\widehat A,\widehat\Phi]_\pi\ =\ 0\
,\label{vas1}\ee\be \widehat K_{ij}\star \widehat \Phi\ =\ 0\
,\label{vas2}\ee
where $\widehat d=d+d'$, with $d$ and $d'$ given by the exterior
derivatives on a \emph{commutative space-time manifold ${\cal M}$}
-- which we shall refer to as the \emph{unfold} -- and a
\emph{non-commutative phase space} ${\cal Z}$, respectively;
$\widehat A$ and $\widehat \Phi$ are \emph{adjoint and strongly
projected twisted-adjoint bi-local master fields} of total degree
$1$ and $0$, respectively, where the total degree is the sum of
form degrees on ${\cal M}$ and ${\cal Z}$; the master-field
components are functions of $(x^\mu,z^A_i)\in {\cal M}\times {\cal
Z}$ taking values in a fiber spanned by functions of
$y^A_i\in{\cal Z}$; $\widehat J'$ is a fixed \emph{intertwiner} of
degree $(0,2)$, and $[\cdot,\cdot]_\pi$ is the twisted-adjoint
representation map; and, $\widehat K_{ij}$ are dressed-up versions
of the $\msp(2)$ generators in \eq{diraccone}.

Originally, the equations were presented in the \emph{weakly
$\msp(2)$-projected form} \cite{Vasiliev:2003ev}
\be (\widehat F+\widehat \Phi\star \widehat J')\star\widehat M\ =\
0\ ,\qquad \widehat D\widehat\Phi\star\widehat M\ =\ 0\
,\label{vas}\ee
where the field strength and covariant derivative are the same as
in \eq{vas1}, and $\widehat M$ is a dressed-up $\msp(2)$-projector
-- or phase-space propagator -- obeying
\be \widehat K_{ij}\star\widehat M\ =\ 0\
.\label{widehatMintro}\ee
The weak projection induces shift symmetries that eliminate the
Lorentz traces in the fiber indices of the zero-form \emph{and}
the one-form, resulting in a non-linear system built on Fronsdal's
doubly traceless free-field equations \cite{Fronsdal:1978rb}. The
strong projection, on the other hand, only reduces the zero-form
letting the gauge fields adjust to the source and leading
\cite{dvh,bb,Sagnotti:2005ns} to a system built on
Francia-Sagnotti's geometric compensator form of the free-field
equations \cite{deWit:1979pe,Francia:2002aa}, containing
$\msl(D)$-tensor gauge fields closely connected to the naive
tensionless limit of flat-space string field theory. The precise
relation between the two types of projections remains to be
uncovered, and the details of the topological open string on the
Dirac hypercone may provide useful clues to this end.

Three salient features of the Vasiliev equations are:

\begin{itemize}

\item[\emph{1.}] \emph{Referring ${\cal M}$ to a $D$-dimensional spacetime}
(see Section \ref{sec:unf}), the equations \emph{yield ghost and
tachyon-free generally covariant field equations}, with a
well-defined weak-field expansion, and a strongly coupled yet
controllable derivative expansion, with fundamental length scale
equal to the radius of an unbroken anti-de Sitter vacuum.

\item[\emph{2.}] The equations are \emph{unfolded, i.e. written
in terms of differential forms only without any explicit
contractions of curved indices using the metric}, in turn inducing
manifest invariance under \emph{homotopy transformations of ${\cal
M}$ preserving the cohomological data contained in the master
fields} \cite{Engquist:2002gy,Vasiliev:2003ar}.

\item[\emph{3.}] The complete perturbative spectrum is contained in
the twisted-adjoint initial condition
$\widehat\Phi(x,z;y)|_{x=z=0}$.

\end{itemize}

The equations exhibit nonetheless a number of tantalizing
properties: first, the structure of the bi-local $\star$-product
algebra, \emph{viz.}
\be \widehat f(y,z)\star \widehat g(y,z)\ =\ \int {d^{2(D+1)}\y
d^{2(D+1)}\xi\over (2\pi)^{2(D+1)}} e^{i\y^{iA}\x_{iA}} \widehat
f(y+\x,z+i\x)\widehat g(y+\y,z-i\y)\ ,\label{intro:osc}\ee
such that $y^A_i$ and $z^A_i$, which commute to each other, have
the \emph{``skew'' mutual contractions}
\be y^A_i\star z^B_j-y^A_i z^B_j\ =\ i\e_{ij}\eta^{AB}\ ,\qquad
z^A_i\star y^B_j-z^A_i y^B_j\ =\ -i\e_{ij}\eta^{AB}\
.\label{skew}\ee
Second, the detailed structure of the \emph{intertwiner},
\emph{viz.}
\be\widehat J'\ =\ v_A v_B dz^{iA}\wedge dz_i^B \kappa\ ,\qquad
\kappa\ =\ \exp{\left(v_A v_B y^{Ai}z^B_{i}\right)}\ ,\qquad v^A
v_A\ =\ -1\ , \label{Jprime}\ee
serves a dual purpose, projecting also the linearized zero-form
$\Phi(x;y)$ to
$\Phi(x;y^{ai},y^i)\star\kappa|_{z=0}=\Phi(x;y^{ai},0)$ that
contains generalized Weyl tensors sourcing the spin
$s=1,2,3,\dots$ curvatures on ${\cal M}$ upon unfolding. Third,
shrinking ${\cal M}$ to a point, denoted here by priming, leads to
\emph{open-string-field-like equations on ${\cal Z}$ with a
consistent ``classical anomaly''}, \emph{viz.}
\bea  (d'\widehat A'+\widehat A'\star \widehat A'+\widehat
\Phi'\star \widehat J')\star\widehat M'&=&0\ ,\label{vas1pr}\\
(d'\widehat \Phi'+[\widehat A',\widehat \Phi']_\pi)\star\widehat
M'&=&0\ ,\label{vas2pr}\eea
with the BRST-like exterior derivative
\be d'\ =\ dz^{Ai}{\partial\over\partial z^{Ai}}\
.\label{dprime}\ee

In the geometric realization, the two end-points of the string
will be coordinatized by $y^A_i$ and $z^A_i$. The topological
gauge symmetries allow observables to depend on the center-of-mass
but not the relative distance between the end-points, which is
equivalent to taking the linearized BRST-operator to be given by
\eq{dprime} with $dz^A_i$ identified as one of the non-zero-mode
oscillators of the shift-symmetry $C^{Ai}$-ghost. Moreover, the
topological Green functions, which are essentially locally
constant phase factors (see Section \ref{sec:osc}), contain
long-range correlations between $y^A_i$ and $z^A_i$ leading to
\eq{skew}. Finally, we propose that the intertwiner $\kappa$
arises in the map taking an operator at $\partial\S$ representing
an initial state of the radial-ordered evolution on $\S$ to a
corresponding operator inserted into the path-ordered evolution
along $\partial\S$ (see Section \ref{sec:kappa}).

We are led to propose in an {\it ad hoc} fashion (see Section \ref{sec:vas}) that \emph{the
internal part of the weakly projected Vasiliev equations follows
from demanding exact marginality of the phase-space observables}
\be \widehat {Tr}_\pm \left[ T \left(\exp \oint_{\partial\S}
\widehat A'\right)~ R\left(\exp i\int\!\!\!\!\int_{\S} v_A v_B
dY^{iA}\wedge dY^B_i\widehat \Phi'\right)\right]\
,\label{observables}\ee
where the master fields are integrated in one of their arguments,
identified on $\partial\S$ with $z^A_i$, while the other argument
is attached to a fixed base-point on $\partial\S$, identified with
$y^A_i$. Thus, the consistent classical anomaly in \eq{vas1pr} is
an ``inflow'' from the bulk, while the non-abelian structures
arise from a Wilson-loop on the boundary. Moreover, the fact that
the Vasiliev equations are built from a one-form and a zero-form
but no higher-rank forms is the result of the world-sheet geometry
and that a Lorentz invariant Wess-Zumino potential can be built
from $d^2 Y$ and the zero-form. In this spirit, it is natural to
expect that the original 4D spinor-oscillator Vasiliev equations
originate from a topological open spinor string, as we shall
discuss in Section \ref{sec:4d}.

Our arriving at \eq{vas1pr} and \eq{vas2pr} starting from
\eq{observables} relies on a number of assumptions: \emph{a)} that
the classically consistent truncation of the $\msp(2)$-triplet
sector in \eq{intro:cfSP} -- which results in a free world-sheet
theory subject to a subsidiary $\msp(2)$ constraint -- can be
implemented at the quantum level by means of an insertion of
$\widehat M$ into the free-field trace; \emph{b)} that the
observables can be constructed entirely in terms of the embedding
field $Y^A_i$; \emph{c)} that it is consistent to drop terms on
$\partial\S$ that are exact with respect to the shift-symmetry
BRST operator -- which contain non-zero-mode oscillators.

Clearly, assumption \emph{(a)} incorporates the weakly projected
formalism from the outset, while the strongly projected equation
relates more naturally to correlators with un-amputated external
legs. Another subtlety resides in the implementation at the full
level of the $\msp(2)$-symmetry of the linearized theory. A
rigorous treatment of the $\msp(2)$-gauging may furthermore result
in a critical dimension. On the one hand, this would be
unexpected, since the model is a reformulation of a
point-particle, but on the other hand the observable involves a
genuinely two-dimensional surface term.

Due to the close analogy with ordinary open-string field theory,
we expect assumption \emph{(b)} to be valid at the classical
level, while the quantization of the open string will require a
thorough understanding of moduli associated with shift-symmetry
ghosts. The truncation \emph{(c)} is ideally also
consistent and simply amounts to the dressing of the
$\msp(2)$-projector described above, as otherwise there would
arise a puzzle, in that the oscillator corrections are not
suppressed by the analog of $\alpha'$ due to the topological
nature of \eq{intro:cfSP}. One possibility is, of course, that
there are many different separately consistent but mutually
inconsistent master field equations that can be built on top of
the linearized content of the open string.

Conceptually speaking, the possibility to define Higher-Spin Gauge Theory
as the ``Theory of Phase-Space Strings'' may provide a powerful constructive
principle. At this moment, however, we are forced to leave this issue open, but we hope eventually to gain such an understanding of the microscopic nature of the open string interactions that we can derive \eq{observables} and its consequences from first principles, {\it i.e.}~an open string vertex and BRST operator. Nonetheless, we think it is clear, and we want to emphasize, that \emph{the quantization
takes place in phase space while the chronologically ordered
presentation of the physical information in the string field is
given in the space-time unfold ${\cal M}$}, and we shall touch
this issue in somewhat more detail in Section \ref{sec:unf}. A
list of challenging problems include: \emph{i)} the formulation of
quantum consistent phase-space and spinor strings, involving full
supersymmetric higher-spin gauge theories in $D>4$ (where
presently only partial results exist
\cite{Sezgin:2001yf,Alkalaev:2002rq,Sezgin:2002rt,misha}) and
incorporation of massive states via topological chiral closed
strings/open membranes (which we shall touch below); \emph{ii)}
the construction and examination of classical solutions to the
master field equations
\cite{Prokushkin:1998bq,Bolotin:1999fa,solutions}, requiring
deformed oscillators \cite{Prokushkin:1998bq,solutions}, charges
and on-shell actions \cite{Vasiliev:2005zu}; \emph{iii)} various
aspects of the doubling approach, such as the manifestly
higher-spin covariant formulation of the unfolded geometry, the
relation between unfolded and Lagrangian quantum corrections, and
holographic aspects.

Our proposal for relating the Vasiliev equation to singleton
deformation quantization, relies on open-string tree-level
amplitudes. At this level, the deformation quantization requires a
well-defined notion of hermiticity of the master fields, while it
does not require unitarity of the underlying singleton
representation. The formalism should therefore apply equally well
to general space-time signatures. Indeed, the vector-oscillator
form of the Vasiliev equations exist in spacetimes with more
general signatures \cite{Vasiliev:2003ev,Sagnotti:2005ns}.

Ultimately, in analogy with ordinary supergravity and open-string
theory, the quantum theory should become part of a fuller theory
with massless two-singleton as well as massive multi-singleton
states -- admitting the Vasiliev equations as a classically
consistent truncation. Indeed, the resemblance between the
spectrum of massless fields \eq{specintro1} and the tensionless
limit of the leading flat-space Regge trajectory
\cite{nocritdim,pashnev,Francia:2002aa,Sagnotti:2003qa} can be
made into a more precise correspondence, which includes massive
multi-singletons and higher trajectories, by using supersymmetry
and holography arguments
\cite{Polyakov:2001af,deser,Sezgin:2002rt,Bianchi:2003wx,Beisert:2004di}.
Clearly, this motivates establishing a more direct link between the phase-space approach
and the tensionless limit of (bosonic) $p$-branes in anti-de Sitter spacetime.


\scss{Tensionless Limits}


To follow extended objects to small tension, it is useful to
picture the brane phase space covariantly as the space of all
classical solutions. As the tension of a classical rotating closed
brane is switched off adiabatically, the centrifugal force causes
the energy density to accumulate at ultra-relativistic
\emph{cusps} connected to the center-of-mass region by thinly
stretched portions of the brane
\cite{Gubser:2002tv,Kruczenski:2004wg}. As we shall see, this
naive argument fails in flat spacetime, while it holds in anti-de
Sitter spacetime, where the background curvature exerts a tidal
force acting together with the centrifugal force to induce an
enhanced accumulation of energy-momentum to the cusps. This leads
to a potential well in the normal-coordinate mass term, as we
shall describe in more detail below. In the case of strings, the
well contains $D-2$ bound states, giving rise to additional
bound-state oscillators in the normal-coordinate field theory, out
of which $D-3$ have protected frequencies and the remaining one
has a small fixed anomalous dimension. Remarkably, this result
extends to membranes, while the potential well contains additional
bound states for $p>2$.

Thus, the negative cosmological constant is forced if one requires
that rotating branes configurations fill a distinct \emph{partonic
region of phase space}, parameterized by
\be \bigcup_{N=2}^\infty \{X^m(\t;\x)\}_{\x=1}^N\ ,\ee
where $N$ runs over the number of partons and $X^m(\t;\x)$ denote
their space-time trajectories. In the ``dilute gas''
approximation, this region is closed under time-evolution
\emph{i.e.} the total Hamiltonian
\be H\ \simeq\ \sum_{N=2}^\infty H_N(\{X^m(\t;\x)\})\ ,\ee
where the omitted subleading interactions are off-diagonal
elements suppressed by inverse powers of semi-classical parameters
followed by multi-body interactions suppressed by powers of the
space-time Planck constant.

Focusing on the leading part, any amount of tension -- no matter
how small it is -- leads to attraction between the cusps keeping
them on large circular orbits. Thus, in addition to
point-particle-like mass-shell conditions, the cusps obey
additional Gauss'-law-like constraints, leaving $D-2$ physical
normal-coordinate oscillators, since a cusp at the end of a long
stretched portion of a $p$-brane is free to move in $D-2$ overall
transverse directions.

The results referred to above point to a \emph{transition from
cusps to singletons as the tension is switched off}. There are
several ways of understanding this: first, if the tension is
switched off adiabatically, then each cusp becomes a partonic lump
following a massless geodesic confined to a $(D-2)$-dimensional
hypersurface, identifiable as a short open $p$-brane attached to a
$(D-2)$-brane in a decoupling limit in which the effective
$p$-brane tension is sent to infinity resulting in massless quanta
on the $(D-2)$-brane at the end-of-the-universe
\cite{Bergshoeff:1988jm,Duff:1989ez,Batrachenko:2002pu}. Lending
the terminology of
\cite{McGreevy:2000cw,Grisaru:2000zn,Hashimoto:2000zp}, the
$(D-2)$-brane solution is a \emph{giant vacuum} with a
\emph{singular conformal field theory} \cite{Seiberg:1999xz}
living on it. The resulting \emph{tensionless spectrum of
space-time one-particle states consists of symmetrized
multipletons\footnote{We use the terminology of
\cite{Bianchi:2003wx,Beisert:2004di}.}}
\be {\cal S}\ =\ \sum_{N=1}^\infty [\sfD^{\otimes N}]_{\rm symm}\
,\label{introcalS}\ee
in agreement with the above interpretation and the fact that the
normal-coordinate realization of the soliton gas obeys Bose
symmetry in the leading order of the semi-classical expansion. In
the tensionless limit, mixed multipletons should arise as
space-time multi-particle states. These are shifted into the
spectrum of one-particle states by tensile deformations, such as
those making up the non-trivial part of stringy spin-chains
\cite{spinchain}. We also note that additional bound states, do
not arise in the maximally supersymmetric cases, where Type IIB
open strings end on $D3$-branes at the boundary of $AdS_5 \times
S^5$, and $M2$-branes end on $M2$ or $M5$-branes at the boundary
of $AdS_4\times S^7$ or $AdS_7\times S^4$, respectively.

Second, the discretized closed $p$-brane provides a manifestly
weakly coupled partonic description in the tensionless limit.
Indeed, taking the \emph{singular tensionless limit}, \emph{viz.}
\be L\mu\rightarrow 0\ ,\qquad T_p\mu^{-p-1}\rightarrow 0\ ,\ee
where $\mu$ is a lattice constant, and restricting to a sector
with a \emph{fixed number $N$ of partons}, we find the
$\msp(2N)$-gauged sigma model \cite{Gomis:1993pp}
\be S\ =\ \frac14 \int Y^{I A}DY_{IA}\ ,
\label{introSN}\ee
where $I=(i,\x)$, $\xi=1,\ldots,N$. In the quantum theory,
we impose each $\msp(2)_{(\x)}$ strongly on physical states leaving the
off-diagonal generators to vanish weakly. Thus, each parton is a scalar
singleton on the hypercone \eq{diraccone}. The global $\msp(2N)$-invariance
enforces the symmetrization leading to \eq{introcalS}.

However, unless $D=3$ mod $4$, the wave-functions exhibit a
\emph{global anomaly} under large $Sp(2N)$ gauge
transformations in the form of reflections in the apex of the
Dirac hypercone. The functions also present subtleties in the form
of $\delta$-function distributions, that we conjecture combine with the analytic
non-normalizable part into
normalizable states -- which we think of as squeezed versions of
normalizable scalar-field mode-functions in ordinary anti-de
Sitter spacetime.


\scss{Singleton Closed String/Open Membrane}


The natural interpretation is that the multi-singleton system
lives in the \emph{phase space} of the Dirac hypercone, where the
two-parton sector yields Vasiliev's equations as discussed above and
more generally \eq{introcalS} has a natural interpretation as a
generalized Chan-Paton factor.

In the continuum limit $N\rightarrow\infty$ the discrete
tensionless $p$-brane \eq{introSN} becomes a 2d chiral
$\msp(2)$-gauged phase-space sigma model \cite{lindstrom}, which we
find to be critical in $D=7$ where it is furthermore dual to a
WZW-model based on $\widehat \mso(6,2)_{-2}$ in turn containing
the coset model
\be \widehat \mso(6,2)_{-2}/(\widehat\mso(6)\oplus
\widehat\mso(2))_{-2}\ ,\ee
with vanishing Virasoro charge, giving rise to a chiral ring
generated by $\sfD$-valued weight-$0$ spin fields. We identify
this topological closed string as the proper framework for
computing with the generalized Chan-Paton factor.

We expect the actual space-time dynamics to arise via deformations
of the closed string governed by a
topological open membrane equipped with a suitable generalization
of the bi-local structures of the open string. Its pursuit is a
truly challenging problem, whose resolution we believe will
contain important new stringy physics.

\scs{FROM CUSPS TO SINGLETONS}\label{sec:cusps}


In this Section we analyze anti-de Sitter analogs of states on low
Regge trajectories, that is, states with spin $S$ and small
excitation energy $E-S$.

We shall first consider the semi-classical representation as long
rotating $p$-branes with energy and spin concentrated to cusps
\cite{Gubser:2002tv,Frolov:2002av,Tseytlin:2002ny,Kruczenski:2004wg},
giving rise to wave-functions depending on finite sets of
oscillators. Roughly speaking, the physical role of the extended
part of the brane is limited to constraining the cusps to angular
motion, resulting in a dilute ``gas'' relatively insensitive to
the bare $p$-brane tension.

To exhibit their ``singletonic'' nature, we shall then examine an
alternative description of the partons as short open $p$-branes
attached to a dual $(D-2)$-brane in $AdS_D$. This system depends
on the asymptotic $p$-brane tension, rather than the bare tension,
and in a decoupling limit \cite{Seiberg:1999xz}, where the bare
tension vanishes, it becomes a $(D-1)$-dimensional conformal field
theory -- living on the brane at the end-of-the-universe -- where
each parton is realized as a proper singleton.

\scss{Partonic Regions of Brane Phase Space}

We are interested in the dynamics of closed $p$-branes described
by the ordinary Nambu-Goto action
\be S_{\rm NG}\ =\ -T_p \int d^{p+1}\sigma \sqrt{-\det{g}}\
,\label{eq:ng}\ee
with ``bare'' tension $T_p$ and induced metric
$g_{\alpha\beta}=\partial_\alpha X^m
\partial_\beta X^n G_{mn}(X)$, where $G_{mn}$ is the metric of anti-de Sitter
spacetime with radius $L$.

The phase-space has a covariant meaning as the space of all
classical solutions, which is suitable for diffeomorphism
invariant field theories, since it does not rely on fixing the
topology. Moreover, it avoids the strong-coupling problems at
small tension that arise from parameterizing phase space using
maps from a smooth $p$-dimensional volume to target space.

As in ordinary field theory, one then considers solutions with
space-time energy-momentum concentrated to portions of the
worldvolume that can be made asymptotically relatively small in
some parametric limit. Solutions containing $N$ solitons can then
be faithfully represented by their positions,
\be \{X^m(\tau;\xi)\}_{\xi=1}^{N}\ .\ee
The resulting partonic regions of phase space give rise to sectors
of wave-functions of the form
\be \Psi_N(\{X^m(\xi)\})\ ,\ee
subject to physical-state conditions and Bose symmetry.

We shall consider partons obtained by minimizing the energy $E$ in
parameter families of rotating branes while keeping the angular
momentum $S$ fixed \cite{Gubser:2002tv,Kruczenski:2004wg}. The
variational parameters represent adiabatic deformations in which
the centrifugal force pushes the energy-momentum towards
relativistic portions of the worldvolume, where it opens up into a
folded shape that we shall refer to as a \emph{cusp}. In general,
the cusps attract each other, which is not a problem for
stability, however, provided that the system has non-vanishing
impact parameters \cite{Kruczenski:2004wg}.

The motion of the cusps is governed by their relativistic inertia
as well as the inward pull from the tension. In global
coordinates, with a radius $r$ vanishing at the center-of-mass and
$D-2$ angles, this yields angular motion with suppressed radial
fluctuations, resulting in that each cusp excites $(D-2)$ degrees
of freedom. This behavior is reminiscent of that of anti-de Sitter
singletons, and the partonic behavior indeed requires an
\emph{enhanced accumulation} of energy-momentum, exhibited in
anti-de Sitter spacetime but not in flat spacetime.

The size of the fluctuations around a solution $\overline X^m$ is
governed by the \emph{effective tension}
\be T_{\rm eff}L^{p+1}\ =\ T_p L^{p+1}\sqrt{-\det \bar g}\ ,\ee
which vanishes on the cusps and is of order $T_pL^{p+1}$ not too
far away from them. Thus, in the limit of large $T_pL^{p+1}$ the
functional integral collapses to an integral over $X^m(\t;\x)$.

This contrasts to ordinary field theories, where non-perturbative
degrees of freedom in general decouple in the free limit. A
related subtlety presents itself in that the size of the cusps
remains finite and of order $L$, essentially due to the fact that
the angular velocity is small and of order $1/L$. However, the
important geometric parameter is the ratio between the size and
the relative separation of the cusps, which can be made
arbitrarily small by considering states with sufficiently large
spin. This leads to a partonic region of the $p$-brane phase space,
consisting of well-defined solitons with small and fixed energy $E-S$,
which should dominate the classical charges as well as the
functional integral.

Having made these preliminary remarks, let us now turn to a more quantitative analysis.


\scss{Rotating Strings}


Let us consider the folded and rotating closed string in
$(AdS_D)_L$ given by \cite{Gubser:2002tv}
\bea ds^2&=& -\cosh^2\frac {r}{L}~ dt^2+dr^2+ L^2
\sinh^2\frac{r}{L} ~d\O_{D-2}^2\ ,\\ t&=&t(\t)\ ,\qquad r\ =\
r(\s)\ ,\qquad \phi\ =\  {\omega_0 t\over L}\ ,\nn\\\qquad
\omega_0\ &=&\ \o(r_0)\ \equiv \ \coth \frac{r_0}{L}\
,\label{eq:omega}\eea
where $\phi$ is an azimuthal angle on $S^{D-2}$; $2r_0$ is the
proper length of the string, and the map $r:\s\mapsto r(\s)$ has
winding number $1$. The induced worldvolume metric reads
\be ds^2(g)\ =\ \mu^2( -dt^2+dy^2)\ ,\qquad dr^2\ =\ \mu^2 dy^2\
,\qquad \mu^2\ \equiv \ \cosh^2\frac{r}{L}-\omega^2_0
\sinh^2\frac{r}{L}\ ,\ee
where $y(r)=\int_r^{r_0} \mu^{-1} dr$, that is
\be {dy\over dr}\ =\ -{1\over \mu}\ ,\qquad e^{{r-r_0\over L}}\ =\
{1\over \cosh {y\over L}}+{\cal O}(e^{-{2r_0\over L}})\ ,\ee
so that $y(r_0)=0$ and $y(0)\ =\ r_0+L\log 2+{\cal
O}(e^{-{2r_0\over L}})$.

The classical energy and spin are given by\footnote{Our choices of
conventions for $\mso(D-1,2)$ are given in Appendix
\ref{sec:rep}.}
\bea E_{\rm cl}&=&4TL \int_0^{r_0+L\log 2}
dy\cosh^2\frac{r}{L}+{\cal O}(e^{-{2r_0\over L}})\ ,\\ S_{\rm
cl}&=&4TL\omega_0 \int_0^{r_0+L\log 2} dy~
\sinh^2\frac{r}{L}+{\cal O}(e^{-{2r_0\over L}})\ .\eea
For $r_0\gg L$, a fraction $fE_{\rm cl}$ arises from a small
interval of $y$-values of width $\Delta\sim -{L\over
2}\log(1-f)\ll r_0$, \emph{i.e.} the classical energy and spin are
dominated by the contributions from two localized cusps,
$\xi=1,2$, of fixed width $\Delta \sim L\ll r_0$ for fixed $f$.
The spin and energy are thus given to leading order by a
\emph{sum} over contributions from each cusp,
\be E_{\rm cl}\ \approx\ \sum_{\xi=1,2} E(\xi)\ ,\qquad S_{\rm
cl}\ \approx\ \sum_{\xi=1,2} S(\xi)\ ,\label{ecl}\ee
with $E(\xi)$ and $S(\xi)$ determined by the \emph{local}
functional form of the energy and spin densities at the cusps.
Since these agree to leading order, it follows that
\be E(\xi)\ =\ S(\xi)\ ,\ee
and hence that
\be E_{\rm cl}\ \approx\ S_{\rm cl}\ .\ee
The proportionality between $E(\xi)$ and $S(\xi)$ implies that the
cusps have generalized angular momenta that are ``light-like'' in
the sense that
\be \frac12 M_{A}{}^C(\xi)M_{BC}(\xi)\ =\ 0\ .\ee
For later reference, we note that
\bea E_{\rm cl}&\approx &S_{\rm cl} \;\;\sim\;\; TL^2~e^{2r_0/L}\
,\eea
and that the leading correction to the energy is logarithmic in
$S_{\rm cl}$, \emph{viz.}
\bea E_{\rm cl}-S_{\rm cl} &\sim & TL^2 \log {S_{\rm cl}\over
TL^2}\ ,\eea
\emph{i.e.} the cusps interact via a linear potential.

More general spiky string solutions have been given in
\cite{Kruczenski:2004wg}. These contain $N$ cusps rotating
co-planarly, each carrying energy $E_{\rm cl}/N$ and spin $S_{\rm
cl}/N$. These highly symmetric configurations have vanishing
impact parameters, and are therefore unstable against
perturbations \cite{Kruczenski:2004wg}. Solutions with randomly
distributed impact parameters will describe stable clusters of
cusps, with total angular momentum obeying the addition rule
\bea M_{AB}&\approx& \sum_{\xi=1}^N M_{AB}(\xi)\
,\label{addrule}\eea
where $M_{AB}(\xi)$ are light-like angular momenta, and the
subleading terms are due to interactions, which we shall comment
on below.

In the flat space-time limit $L\rightarrow \infty$, one finds
scale-invariant energy and spin densities, proportional to
$(1-(r/r_0)^2)^{-1/2}$ and $(r/r_0)^2(1-(r/r_0)^2)^{-1/2}$,
respectively. Hence $\Delta\sim r_0$, so the densities do not form
lumps, the energy and spin are related non-linearly as fixed by
dimensional analysis, and the partonic picture is lost.

To examine the cusps in more detail, we consider the expansion of
$S_{\rm NG}[X]$ in inverse powers of $TL^2\gg 1$ around the folded
string solution $\overline X^m$,
\be S_{\rm NG}[X]\ =\ \bar S_{\rm NG}+\sum_{p=2}^\infty
S_p[\varphi]\ ,\qquad \varphi^{a}\ =\ T^{1/2}(X^m-\overline
X^m)\overline E_{m}{}^a\ ,\label{nce}\ee
where $\varphi^a$ is a canonically normalized fluctuation field.
The quadratic part can be expressed using normal coordinates as
\bea S_2[\varphi]&=&-\frac12\int d^2\s \sqrt{-\bar
g}\left(\nabla^\a\varphi^a\nabla_\a\varphi_a- E^{\a a} E_{\a}{}^b
R_{ac,bd}\varphi^c\varphi^d\right.\\\nn&&\left.+
\nabla^\a\varphi^a\nabla^\b\varphi^b(2E_{\a [a} E_{\b b]}-\bar
g_{\a\b}  E^\c{}_a E_{\c b})\right)\ ,\label{eq:normcoord}\eea
where $\nabla_\alpha$ contains the space-time Lorentz connection
and $E_\a{}^a=\partial_\a \overline X^m \overline E_m{}^a$. The
fluctuations $E_\a{}^a\varphi_a$ contain pure-gauge fluctuations
and a zero-mode describing changes in the physical length of the
string, that we shall not consider here. One finds that
$S_2=S_2[\varphi_{i'}]+S_2[\varphi^{\underline
t},\varphi^{\underline r},\varphi^{\underline \phi}]$, with
$\varphi_{i'}$, $i'=1,\dots,D-3$, containing the fluctuations tangent to $S^{D-3}$,
for which $\nabla_\a\varphi_{i'}=\partial_\alpha\varphi_{i'}$.
The $(\varphi^{\underline t},\varphi^{\underline
r},\varphi^{\underline \phi})$-sector contains one physical
fluctuation field, $\tilde\varphi$.

The end-result reads \cite{Frolov:2002av,Tseytlin:2002ny}
\be S_2[\varphi]\ =\ - \frac 12\sum_{i=1}^{D-2} \int dt
 dy \left((\partial \varphi_i)^2+
{q_i\mu^2+q'_i\mu^{-2}\over L^2}(\varphi_i)^2\right)+S'_2\ ,\ee
where $\varphi_i=(\varphi_{i'},\tilde\varphi)$ ($i=1,\ldots,D-2$); the parameters $q_i$ and $q'_i$ are given by
\bea q_{i'}&=&2\ ,\qquad q'_{i'}\ =\ 0\ ,\\
\tilde q&=&4+\tilde q'\ ,\label{qprimed-2}\eea
with $\tilde q=q_{D-2}$ and $\tilde q'=q_{D-2}'$; and $S_2'$ contains gauge artifacts and the radial zero-mode. The
parameter $\tilde q'$ is the coefficient of the ${\cal
R}_{(2)}$-term, which receives corrections from the conformal
anomaly of the Nambu-Goto action, as explained in
\cite{Frolov:2002av}, where it was also proposed that $\tilde
q'=0$ in the case of critical strings.

Expanding the physical fields as
\be \varphi_{i}\ =\ \sum_{k}e^{{i\nu_{i,k} t\over L
}}\varphi_{i,k}(y){a^\dagger_{i,k}\over \sqrt{\nu_{i,k}}}+{\rm
h.c.}\ ,\ee
where we have assumed that all frequencies are non-vanishing, one
finds that the mode functions obey the Schr\"odinger problem
($-r_0\leq r\leq r_0$)
\be \left(-L^2{d^2\over dy^2}+
q_i\mu^2+q'_i\mu^{-2}\right)\varphi_{i,k}\ =\
\nu^2_{i,k}\varphi_{i,k}\ ,\qquad \left.{d\varphi_{i,k}\over
dy}\right|\ =\ 0\ .\ee
In the case $q'_i=0$, there are two small potential wells of width
$L$ centered around $r=\pm r_0$, that become well-separated for
$r_0/L\gg 1$, and reduce to exactly solvable \emph{P\"oschl-Teller
potentials} \cite{Cooper:1994eh}\footnote{The Hamiltonian has a
superpotential, $H=A^+_{\a_0}A^-_{\a_0}+C_0$ with
$A^{\pm}_\a=\mp{d\over dy}+W_\a$, $W_\a=\a\tanh y$,
$\a_0=C_0=(\sqrt{1+4q}-1)/2$, where the oscillators obey the
deformed Heisenberg relation $A^-_\a
A^+_\a=A^+_{f(\a)}A^-_{f(\a)}+R_\a$, with $f(\a)=\a-1$ and
$R_\a=2\a-1$. The normalizable solutions to $(H-\l)\varphi=0$ are
given by $\varphi_{(n)}={\cal N}_{n} A^+_{\a_0}A^+_{f(\a_0)}\cdots
A^+_{f^{n-1}(\a_0)}\exp(-\int^y W_{f^n(\a_0)})$ and
$\l_n=C_0+R_{\a_0}+\cdots +R_{f^{n-1}(\a_0)}$ for
$n=0,1,\dots,[\a_0-1]$.} in the limit $r_0/L\rightarrow \infty$,
\be q_i\mu^2\ =\ q_i\Big(1-{1\over \cosh^2 \frac{y}{L}}\Big)+{\cal
O}(e^{-2r_0/L})\ .\label{eq:mu}\ee

\begin{figure}[t]
    \begin{center}
        \includegraphics[width=10cm,bb=0 0 700 500]{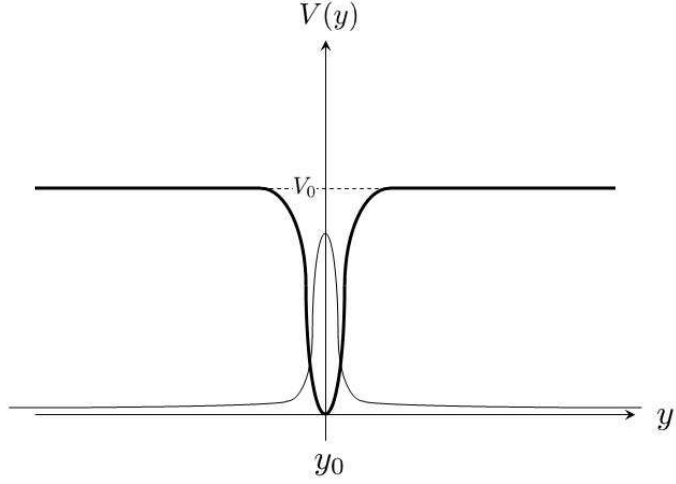}
    \end{center}
                \vspace{-1cm}
        \caption{{\it The P\"oschl-Teller potential}: The $\sigma$-dependent
    mass-term on the rotating string has height $\sim 1/L$ a small
    ``well'' of width $\sim L$ which fits precisely one even bound state.}
    \la{boundstate}
\end{figure}

For $q\leq 6$ this potential admits precisely one even bound
state, given by
\be \varphi\ =\ {\cal N}\left({1\over \cosh
\frac{y}{L}}\right)^{\nu^2}\ ,\qquad \nu^2\ =\
\frac12(\sqrt{1+4q}-1)\ ,\ee
where the normalization ${\cal N}\sim L^{-1/2}$. Thus, the
frequency of the \emph{overall-transverse bound-state modes} is
given by
\bea\nu_{i'}&=& 1\ ,\eea
The longitudinal bound-state problem for $i=D-2$ is highly
sensitive to the precise value of the parameter $\tilde q'$: a
negative value deepens the potential well so that there may arise
additional bound states, while a positive value makes the well
more shallow so that bound states may be lost. Assuming
$\tilde q=4$ we find the \emph{longitudinal bound-state mode}
\be \tilde \nu^2\ =\ \frac12(\sqrt{17}-1)\ .\label{nud-2}\ee

For finite $r_0/L$, the bound-state wave functions $\varphi(\x)$
($\x=1,2$) have finite small overlaps. As a result, the
Hamiltonian is diagonalized by wave functions of the approximate
form $\varphi_\pm=\frac1{\sqrt 2} (\varphi(1)\pm\varphi(2))$,
while there are only small corrections to the frequencies.

The \emph{wave modes} are approximately given by
\be \varphi_{i,k}\approx {\cal N}_{i,k}~\cos{ky\over r_0}\ ,\qquad
\nu^2_{i,k}\ =\ q_i+{k^2L^2\over r_0^2}\ ,\qquad k=0,1,\dots\ ,\ee
with ${\cal N}\sim r_0^{-1/2}$, and can be divided into
\emph{short waves} with $k\gg r_0/L$, $\nu_k\approx kL/r_0$, that
decouple in the limit $r_0/L\rightarrow \infty$; \emph{long waves}
with $k\ll r_0/L$, $\nu_k\approx \sqrt{2}(1+k^2L^2/4r_0^2)$; and
\emph{intermediate waves} with $k\sim r_0/L$, $\nu_k\sim 1$. We
note that both long and intermediate waves have frequencies
approximately equal to the excitation energy of the
P\"oschl-Teller bound states.

Analogously, the normal-coordinate fluctuations around the
symmetric $N$-cusp solutions \cite{Kruczenski:2004wg} give rise to
$N(D-2)$ bound-state oscillators $a^\dagger_{i}(\xi)$,
$\xi=1,\dots,N$, splitting into $N\times (D-3)$ overall-transverse
states with frequency $1$, and $N$ longitudinal states with
frequency $\tilde\nu$. Thus the $N$-cusp ground-state energy,
$E=\sum_{n=0}^\infty E_n$, with $E_0=E_{\rm cl}$ and
$E_n=(TL^2)^{1-n}{\cal E}_n$, receives a $1$-loop cusp correction,
$E_1=E_1({\rm waves})+E_1({\rm cusps})$, given by the zero-point
energy contribution
\be E_1({\rm cusps})\ =\ N\times \left( D-3 +
\tilde\nu\right)\times {1\over 2}\ .\ee
Remarkably, even though we are examining the string
in the limit $TL^2\gg 1$, the contribution from each cusp is given by the
singleton ground-state energy $\e_0$ given in \eq{sfD}, plus a
finite anomalous part\footnote{From the holographic point-of-view, this
indicates that insertions of gauge-covariant $D_\perp$ derivatives
into light-like bilinear Wilson lines $W$ receive only very mild
anomalous corrections at strong coupling: if $W$ is built from two
scalar fields, $S$ light-like $D_+$, $K$ $D_\perp$, and $\widetilde K$
$D_-$, then the string result indicates that
$\D(W)-(D-3+S+K+\widetilde K)=\tilde \nu+(\tilde\nu-1)\widetilde K
+(f_1(\l)+f_2(\l){1\over S}+\cdots)\log S+g_1(\l;K){1\over
S}+g_2(\l;K){1\over S^2}+\cdots$.}.

\scss{Dilute-Gas Approximation}\label{sec:gas}

In this section we provide a heuristic interpretation of the results
obtained in the previous section. To begin with, in \eq{nce} the
interactions contained in $S_p$ for $p>2$ are
built from multiple gradients $\partial_\a \overline X^m$ and
$\overline\nabla_\a \varphi_i$ contracted with $\bar g^{\a\b}$,
and blow up inside the P\"oschl-Teller potential wells. Combined
with the variational principle, this results in modified boundary
conditions, so that actual physical quantities remain finite in
the classical theory.

\begin{figure}[t]
\begin{center}
        \includegraphics[width=10cm,bb=0 0 700 500]{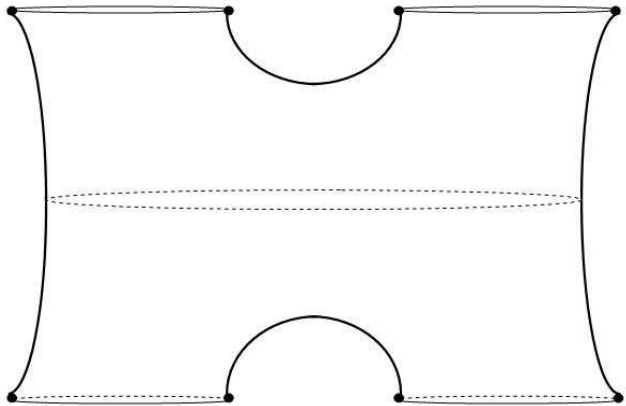}
\end{center}
\vspace{-1cm}
\caption{{\it Long-string interactions}: Two long strings with
cusps at their ends interact. Two cusps annihilate while emitting
waves that later recombine into a new pair of
cusps.}\label{fig:long}
\end{figure}

Alternatively, the interactions\footnote{The normal-coordinate
Hamiltonian does not incorporate splitting and joining of strings,
which are processes suppressed by powers of the closed-string
coupling $g_s$. In the case of spiky strings, the underlying
``mechanical'' forces are largest in the interior, which is where
they tend to break most easily \cite{Russo,Kruczenski:2004wg}.
Thus, the first subleading order in $g_s$ contains processes in
which $N$-cusp solitons $\leftrightarrow$ $(N_1,N_2)$-cusp
two-body solitons with $N_1+N_2=N+2$ (see Fig.~\ref{fig:long}).}
can be obtained using the variational parameters, \emph{e.g.} by
considering folded strings of length $l$ rotating with frequency
$\o<\o(l)$ such that $\mu(r,\o)|=\mu(l,\o)>0$. We note that fixing
$S_{\rm cl}$ yields a line $\o=\o(l,S_{\rm cl})$ in the region
$\o>\o_0$ and $l<r_0$, terminating at the point $\o=\o_0$ where
$E_{\rm cl}$ becomes minimal. We shall assume that $S_p$ collapses
in the limit $TL^2\rightarrow \infty$ and $\mu(l,\o)\rightarrow
0$, to an integral \emph{localized to the cusps} resulting in
cusp-wave, wave-wave and cusp-cusp interactions with couplings
whose $(r_0/L)$-dependence is governed by localized integrals of
products of mode functions.

The resulting cusp-wave and wave-wave couplings scale like
negative powers of $r_0/L\sim \log S_{\rm cl}$, such that at fixed
order in number of oscillators, the wave-wave vertices are
suppressed by larger powers than wave-cusp vertices. The cusp-cusp
couplings scale like $\exp(-d/L)\sim (S_{\rm cl})^{-d/r_0}$, where
$d$ is the distance $d$ between the cusps. The normal-coordinate
theory therefore contains a multi-cusp system with Hamiltonian
\be H_N[\{a_i(\x),a^\dagger_i(\x)\}_{\x=1}^N]\ ,\label{HN}\ee
consisting of kinetic terms and an effective long-range potential
representing the exchange of long and intermediate wave quanta,
and giving rise to a dispersion relation of the form \eq{addrule}.

Let us compare the corresponding multi-cusp states,
\be \ket{\Psi}_N\ =\ \Psi[\{a^\dagger_i(\xi)\}_{\x=1}^N]\ket{0}\
,\ee
with the tensor products of scalar-singleton states belonging
to a sector carrying one large spin component, say
$\ket{\Psi}\in\sfD$ obeying
\be \bra{\Psi}M_{D-2,D-1}\ket{\Psi}\sim S_{\rm cl}\gg 1\ .\ee
Defining $A_{i'}\equiv L^-_{i'}/\sqrt{2S_{\rm cl}}$,
$i'=1,\dots,D-3$, and $A_{\pm}\equiv(L^-_{D-2}\pm i
L^-_{D-1})/\sqrt{8S_{\rm cl}}$, where $L^\pm_r$ ($r=1,\dots,D-1$)
are the spin-boost generators defined in \eq{boosts} and
\eq{lpllmi}, then the scalar-singleton mass-shell condition
$L^+_rL^+_r\ket{\Psi}=0$ corresponding to the singular vector
\eq{singv}, yields
\bea \Big(4A_-^\dagger A_+^\dagger +\sum_{i'=1}^{D-3}
(A_{i'}^\dagger)^2\Big)\ket{\Psi}&=&0\ .\label{mscond}\eea
Thus, acting in the large-spin sector, $A_i\equiv (A_{i'},A_+)$
behave as oscillators,
\be [A_i,A^\dagger_j]\approx \d_{ij}\ ,\ee
while $A_-$ behaves classically
\bea [A_-,A_-^\dagger]&\sim& 1/S_{\rm cl}\ ,\eea
and decouples in the semi-classical limit via the mass-shell
condition \eq{mscond}. In the normal-coordinate expansion,
$(E-M_{D-2,D-1})/L$ is identified as the world-sheet Hamiltonian,
although $E$ and $M_{D-2,D-1}$ are not quantized separately
\cite{Frolov:2002av,Tseytlin:2002ny}. The energy and spin
eigenvalues carried by $A^\dagger_{i'}$ and
$A^\dagger_+$ correspond to frequencies $\nu_{i'}=1$ and $\nu_+=0$, respectively.
Thus, in going from strong to weak tension, we may identify $A_{i}^\dagger$ as the limit of
$a^\dagger_{i}$, where we note that the frequency $\nu_{i'}$ of the overall tranverse
set of  oscillators remain constant while the frequency of the longitudinal oscillator
changes from $\tilde \nu$ to $\nu_+$.

Hence, the state space of the multi-cusp system can be identified
as a deformed corner of a multi-singleton system\footnote{Pulsating strings, with small $S_{\rm cl}$ an
$\s$-independent effective tension $T_{\rm eff}$, should
correspond to deformations of multi-singleton states, though here
the anomalous corrections to the energy will be much larger.}, schematically
\be {\cal S}_{\rm cusps}\ \subset\ \bigoplus_{N=2}^\infty
m_Y\left[\sfD^{\otimes N}\right]_{Y}\ ,\label{multisingleton}\ee
where $Y$ denote Young tableaux and $m_Y$ multiplicities. At the
linearized level, the Bose symmetry of the normal-coordinate
quantum field theory implies that products of $a_i^\dagger (\xi)$
are fully symmetrized, \emph{i.e.} at this level $m_{\rm symm}=1$ while all other
Young projections are absent. At finite tension and including interactions, or in
the corresponding discretized quantum mechanical formulation, which we shall study below,
it is natural to expect further states that break $S_N$ to $\integ_N$.
Indeed, the spectrum of the $10D$ tensile $IIB$ closed string in flat spacetime appears to cross over into a tensionless spectrum on AdS$_5\times S^5$ given by $\integ_N$-invariant tensor products of $\psu$ supersingletons (corresponding holographically to single-trace operators) \cite{Polyakov:2001af,Sundborg,Sezgin:2002rt,Bianchi:2003wx}.
However, as we shall see, the cyclic invariance seems to be associated to the stringy nature of the tensile deformation, rather than to any intrinsic property of the undeformed tensionless theory, which we shall find consists of fully symmetrized states -- at least in the scheme for discretization that we set up.

The fact that each string-cusp yields $D-2$ oscillators is a
direct result of the singleton-like rotational motion with heavily
suppressed radial fluctuations, rather than of the fact that the
string has $D-2$ transverse directions. This suggests that
multi-singleton states arise also for higher $p$-branes.


\scss{Rotating Membranes and $p$-Branes}\label{sec:rotp}


A closed membrane with topology $T^2\times \Real$, parameterized
by $(\t;\s,\r)\in \Real\times [0,2\pi[\times [0,2\pi[$, can be
folded and embedded into the following rotating configuration in
$(AdS_D)_L$ \cite{Sezgin:2002rt}
\be t\ =\ t(\t)\ ,\qquad r\ =\ r(\s,\r)\ ,\qquad \th\ =\ \th(\r)\
,\qquad \phi\ =\ \omega_0 t\ ,\ee
where $\th$ is a polar angle in $S^{D-2}$, defined by
$d\O^2_{D-2}=d\th^2+\sin^2\th d\phi^2+\cos^2\th d\O^2_{D-4}$, and
the maps $r(\cdot,\r):\s\mapsto r(\s,\r)$ and $\th:\r\mapsto
\th(\r)$ have winding number $1$. The embedding depends on two
variational parameters, parameterizing the turning points of
$\theta$ and $r$, so that $\th\in[{\pi\over 2}-{\ell\over
2L},{\pi\over 2}+{\ell\over 2L}]$ and $r\in [-r_0,r_0]$ when
$\theta=\pi/2$. These parameters are fixed by minimizing the
energy at fixed spin under the assumption that the membrane has
two long edges at $\theta={\pi\over 2}\pm {\ell\over 2L}$ and two
short edges at $r=\pm r_0$, \emph{viz.}
\be r_0\gg L\ ,\qquad \ell\sinh {r_0\over L}\ll L\ ,\ee
and that the induced volume element, given by
\be -\det{g}=\left({\partial r\over\partial
\s}\right)^2\left(\cosh^2{r\over L}-\o^2\sinh^2{r\over
L}\sin^2\th\right)\left(L^2\sinh^2{r\over
L}\left({\partial\th\over\partial\r}\right)^2-\left({\partial r\over\partial
\r}\right)^2\right)\ \ee
vanishes on the short edges. The latter condition requires $\coth
{r|\over L}=\o\sin\th$, where $r|$ is the restriction of $r$ to
the short edges, that in turn implies ${\partial
r|\over\partial\r} \ =\ -\o L\sinh^2 {r|\over L} ~\cos\th
~{\partial \th\over\partial\r}$, which together with the shortness
condition yields
\be \det g\approx -\mu^2\left({\partial r\over\partial
\s}\right)^2 L^2\sinh^2{r\over
L}\left({\partial\th\over\partial\r}\right)^2\ .\ee
Thus, to the leading order one can set
\bea r&\approx& r(\sigma)\ ,\eea
and expand around $\th={\pi\over 2}$, resulting in
\bea E_{\rm cl}(\ell,r_0)&=&8T\ell L^2 \int_0^{r_0+L\log 2}
dy~\sinh\frac{r}{L}\cosh^2\frac{r}{L}\ ,\\ S_{\rm cl}(\ell,r_0)&=&8T\ell L^2\omega
\int_0^{r_0+L\log 2}dy~ \sinh^3\frac{r}{L}\ .\eea
The leading contributions scale like
\be E_{\rm cl}(\ell,r_0)\ \approx\ S_{\rm cl}(\ell,r_0)\sim
TL^2~\ell~ e^{{3r_0\over L}}\ ,\ee
and the first ``anomalous'' correction to the energy at fixed spin
and variational parameter $\ell$ scales like
\be E_{\rm cl}(\ell)-S_{\rm cl} \ \sim\ (T\ell L^2 )^{\frac23}
S_{\rm cl}^{\frac13}\ .\label{m2anomaly}\ee
It follows that $E_{\rm cl}(\ell)$ is minimized at fixed $S_{\rm
cl}$ in the limit\footnote{From the holographic point of view,
$r_0\rightarrow \infty $ a UV limit, and $\ell\sinh{r_0\over
L}\rightarrow 0$ is limit that localizes the Wilson surface
coupling to the boundary of the open membrane.}
\be {\ell\over L}\rightarrow 0\ ,\qquad {r_0\over L}\rightarrow
\infty\ ,\ee
where we note that the shortness condition is indeed obeyed, that
is
\be \ell \sinh {r_0\over L} \sim S_{\rm cl}
\exp(-2r_0/L)/TL^3\rightarrow 0\ .\ee
Hence, the membrane
collapses to an infinitely long string-like configuration with
\be E_{\rm cl}\ =\ S_{\rm cl}\ .\ee
Thus, the two-dimensional gas of cusps
on the membrane has a vanishing potential to the leading order.
This is to be contrasted with the linear potential that arise to the same order
in the case of the one-dimensional gas of cusps on the string.

Turning to the fluctuation analysis, the sector of
$\r$-\emph{independent} fluctuations $\phi^{(0)}_a$, which are
normalized with string tension $T\ell$, contains $D-2$ physical
fields: $D-4$ overall transverse fields $\varphi^{(0)}_\perp$; an
admixture $\tilde\varphi^{(0)}$ arising in the
$(\varphi^{\underline t},\varphi^{\underline
r},\varphi^{\underline \phi})$-sector; and
$\varphi^{(0)}_{\underline\th}$, which describes an actual
deformation of the embedding in target space that cannot be gauged
away. To show this one observes that the fluctuation field
$\varphi^{\underline\th}$ transforms under world-volume
diffeomorphisms with parameters $V=V^\a\partial_\a$ as
$\delta\varphi^{\underline{\th}}=V^\r\partial_\r \th L\sinh
{r\over L}$. There is no $\r$-independent mode in
$\partial_\r\th$, since natural boundary conditions imply that
this field vanishes on the long edges. Thus, the variation
$\delta\varphi^{\underline{\th}}$ does not contain any
$\r$-independent mode, so that $\varphi^{(0)}_{\underline{\th}}$
cannot be gauged away.

The quadratic action for $\varphi^{(0)}_\perp$ reads
\be S_2[\varphi_\perp]\ =\ -\frac12 \int dt dy \sinh {r\over
L}\left((\partial \varphi^{(0)}_\perp)^2+{3\mu^2\over
L^2}(\varphi_\perp^{(0)})^2\right)\ .\ee
Rescaling, $\varphi^{(0)}_\perp=\hat\varphi/\sqrt{\sinh(r/L)}$,
expanding $\hat\varphi=\sum_k e^{{i\nu_k t\over L}
}\hat\varphi_k(y)a^\dagger_k /\sqrt{\nu_k}$, and taking the limit
$r_0/L\rightarrow\infty$ using that $\sinh(r/L)\approx
\exp(r_0/L)\cosh^{-1}(y/L)$ and $\coth(r/ L)\approx
1 + 2\exp(-2r_0/L)\cosh^{-2}(y/L)$, one finds the P\"oschl-Teller
potential problem
\be \Big(-L^2{d^2\over dy^2}+q\mu^2\Big)\hat\varphi_k\ =\
\big(\nu^2_k+\ft12\big)\hat\varphi_k\ ,\qquad
\left.{d\hat\varphi\over dy}\right|_{y=0}\ =\ 0\ ,\qquad
q={15\over 4}\ ,\ee
that admits precisely one bound state with protected frequency
\be \varphi^{(0)}_\perp\ =\ {\cal N} {1\over \cosh{y\over L}}\
,\qquad \nu^2_\perp\ =\ 1\ ,\ee
where the power of $e^{-{r_0\over L}}$ in $(\sinh{r\over
L})^{-{1\over 2}}$ has been absorbed into the normalization. Not
so surprisingly, this solution is identical to that found in the
string case. We expect that also $\varphi^{(0)}_{\underline\th}$
contains the same bound state, so that there is a total of ${D-3}$
bound states at each cusp with frequency $\nu=1$, while
$\tilde\varphi^{(0)}$ should contain bound states with anomalous
frequency, \emph{c.f.} eqs.~\eq{qprimed-2} and \eq{nud-2}.

In the case of a folded and rotating $p$-brane with $p\leq D-2$,
the topology is $T^p\times \Real$, and the folded world-volume
extends in one radial direction and $p-1$ directions transverse to
the plane of rotation. As for $p=2$, the classical energy is
minimized on a stringy configuration, with
\be E_{\rm cl}\ =\ S_{\rm cl}\sim T_pL^2\ell^{p-1}~ e^{{(p+1)r_0\over
L}}\ ,\qquad E_{\rm cl}(\ell)-S_{\rm cl}\ \sim\ (T_p
\ell^{p-1}L^2)^{2\over p+1} S_{\rm cl}^{p-1\over p+1}\ . \ee
The quadratic action for the $D-p-2$ overall transverse
``stringy'' fluctuations read
\be S_2\ =\ -{1\over 2}\int dt dy \sinh^{p-1}{r\over
L}\left((\partial \varphi^{(0)}_\perp)^2+{(p+1)\mu^2\over
L^2}(\varphi^{(0)}_\perp)^2\right)\ .\ee
This is a P\"oschl-Teller problem for $(\sinh{r\over
L})^{{p-1\over 2}}\varphi^{(0)}_\perp$ with $q={1\over
4}(p^2+4p+3)$ and eigenvalue $\nu^2+{p-1\over 2}$. There are
$[(p-1)/2]$ bound states in the potential well. Thus, for $p\geq
3$, there is at least one extra bound-state oscillator, over and
above the ground state, making the interpretation of the cusps
more problematic. We shall address this issue below in Sections
\ref{sec:eou}, in relation to branes at the end-of-the-universe,
and in Section \ref{sec:M2def}, by examining tensile deformations
of the tensionless discretized brane.

Thus the covariant phase-space of $p$-branes in anti-de Sitter
spacetime contains partonic regions giving rise to wave functions
that fit into large-spin corners of multi-singleton spaces.
Remarkably, the quantum numbers remain protected in the bosonic
case, although supersymmetry should ultimately be of importance.


\scss{Remarks on The Superstring and
Supermembrane}\label{sec:susy}


In the case of the superstring on $AdS_5\times S^5$
\cite{Gubser:2002tv,Frolov:2002av,Tseytlin:2002ny}, the scalar
fluctuations in $S^5$ have vanishing mass and thus do not give
rise to any bound states.
The fermionic fluctuations are $4+4$ Majorana spinors $\Theta$
with mass $\e{\mu\over L}$, $\e=\pm 1$, \cite{Frolov:2002av},
\be S_2[\Th]\ =\ i\int dt dy \Big(\bar \Th
\c^\a\partial_\a\Th+{\mu\over L} \bar\Th\Th\Big)\ .\ee
Taking $\c^\a=(i\s^2,\s^1)$ and $C=-i\s^2$ and real $\Th$, the
natural boundary conditions $\bar\Th \gamma^\a n_\a\delta\Th|=0$
at $r=\pm r_0$ imply $(U\delta U-V\delta V)|=0$ where $U=\Th_1$
and $V=\Th_2$. This leaves two possibilities that do not violate
the discrete $\integ_2$ symmetry under exchange of cusps, namely
$U| =\s V|$ for $\s=\pm 1$. Expanding into mode-functions
$(e^{{i\nu_k t\over L}}U_k+{\rm h.c.})$ and $(e^{{i\nu_k t\over
L}}V_k+{\rm h.c.})$, the Dirac equation reads
\be \Big(L{d\over dy}+i\nu\Big)U_k+\e\mu  V_k\ =\ 0\ ,\qquad
\Big(L{d\over dy}-i\nu\Big)V_k+\e\mu U_k\ =\ 0\ .\ee
Using $dr^2=\mu^2 dy^2$, one finds bound states peaked at $r=\s
r_0$ given by
\be U\ =\ {\cal N} e^{ (\sigma r-r_0)\over L}\ ,\qquad \nu\ =\ 0\
,\ee
where ${\cal N}\sim L^{-1/2}$. Thus, there are $4+4$ \emph{real}
Clifford-algebra elements localized at each cusp, that can be
combined into $2+2$ fermionic oscillators\footnote{The wave-modes
with $\nu>0$ gives rise to $4+4$ fermionic creation operators for
each value of $\sigma$.} $\a^\dagger_I$ and $\beta^\dagger_{I'}$,
$I,I'=1,2$. Since $\nu=0$, they do not contribute to the
zero-point energy.

The wave functions at each cusp are thus in rough agreement with a
large-spin sector of the $\psu$ singleton, in which the
spin-boosts and the supercharges behave asymptotically as the
bound-state oscillators.

In the case of the supermembrane on $(AdS_{7})_L\times
(S^4)_{L/2}$ the $\r$-independent fluctuations in $S^4$ are
described by the action $-{1\over 2}\int dtdy \sinh{r\over
L}~(\partial \varphi^{(0)})^2$, which corresponds to a
P\"oschl-Teller problem for $(\sinh{r\over L})^{1\over 2}
\varphi^{(0)}$ with $q={3\over 4}$ and eigenvalue $\frac 1{L^2}(
\nu^2+{1\over 2})$. This problem admits the solution $\nu=0$ and
$\varphi=1$, which does not localize on the cusps, as expected. We
therefore expect that the fermionic fluctuations contain bound
states with $\nu=0$ corresponding to a total of $2+2$ fermionic
creation operators, so that the bound-state wave function can be
interpreted as the large-spin limit of the $\mosp(8^*|4)$
supersingleton.


\scss{Branes at the End-of-The-Universe}\label{sec:eou}


While it has been assumed throughout the above analysis that
$TL^2\gg1$, it seems natural -- drawing on analogies with ordinary
soliton quantum mechanics --  that the complete string spectrum
should be identified with a singleton ``gas'' in the tensionless
regime $TL^2\ll1$. This physical picture seems clearer in the case
of the membrane, where there is no linear potential and Bose
symmetry is more appropriate. An alternative way of thinking of
the tensionless limit, is to view the cusps as ``mechanical''
lumps, and seek an interpretation in terms of short open
$p$-branes ending on a dual $(D-2)$-brane. This system can then be
followed to the boundary region of AdS spacetime, where the
effective open-brane tension diverges while the ``bare'' $p$-brane
tension vanishes, as we shall demonstrate next using some sample
calculations.

\scsss{Giant Vacuum and Singular CFT}


The spiky strings may be thought of as bound states formed by
\emph{long} open strings carrying $U(1)$ Chan-Paton factors. When
the physical sizes of the bound states increase -- as a result of
increasing the ratio between spin and tension -- it is likely for
them to disintegrate, either by formation of cusp-anti-cusp pairs
leading to multi-body interactions of the type depicted in
Fig.~\ref{fig:long}, or via processes in which a single cusp is
shredded off while the remaining stretched junction to the
interior of the string retracts smoothly. Drawing on the classical
mechanical properties of branes, one may argue that the latter
processes should start to dominate as the bare tension decreases,
since then the ``snapping'' associated with breaking of strings,
which is the mechanism responsible for forming cusp-anti-cusp
pairs, becomes less distinct.

The released cusp is a lump of relativistic energy moving along a
light-like geodesic line. In Poincar\'e coordinates, \emph{viz.}
\be ds^2\ =\ L^2 \Big(u^2dx^\mu dx_\mu+{du^2 \over u^2}\Big)\ ,\ee
the geodesic can be chosen to be a curve with constant $u=u_0$
along the null-direction in $x$-space. Furthermore, it is natural
to assume that the number of degrees of freedom that go into the
cusp wave-function should not change in the process. Assuming the
wave-function to be built from $(D-2)$ oscillators $a^\dagger_i(\x)$,
this suggests that the released cusp be identified as a
\emph{short} open string on a stable $(D-2)$-brane at $u=u_0$.
This requires an asymptotically large effective open-string
tension
\be T_{\rm s,eff}\ =\ TL^2u_0^2\ \gg\ {1\over L^2}\ ,\ee
and that the $(D-2)$-brane couples magnetically to the
cosmological constant $\L$, that is $\Lambda={1\over
2D!}|dC_{D-1}|^2$ and
\be S_{D-2}\ =\ -T_{D-2}\int d^{D-1}\sigma\sqrt{-\det
g}+Q_{D-2}\int C_{D-1}\ , \ee
with charge and tension given by\footnote{In the purely bosonic
case, we dualize the cosmological constant using the convention
$S_{\rm grav}=-\int d^D X\sqrt{-G}(R+\L)$ with $\Lambda={1\over
2D!}|dC_{D-1}|^2=-{(D-1)(D-2)\over L^2}$. In the supersymmetric
extensions this relation is modified.}
\be Q_{D-2}\ =\ \left({D-1\over 2(D-2)}\right)^{1\over 2}~T_{D-2}\
,\qquad K\ \equiv \ T_{D-2} L^{D-1}\gg (TL^2)^{D-1\over 2}\
.\label{d-2charge}\ee
The open-string corrections are dressed with positive powers of
$1/T_{\rm s,eff}$ and a suitable open-string coupling $G_{\rm
OS,eff}^2$, assumed to be proportional to the closed-string
coupling $g_s$.

In static, or ``Monge''\footnote{Coined by P. Howe.}, gauge
\be x^\mu\ =\ \sigma^\mu\ ,\qquad \phi\ \equiv \ w-w_0\ =\
\phi(x)\ ,\qquad w\ \equiv \ u^{D-3\over 2}\ ,\ee
the limit
\be TL^2\rightarrow 0\ ,\qquad \mbox{at fixed $\phi$,
$\partial_\mu\phi$, $T_{\rm s,eff}L^2$}\ ,\label{limit1}\ee yields
\be S_{D-2}\ =\ -{\widetilde K\over 2}\int d^{D-1}x ~(\partial
\phi)^2\ ,\qquad \widetilde K\ =\ {4K\over (D-3)^2}\ ,\ee
with evanescent terms given by positive powers of
$\partial_\mu\phi/(u_0w_0)$. Sending also
\be P_\mu/T_{\rm s,eff}\rightarrow 0\ ,\quad G_{\rm OS}\rightarrow
0\ ;\qquad \mbox{at fixed $P_\mu$ and $K\gg 1$}\
,\label{limit2}\ee
where $P_\mu$ is the scalar field UV cut-off, yields a decoupled \emph{singular conformal field theory}
\cite{Seiberg:1999xz}. This theory is based on the scalar singleton $\sfD$
arising from the quadratic action, and can be used to build
composite operators (vertices) filling the spectrum
\be {\cal S}_{\rm giant}\ =\ \sum_{N=1}^\infty (\sfD^{\otimes
N})_{\rm symm}\ ,\label{1elbrane}\ee
which we identify as \eq{multisingleton} with an additional single
singleton sector.

On the closed-string side, the limit corresponding to \eq{limit2}
is
\be TL^4u^2\rightarrow \infty\ ,\quad g_{\rm s}\rightarrow 0\
;\qquad \mbox{with fixed $E$, $T_{D-2}L^{D-1}\gg 1$}\ .\ee
Drawing on the relation between tension and dilaton in
$AdS_5\times S^5$, this indeed requires $TL^2\rightarrow 0$, which
together with the insensitivity of the singular theory to the bare
tension $TL^2$ provides further evidence for that there is a
smooth transition from the tensile string to a multi-singleton
theory.

As already indicated, supersymmetry will be crucial in order to
make the above parametric relations precise. Moreover, the
identification of the singleton and the massless sector of the
open string on the $(D-2)$-brane in $AdS_D$ is quite cumbersome.
The $U(1)$ Chan-Paton factors yield even and odd spins in ordinary
flat space, and one might speculate that the tachyon becomes
massless as an effect of finite or even vanishing $L$. In this
respect, the identification of the $\psu$ supersingleton starting
from a $D3$-brane in $AdS_5\times S^5$ should be is simpler.
Finally, from the analysis of Section \ref{sec:rotp}, it is clear
that the assumption that the wave-functions of the cusps are built
from $D-2$ oscillators becomes problematic for $p>2$.

The above considerations motivate a closer look at the maximally supersymmetric
cases where open Type IIB strings end on D3-branes and open M2-branes end on
M2 or M5-branes.


\scsss{Maximally Supersymmetric Cases}


Let us discuss the singular limit of maximally supersymmetric M2,
D3 and M5 branes in $(AdS_D)_{L}\times (S^{\widetilde D})_{\widetilde L}$
with
\be \qquad {\widetilde L\over L}\ =\ {D-1\over\widetilde D-1}\ =\
{D-3\over 2}\ .\ee
We start from the potentials
\bea C_{D-1} &=& \widetilde L^{D-1} u^{D-1} d^{D-1}x\ ,\qquad d^{D}x\
\equiv\ dx^0\wedge\cdots\wedge dx^{D-1}\ ,\label{eou1}\\
\widetilde C_{\widetilde D-1} &=& L^{\widetilde D-1} \o_{\widetilde D-1}\
,\label{eou3} \eea
where $d\omega_{\widetilde D-1}$ is the volume form on the unit
$S^{\widetilde D}$. In Monge gauge, the bosonic fields are $\widetilde
D+1$ scalar fields $w^I$ defined by
\be \sum_{I=1}^{\widetilde D+1}(dw^I)^2\ =\ dw^2 + w^2 d\O^2_{\widetilde
D}\ ,\qquad w\ =\ u^{\widetilde L\over L}\ ,\label{eou5} \ee
and the fields strengths
\bea D3\ &:&\qquad f\ =\ da\ ,\qquad\qquad\qquad a\ =\ dx^\m a_\m\ ,\qquad \\
M5\ &:&\qquad h\ =\ db+\sqrt{T_5} \widetilde C_3\ ,\qquad b\ =\
{1\over 2}dx^\m\wedge dx^\n b_{\m\n}\ .\eea
The Lagrangians are given by
\be \cL_{D-2}\ =\ \cL_{\rm NG}+\cL_{\rm tensor}+\cL_{\rm
el}+\cL_{\rm magn}\ ,\ee
where $g_{\mu\nu}=\widetilde L^2 w^{2L\over \widetilde L}\eta_{\mu\nu}+
{L^2\over w^2}\partial_\mu w^I\partial_\nu w^I$, so that
\be \cL_{\rm NG}\ =\ -T_{D-2}\ \widetilde L^{p+1}(w^{DL\over\widetilde
L}+{ L^2\over 2\widetilde L^2}\eta^{\mu\nu}\partial_\mu
w^I\partial_\nu w^I+\cdots) \ ,\label{eou6} \ee
and the tensor-field Lagrangians and Wess-Zumino terms are given
by \footnote{For the M5 brane we use the formalism of
\cite{Cederwall:1997gg,Sezgin:1998tm}, with
\bea \Phi(g,h)&=&{T_5^{-1}\over 12} g^{\mu_1\nu_1}g^{\mu_2\nu_2}
g^{\mu_3\nu_3}h_{\mu_1\mu_2\mu_3}h_{\nu_1\nu_2\nu_3}\label{M5kin}\\
&&+{T_5^{-2}\over 288}g^{\mu_1\nu_1}g^{\mu_2\nu_2}
g^{\mu_3\nu_3}g^{\mu_4\nu_4}g^{\mu_5\nu_5}
g^{\mu_6\nu_6}\nn\\&&\times(h_{\mu_1\mu_2\mu_3}h_{\nu_1\nu_2\nu_3}
h_{\mu_4\mu_5\mu_6}h_{\nu_4\nu_5\nu_6}-3h_{\mu_1\mu_2\mu_3}
h_{\nu_2\nu_3\nu_4}h_{\mu_4\mu_5\mu_6}h_{\nu_5\nu_6\nu_1})\
.\nn\eea}
\bea \mbox{D3}&:& \cL_{\rm tensor} = \sqrt{\det{\delta_\nu^\mu+
2\pi\alpha'g^{\mu\rho}f_{\rho\nu}}} = 1+{(2\pi\alpha')2\over
4 \widetilde L^4 w^4}f^{\mu\nu}f_{\mu\nu}+ \cdots ,\label{vector}\\
\mbox{M5}&:& \cL_{\rm tensor}\ =\ \sqrt{1+\Phi(g,h)}\ =\
1+{T_5^{-1}\over 24 \widetilde L^6 w^3} h^{\mu\nu\rho}
h_{\mu\nu\rho}+\cdots\ ,\label{tensor}\eea
\be \cL_{\rm el}\ =\ T_p\widetilde L^{D} w^{DL\over \widetilde L}d^{D-1}x\
,\qquad d^{D}x\ =\  dx^0\wedge\cdots \wedge dx^{D-1}\ ,\ee
\bea \mbox{D3} &:& \cL_{ magn}\ =\ -T_3 L^4 \o_4\ =\ -{T_3
L^4\over 4!} \epsilon^{\mu_1\mu_2\mu_3\mu_4}
\omega_{\mu_1\mu_2\mu_3\mu_4}d^{D-1}x\ ,\label{magnD3}\\
\mbox{M5} &:& \cL_{ magn}=-\sqrt{T_5} L^3 \omega_3\wedge db
=-{\sqrt{T_5} L^3 \over 12}
\epsilon^{\mu_1\mu_2\mu_3\mu_4\mu_5\mu_6} \omega_{\mu_1\mu_2\mu_3}
\partial_{\mu_4}
b_{\mu_5\mu_6}d^{D}x\ ,\nn\\ \label{magnM5}\eea
where the Lagrangians are expanded in positive powers of
$\partial_\mu w^I/(uw)$, $f_{\mu\nu}/(uw)$ and $h_{\mu\nu\r}/(uw)$
contracted with $\eta^{\mu\nu}$. In the IR limit, the electric WZ
term cancel against the leading contribution from the kinetic
term, while the magnetic terms, including the CS modification in
$h=db+\sqrt{T_5}L^3 \omega_3$, are fixed, which yields
\bea S_{M2}&=&-{T_2 L^3} \int d^3x {1\over 4}
\partial^\mu w_I \partial_\mu w_I\ ,\label{SM2}\\
S_{D3}&=&=-{T_3 L^4} \int d^4x ({1\over 2}\partial^\mu w^I
\partial_\mu w^I +{1\over 4Ng} f^{\mu\nu}
f_{\mu\nu})- {T_3 L^4}\int \omega_4\ ,\label{SD3}\\
S_{M5}&=&-{T_5 L^6}\int d^6x ({1\over 16}
\partial^\mu w_I \partial_\mu w_I+{1\over 24 T_5 L^6}
h^{\mu\nu\rho}h_{\mu\nu\rho})\nn\\&&- \sqrt{T_5}L^3\int
\omega_3\wedge db\ ,\qquad h\ =\ db+\sqrt{T_5}L^3 \omega_3\ .
\label{SM5}\eea
Hence, the singular M2-brane is a free $\mosp(8|4)$ supersingleton
\cite{Duff:1989ez}, while the singular D3 and M5-branes are
described by interacting $\psu$ and $\mosp(8^*|4)$
supersingletons, respectively. The purely bosonic sector of the
interactions are given above, and thus governed by the
scale-invariant scalar-field construct $\o_{\widetilde D-1}$.
Moreover, the singular M5-brane tensor-field equation is the
non-linear self-duality condition $f_3=\star f_3$ containing
tensor-scalar interactions \cite{Cederwall:1997gg,Sezgin:1998tm}.
The superconformal completions are given in \cite{Claus:1998mw},
although they have not been worked out in detail in the singular
limit.


\scs{THE TENSIONLESS LIMIT}  \label{sect3}


In the previous section we have argued that the covariant phase
space of tensile $p$-branes in anti-de Sitter spacetime contains
partonic subspaces, represented semi-classically by singleton-like cusps on
rotating $p$-branes in the limit $S\gg T_pL^{p+1}\gg1$ and fixed $E-S$.
Moreover, we have proposed an alternative description of the partonic
degrees of freedom, interpolating smoothly from $TL^2\gg1$ to $TL^2\ll1$, in
terms of decoupled massless open-$p$-brane excitations of $(D-2)$-branes at the boundary
of anti-de Sitter spacetime. These considerations suggest a partonic
spectrum given by symmetrized singletons (see \eq{1elbrane}), raising
the question whether these can be seen directly at the level of the
tensionless limit of the closed $p$-brane in anti-de Sitter spacetime.

In this section we shall examine the limit $TL^2\rightarrow 0$ of
the bulk $p$-brane assuming that: 1) the \emph{entire} $p$-brane
has a faithful description in terms of the $(0+1)$-dimensional
discretized Nambu-Goto action; 2) each fixed number, $N$, of discretized
degrees of freedom, or fundamental partons, is an exact sector of the
theory up to $1/N$ corrections\footnote{This treatment harmonizes with that
of the membrane in flat spacetime, with the crucial difference that the
partonic spectrum in anti-de Sitter spacetime is discrete, with
a well-defined interpretation in terms of space-time one-particle states.}.
We note that the partons carry space-time energy-momentum and spin
but no other ''internal'' quantum numbers,
as opposed to Thorn's colored string bits \cite{thorn} (see also \cite{gopakumar}).

We stress that in taking the tensionless limit at the level of the discretized action
-- where it is a weak coupling limit -- we will end up with a model based on a Lagrangian
quite different from the $(p+1)$-dimensional $p$-brane
Lagrangian -- where the limit is of course a strong coupling limit. The two models should
therefore be compared at the level of the covariant phase space.
To be more precise, we shall demonstrate a
correspondence between the partonic subspace of the covariant phase space of the tensionful
brane and the fundamental \emph{quantum} states of the discretized tensionless model.

Our two main findings are: first, if $\mu$ denotes the inverse
lattice spacing, the proper tensionless limit is given by
\be L \mu\rightarrow 0\ ,\qquad T_p \mu^{-p-1}\rightarrow 0\ ,
\label{eq:L0}\ee
resulting in an $\msp(2N)$-gauged sigma model (see
Section \ref{sec:matrix}) on $N$ copies of the Dirac
hypercone.
Second, the model has a well-defined continuum limit in
$D=7$ giving rise to a topological closed string (see Section
\ref{sec:con}) containing a singleton spin field.
We also provide arguments why both systems reproduce the
symmetrized-singleton spectrum \eq{1elbrane}.

A subtlety presents itself in that the spatial directions of the
$p$-brane become blurred in the tensionless limit. Instead it
makes sense to expect that the $p$-dependence enters via tensile
deformations. Indeed, in the previous section we found that the
cusps have $p$-dependent quantum properties, such that those on
strings and membranes carry the same number of quantum degrees as
actual singletons, while those on branes with $p>2$ carry
additional quantum degrees of freedom. The view that cusps on
strings and membranes are deformed singletons, while those on
higher $p$-branes are more complex, was lended further support by
the fact that open $p$-branes ending at the branes at the boundary
of anti-de Sitter spacetime can be made maximally supersymmetric
maximally supersymmetric in setups with strings and membranes.
Here we shall provide further evidence for the special status of
$p=1$ and $p=2$, by showing that a simple tensile deformation that
preserves $S_N$-invariance indeed lead to a dispersion relation
between $E$ and $S$ of the same type as for the tensile membrane.

Moreover, we shall find that the singleton wave-functions involve
subtleties and actual $\msp(2N)$-anomalies that are remedied
quantizing in phase space (see Sections \ref{sec:phase} and
\ref{sec:alg}), in turn facilitating the above continuum limit as
well as the treatment in Section \ref{sec:tos} of the massless
two-parton system giving rise to Vasiliev's equations.

Before we turn to the discretization and continuum limit, we shall first discuss
the issue of inequivalent massless limits of an ordinary
point-particle in anti-de Sitter spacetime, focusing on the singular limit corresponding to
\eq{eq:L0}, leading to the vector-oscillator realization of the scalar singleton.
The different massless limits are discussed in Section \ref{sec:phases}, while
the vector-oscillator realization of the singleton is described in Sections \ref{sec:geom} and
\ref{sec:singgs}.


\scss{Dirac's Hypercone and The Singleton}\label{sec:pp}


\scsss{Phases of the $\msp(2)$-Gauged Sigma Model}\label{sec:phases}


To describe the point particle in anti-de Sitter spacetime, it is
most natural to start from the formulation in the ambient
$2(D+1)$-dimensional phase space ${\cal Z}$ based on the
first-order action
\be S=\int(\frac14 Y^{Ai}DY_{Ai}-\frac12\Lambda^{ij}V_{ij})\
,\label{eq:psa}\ee
with $Y^{A i}\equiv \sqrt{2}(X^A,P^A)$, $DY^{Ai}\equiv
dY^{Ai}+\L^{ij}Y^A_j$, where $\L^{ij}$ is an $\msp(2)$ gauge
field, and $V_{ij}$ a fixed function of $\tau$. The classical
field equations read
\be DY^{Ai}\ =\ 0\ ,\qquad K_{ij}+V_{ij}\ =\ 0\
,\label{eq:clkij}\ee
where $K_{ij}$ is the $\msp(2)$ generator
\bea K_{ij}&\equiv& {1\over 2} Y^A_i Y_{jA}\ .\eea
As a result, the angular momentum
\be M_{AB}\ =\ {1\over 2}Y^i_A Y^{\phantom{[]}}_{Bi}\ee
obeys the mass-shell condition
\be \frac12 M^{AB}M_{AB}\ =\ \frac12 V^{ij}V_{ij}\ .\ee
With these normalizations, the underlying ungauged first-order
system $\frac14 \int Y^{Ai}dY_{Ai}$ has Dirac bracket
$\{Y^A_i,Y^B_j\}_D=2\eta^{AB}\epsilon_{ij}$. For the choice
\bea V_{ij}&=&\mx{(}{cc}{M^2&0\\0&L^2}{)}\ .\eea
the first-order equations of motion are equivalent to those of the
point-particle on $(AdS_D)_L$ with mass $M$, described by the
configuration space action
\be S[X^A]\ =\ \ft12 \int d\t (e^{-1} \dot X^2 - eM^2+ \lambda
(X^2+L^2))\ ,\label{STL}\ee
where $e$ and $\l$ are Lagrange multipliers.

For $(M,L)\neq (0,0)$, the constraint \eq{eq:clkij} breaks
$\msp(2)$ to the $\integ_2$ acting as
\be \integ_2\ :\quad M\leftrightarrow L\quad \Leftrightarrow
X^A\leftrightarrow -P^A\ .\label{Z2}\ee
Already at the classical level, it is clear that the
masslessness condition $ML=0$ admits two branches:
\begin{itemize} \item[\emph{i)}] the \emph{massless particle}
\be M\ =\ 0\ ,\quad L>0\quad \leftrightarrow\quad M>0\ ,\quad L\
=\ 0\ ,\label{casei}\ee
moving along straight lines in the $(D+1)$-dimensional ambient
spacetime intersecting $(AdS_D)_L$ defined by the hypersurface
$X^2+L^2=0$;
\item[\emph{ii)}] the \emph{singleton}
\be M\ =\ L\ =\ 0\ ,\label{caseii}\ee
living on the $D$-dimensional \emph{Dirac hypercone}
\be X^2\ =\ 0\ ,\qquad X^A\ \sim \ -X^A\ .\label{cone}\ee
Removing the apex gives a smooth manifold with two boundaries
which is homotopic to a cylinder and has an
$\mso(D-1,2)$-invariant $(D-1)$-bein $e^a$ obeying
\be e^a(\partial_R)\ =\ 0\ ,\label{nullgeom}\ee
where $\partial_R$ is a the vector field pointing in the null
direction of the cone. We shall propose a holographic
interpretation of this geometry in Section \ref{sec:unf}. The
corresponding singular line-element is defined by
\be ds^2\ =\ e^a\otimes e^b\eta_{ab}\ .\ee
Splitting
\be X^A\ =\
(\overbrace{X^{0'}\phantom{,X^0}}^{X^\alpha}\hspace{-20pt}
,\underbrace{X^0,\overbrace{X^1,\dots,X^{D-1}}^{X^r}} _{X^a})\ =\
R\widehat X^A\ ,\quad \mbox{\small $\widehat X^\alpha \widehat
X^\alpha=\widehat X^r\widehat X^r=1$}\ ,\label{XA}\ee
yields the following parameterization of the singular
line-element\footnote{This form can also be obtained starting from
the global non-singular coordinates $ds^2=-\cosh^2{r\over
L}dt^2+dr^2+L^2\sinh^2{r\over L}d\O_{D-2}^2$, via the
reparameterization $r=\e\widetilde r-L\log \epsilon$ followed by
sending $\epsilon\rightarrow 0$ keeping $\widetilde r$,
$\widetilde L=L/\epsilon$ and $R={\widetilde L\over 2}
e^{\widetilde r\over \widetilde L}={L\over 2} e^{r\over L}$ fixed.
In Poincar\'e coordinates, $ds^2=L^2(u^2dx^2+{du^2\over u^2})$,
the singular limit can taken by keeping $\widetilde u\widetilde
L=uL$ fixed, resulting in $ds^2=\widetilde u^2 \widetilde
L^2dx^2$.}
\be ds^2\ =\  R^2 (-d\hat t^2+d\widehat\Omega^2_{D-2})\ ,\qquad
R>0\ , \label{conemetric}\ee
where we note the absence of a $dR^2$ term.
The singleton can perform many types of classical motion depending
on the $\msp(2)$ gauge choice \cite{Barsreview}. In particular,
constant $\o^2=\frac12 \L^{ij}_\t\L_{\t,ij}$ gives rise to five
types of \emph{singletonic motion}:
\begin{itemize}
\item[$\o^2<0$~:] hyperbolic curves turning at a finite distance from
the apex;
\item[$\o^2=0$~:] massless trajectories in the form
of straight lines passing through the apex;
\item[$\L_{ij}=0$~:] vacuum expectation values obeying $\dot Y^{iA}=0$;
\item[$\o^2>0$~:] pulsations along finite straight intervals passing through the apex;
\item[$\o^2>0$~:] rotating trajectories forming closed loops in the
time-like plane.
\end{itemize}
\end{itemize}

Upon quantization one introduces oscillators $y^{iA}$ associated to the
(constant) modes solving the classical equation of motion for $Y^{iA}$. The resulting
non-commutative structure of the ambient phase space ${\cal Z}$ is given by
\be y_i^A\star y_j^B\  =\ y_i^A y_j^B+i\e_{ij}\eta^{AB}\ ,\qquad
y_A^i=(y^i_A)^\dagger\ ,\label{yystar}\ee
where the $\star$ and
juxtaposition denote the non-commutative and Weyl-ordered
products, respectively, giving rise to the associative algebra
${\cal W}[{\cal Z}]$ based on the Moyal $\star$-product formula
\be f_1(y)\star f_2(y)\ =\ \int_{{\cal Z}\times {\cal Z}}
{d^{2(D+1)}S d^{2(D+1)}T\over (2\pi)^{2(D+1)}} e^{iT^{Ai}S_{Ai}}
f_1(y^A_i+S^A_i) f_2(y^B_j+T^B_j)\ .\label{star}\ee
We note that the hermitian conjugation acts as
\be (f(y))^\dagger\ \equiv \bar f(y^\dagger)\ =\ \bar f(y)\
,\qquad (f\star g)^\dagger\ =\ g^\dagger\star f^\dagger\
.\label{daggerosc}\ee

The $\mso(D-1,2)$ and $\msp(2)$ generators
\bea M_{AB}&=&{1\over 2}y^i_{[A} \star y^{\phantom{[]}}_{B]i}\ =\
{1\over 2}y^i_A y^{\phantom{[]}}_{Bi}\
,\qquad (M_{AB})^\dagger\ =\ M_{AB}\label{MAB}\ ,\\
K_{ij}&=&\frac12 y^A_{(i}\star y_{j)A}\ =\ \frac12 y^A_i y_{jA}\
,\qquad (K_{ij})^\dagger\ =\ K_{ij}\label{Kij}\ ,\eea
obey the commutation rules\footnote{For $D=4$ our conventions give
$[M_{12},M_{23}]=iM_{13}$, that differ by a sign from the standard
ones for $\mso(3)$. The canonical $\mso(2,1)$ generators are given
by $K_I =-{1\over 4}(\s_I)^{ij}K_{ij}$, obeying $[K_I,K_J]_\star =
i\e_{IJK}K^L$, where $(\s^I)_{ij}$ are real and symmetric van der
Waerden symbols defined by $(\s_I)_i{}^k(\s_J)_{kj} =
\eta_{IJ}\e_{ij}+\e_{IJK}(\s^K)_{ij}$ with $\e^{ij}\e_{kj}
=\d^{i}_{k}$, $\eta_{IJ}={\rm diag}(++-)$ and
$\epsilon^{IJK}\epsilon_{MNP}=-3!\d^{IJK}_{MNP}$.}
\bea [M_{AB},M_{CD}]_\star&=&4i\eta_{[B|[C}M_{A]|D]}\
,\label{adsalg}\\[4pt]
{}[K_{ij},K_{kl}]_\star&=&4i\e_{(i|(k}K_{l)|j)}\
.\label{sp2alg}\eea
This single-oscillator realization leads to $\mso(D-1,2)$
representations carrying weights restricted by
\be M_{AB}\star M_{CD}\eta^{BD}=K^{ij}\star L_{ij,AC}-{D-3\over
2}\eta_{AC}-{i(D-1)\over 2}M_{AC}\ ,\label{id}\ee
where $L_{ij}^{AB}=\frac12 y^A_{(i}y^B_{j)}$. Taking further
traces yields
\bea C_2[\mso(D-1,2)]&=&-4C_2[\msp(2)]-{1\over 4}(D+1)(D-3)\
,\label{id1}\\
C_2[\mso(D-1)]&=&K^I\star L_{I,rr}+{1\over 2}(D-1)\Big(E-{D-3\over
2}\Big)\
,\label{id2}\\
L^+_r L^+_r&=&K^I\star(L_{I,00}+2iL_{I,0}+L_I)\ ,\label{id3}\eea
where $L_{I,0}=v^A L_{I,0A}$, $L_I=v^Av^B L_{I,AB}$, and
$C_2[\msp(2)] = K^I\star K_I=-\frac18 K^{ij}\star K_{ij}$.

The massive-particle states are defined by the Casimir constraint
\be (\ft12 K^{ij}\star K_{ij}-M^2L^2)\star\ket{\Psi}\ =\ 0\
,\qquad ML>0\ , \ee
which gives the AdS mass-shell condition, and one additional constraint
that is linear in the $\msp(2)$ generators, which can be taken to be either the embedding
condition
\be (X^2+L^2)\ket{\Psi}\ =\ 0\ ,\label{finiteTL}\ee
or, equivalently, via the $\integ_2$-symmetry \eq{Z2}, the
``ambient'' mass-shell condition
\be (P^2+M^2)\ket{\Psi}\ =\ 0\ . \label{finiteTL2}\ee
These yield the $\mso(D-1,2)$ weight space $\sfD(E_0,(0))$ with
lowest energy $E_0=\frac{D-1}2+\sqrt{1+(ML)^2}$. The embedding
condition \eq{finiteTL} yields the wave-function
\be \Psi(X^A)\ =\ \delta(X^2+L^2)\Phi(X^A)\ ,\label{eq:wave}\ee
with $\Phi(X^A)$ obeying the massive Klein-Gordon equation, giving
rise to a harmonic expansion with mode functions in
$\sfD(E_0,(0))$.

There are thus two \emph{inequivalent} ways of sending
$M\rightarrow 0$ at fixed $L>0$: using \eq{finiteTL}, which leads
to
\be (X^2+L^2)\ket{\Psi}\ =\ 0\ ,\qquad K^{ij}\star
K_{ij}\ket{\Psi}\ =\ 0\ ,\label{p1}\ee
describing $\sfD(E_0,(0))$ with $E_0=\frac{D+1}2$ constituting the
\emph{massless scalar field on $(AdS_D)_L$}; or, using
\eq{finiteTL2}, which yields
\be P^2\ket{\Psi}\ =\ 0\ ,\qquad K^{ij}\star K_{ij} \ket{\Psi}\ =\
-\frac12\left(\{X^A,P_A\}_{{\star}}+4i\right)\star
\{X^A,P_A\}_{\star}\ket{\Psi}\ =\ 0 \ ,\label{p2}\ee
in turn $\integ_2$-equivalent to a wave function on the Dirac
hypercone, given by
\be \Psi(X)\ =\ \d(X^2)\left(R^{-\D_-}\varphi_{-}(\widehat
X)+R^{-\D_+}\varphi_+(\widehat X)\right)\ ,\label{singWF}\ee
with $\D_-=\e_0=\frac{D-3}2$ and $\D_+=E_0={D+1\over 2}$, and
where we use the coordinates defined in \eq{conemetric}. Here
$\varphi_{\pm}(x)$ are off-shell scalar fields living on the
conifold $S^1\times S^{D-2}$. Their harmonic expansions yield the
Verma modules $V(E_0,(0))\simeq \sfD(E_0,(0))$, \emph{i.e.} the
$D$-dimensional massless scalar, and $V(\e_0,(0))$, which
contains the ideal
\be {\cal N}(\epsilon_0,(0))\ \simeq\ V(E_0,(0))\ .
\label{factor1}\ee
Its elimination leaves the scalar singleton
\be \sfD(\e_0,(0))\ =\ V(\epsilon_0,(0))/{\cal N}(\epsilon_0,(0))\
,\label{factor2}\ee
containing the on-shell modes of the conformally coupled scalar on
the conifold,
\be \left(\widehat \nabla^2_{S^1\times
S^{D-2}}-\frac{(D-3)^2}4~\right)\varphi_-(\widehat X)\ =\ 0\
.\label{eq:bfe}\ee
This mass-shell condition is $\integ_2$-equivalent to augmenting
\eq{p2} with $X^2\ket{\Psi}=0$, as can be seen from \eq{id3},
implying the strong $\msp(2)$-invariance condition
\be K_{ij} \ket{\Psi}\ =\ 0\ ,\label{eq:sph}\ee
which thus singles out the singleton in the singular wave function
\eq{singWF}.

\scsss{Some Geometric Aspects of the Singleton}\label{sec:geom}

The constraint \eq{eq:sph}, which corresponds to the limit
\eq{caseii}, describes the ``unbroken phase'' of the
$\msp(2)$-gauged sigma model, with action
\be S\ =\ \frac14 \int Y^{A i}D Y_{A i}\ .\label{S0}\ee
The large gauge transformation
\be \rho:\ Y^A_i\mapsto -Y^A_i\label{rho}\ee
descends to a reflection of the hypercone through the apex
preserving the time orientation,
\be \rho\ :\quad (R,\widehat X)\mapsto (-R,\widehat X)\ .\ee
The corresponding transformation of the singleton states read,
\be \rho\ket{\Psi}\ =\ (-1)^{\epsilon_0}\ket{\Psi}\
,\label{glan}\ee
that is, the singleton exhibits a global anomaly unless
$\epsilon_0=0$ mod $2$, \emph{i.e.}~$D=3$ mod $4$.

\begin{figure}[t]
\begin{center}
        \includegraphics[width=10cm,bb=0 0 700 500]{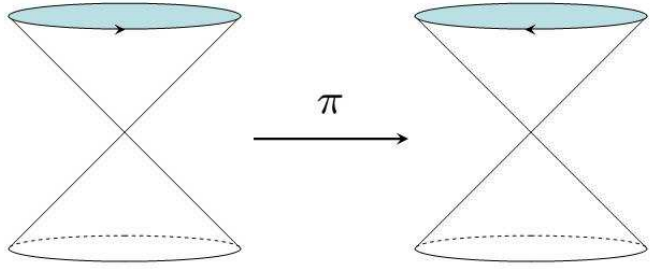}
\end{center}
\vspace{-2cm}
\caption{{\it Parity and time-reversal in ambient spacetime}: The
parity transformation $v_Ax^A_i\rightarrow -v_Ax^A_i$ reverses the
orientation of the worldline of a singleton rotating in ambient
space. }\label{fig1}
\end{figure}

Two other discrete maps are the combined space-time time
reversal and parity transformation $\pi$, defined in \eq{pi} and
\eq{pidpm}, and the related $\tau$-map defined in \eq{tau}. Thus,
if we let $\ket{\a}\equiv e^{i\pi\a/2}|\a+\e_0,(\a)\rangle$ and
$\ket{\widetilde\a}\equiv e^{i\pi\a/2}|-\a-\e_0,(\a)\rangle$ ($\a=0,1,2,\dots$)
denote real bases of the singleton $\sfD$ and the anti-singleton $\widetilde
\sfD$, respectively, and $\bra \a \equiv (\ket\a)^\dagger$ and
$\bra{\widetilde\a}\equiv(\ket{\widetilde\a})^\dagger$ denote the
corresponding bases of the dual spaces $\sfD^\star$ and $\widetilde\sfD^\star$,
with normalization chosen to be
\be \langle\a|\b\rangle\ =\
\langle\widetilde\a|\widetilde\b\rangle\ =\ \d_{\a\b}\
,\label{finorm}\ee
then
\be \pi(\ket{\a})\ =\ \ket{\widetilde\a}\ ,\qquad \pi(\bra{\a})\
=\ \bra{\widetilde\a}\ .\label{piket}\ee
Furthermore,
it can be seen from $\pi(L^\pm_r)=L^\mp_r$, $(L^\pm_r)^\dagger=L^\mp_r$ and
$\t(L^\pm_r)=-L^\pm_r$ that
\bea \t(\ket{\a})&=&\bra{\widetilde\a}\ ,\qquad \t(\bra{\a})\ =\
\ket{\widetilde\a}\ .\label{tauket}\eea
The $\pi$-map lifts to a parity transformation acting in ${\cal
W}[{\cal Z}]$ as
\be \pi(f(y^a_i,y_i))\ =\ f(y^a_i,-y_i)\ ,\qquad y_i\ \equiv \ v_A
y^A_i\ ,\label{pimap}\ee
inducing a reversal in the orientation of a singleton world-line,
as drawn in Fig.~\ref{fig1}. The $\tau$-map, which is a linear
anti-automorphism, lifts to a map acting in ${\cal W}[{\cal Z}]$
as
\be \t(f(y))\ =\  f(i y)\ ,\qquad \t(f\star g)\ =\ \t(g)\star
\t(f)\ ,\label{tauosc}\ee
which we also identify as a reversal of the orientation of the
worldline. Thus, this geometric transformation can be represented
either as an automorphism $\pi$, which separately exchanges
incoming singletons and anti-singletons and out-going dittos, or
an anti-automorphism $\t$, which exchanges incoming singletons
with out-going anti-singletons and vice versa.

The singleton state space,
\be \cH\ =\ \{\ket{\Psi}\,:\ K_{ij}\ket{\Psi}=0\}\
,\label{zeroTL}\ee
is characterized by the Casimir relations \eq{id1} and \eq{id2},
which imply
\bea C_2[\mso(D-1,2)|\cH] &=& -\ft14 (D+1)(D-3)\ ,\label{id4}\\
C_2[\mso(D-1)|\cH]&=& {1\over 2}(D-1)\Big(E-{D-3\over 2}\Big)\ .
\label{id5}\eea
These equations have two roots, given by
\bea &&E_0\ =\ \epsilon_0\ ,\qquad C_2[\mso(D-1)|S_0]\ =\ 0\
,\label{single0}\\ && E_0\ =\ 1\ ,\qquad C_2[\mso(D-1)|S_0]\ =\
{1\over 4}(D-1)(5-D)\ .\label{single1}\eea
The first root obeys the unitarity bound for $D\geq 3$. For $D=3$
it corresponds to the trivial representation, while for $D\geq 4$
it leads to the \emph{extended scalar singleton in $D\geq 4$}:
\be D\geq 4\ :\qquad \sfD_0 \equiv  \sfD(\epsilon_0,(0))\oplus
\widetilde \sfD(-\epsilon_0,(0))\ , \label{singleton}\ee
which is irreducible under $(\mso(D-1,2),\pi)$.

The second root \eq{single1} obeys the unitary bound only for
$D=3,4,5$. For $D=5$ it coincides with \eq{single0}. For $D=4$, it
corresponds to the spinor singleton $\sfD(1,1/2)$ of $\mso(3,2)$,
which cannot be realized using vector oscillators\footnote{The
spinor-singleton root ultimately does give rise to a physical
sector of the higher-spin gauge theory \cite{On}.}. For $D=3$, the
second root corresponds to a quasi-ground state of $\mso(2,2)$
with energy $E_0=|S_0|=1$, that combines with the trivial
representation into the indecomposable \emph{extended 3D scalar
singleton}
\be D=3\ :\qquad \sfD_0 \ =\  \sfD(0,0)\oplus \sum_{\l=\pm 1}
\left(\sfD'(1,\l) \oplus \widetilde \sfD'(-1,\l)\right)\
,\label{3dsingleton}\ee
where $\sfD'(\s,\l)=\{\ket{\s,\l;E,S}:E=\l S=\s,2\s,\dots\}$, and
$L^\pm_{\l'}\ket{\sigma,\lambda;E,S}=c_{\pm,\l';\s,\l;E,S}\times\ket{\s,\l;E\pm1,S+\l'}$
with $c_{\pm,\l';E,S;\s,\l}=0$ iff $E=S=0$ or $\l\l'=\pm 1$ (the
Casimir formula applies since $L^\pm_r L^\mp_r \ket{\pm 1,\pm
1}=0$). As a result, the \emph{extended chiral 3D scalar
singletons}:
\be \sfD_0^\pm\ =\ \sfD(0,0)\oplus \sfD'(1,\pm1)\oplus \widetilde
\sfD'(-1,\pm1)\ .\label{3dchiral}\ee
are irreducible under $\{\mso(2,2),\pi\}$. The special character
of $D=3$ stems from the fact that the conformally coupled scalar
field on $S^1\times S^1$ has vanishing critical mass, which yields
an unpaired zero mode corresponding to $\sfD(0,0)$.

The singleton wave-function,
\be \Psi(X)\ =\ R^{-\D_-}\d(X^2)\varphi_-(\widehat X)\
,\label{singWFsing}\ee
has divergent norm, unlike ordinary particles that have normalizable
wave-functions. Since the spectrum of the
anti-de Sitter energy operator acting in a lowest weight space
$\sfD(E_0;S_0)$ is discrete, ordinary particles are quantized as
if they were in a box, whereby wave-functions normalized to Kronecker
$\d$-functions couple to well-defined second-quantized creation and
annihilation operators. The wave-functions \eq{singWFsing} do not lend themselves
to this interpretation.

Let us therefore digress more into the details of the nature of
the singleton wave-functions following
\cite{Shaynkman:2001ip,misha}.


\scsss{Digression: Singleton Ground States}\label{sec:singgs}


Instead of using the wave-function representation, it is
convenient to work in the generalized Fock-space representation
\cite{Shaynkman:2001ip,misha} of the oscillators
\be w^A\ =\ \frac12 (y^A_1+iy^A_2)\ ,\qquad \bar w^A\ =\
(w^A)^\dagger\ ,\qquad [w^A,\bar w^B]\ =\ \eta^{AB}\ \
.\label{za1}\ee
The compact Fock-space vacuum defined by ($r=1,\cdots,D-1$,
$\a=0,0'$)
\be w_r\ket{0}\ =\ \bar w_\alpha \ket{0}\ =\ 0\ ,\label{za2}\ee
yields the following normal-ordered expressions for the
$\mso(D-1,2)$ generators $M_{AB}=2i\bar w_{[A}w_{B]}$,
\bea E &=& u^\dagger \star u- v^\dagger\star v\ ,\qquad M_{rs}=2i
a^\dagger_{[r}\star a_{s]}\ ,
\label{EMrs}\\
L^{-}_r&=&\sqrt{2}\left(u a_r + v^\dagger a^\dagger_r\right)\
,\qquad L^{+}_r = \sqrt{2}\left(v a_r+ u^\dagger
a^\dagger_r\right)\ , \label{lminusr}\eea
where $a^\dagger_r\equiv i \bar w_r$ necessarily create states
with integer $\mso(D-1)$ weights, $(s)\equiv (s0\dots 0)$,
$s=0,1,\dots$, and $u^\dagger\equiv (w_{0}- i w_{0'})/\sqrt{2}$
and $v^\dagger\equiv (w_{0}+i w_{0'})/\sqrt{2}$ are oscillators
with integer $\mso(2)_E$-helicity, that can be used to create
states with integer as well as half-integer energies. Thus, in
view of \eq{single0} and \eq{single1}, the ordinary Fock space
\be \cF\ =\ \left\{(u^\dagger)^{m}(v^\dagger)^{n} (a_r^\dagger
a_r^\dagger)^p a^\dagger_{\{r_1}\cdots a^\dagger_{r_n\}}\ket{0}\
,\quad m,n,p=0,1,\dots\right\}\ ,\label{Fockspace}\ee
suffices in odd dimensions, while even dimensions require a
representation of $u$ and $v$ in a non-standard generalized Fock
space, taken to be
\be \cF'\ =\ \cF_{1/2,0}\oplus \cF_{0,1/2}\
,\label{genFockspace}\ee
where
\be \cF_{1/2,0}\ =\ \left\{(u^\dagger)^{m+1/2}(v^\dagger)^{n}
(a_r^\dagger a_r^\dagger)^p a^\dagger_{\{r_1}\cdots
a^\dagger_{r_n\}}\ket{0}\ ,\quad m\in{\integ},\
n,p=0,1,\dots\right\}\ , \label{wavefunctions}\ee
and the states $(u^\dagger)^{m+1/2}\ket{0}$ ($m\in{\integ}$)
obey\footnote{In general, the state
$(u^\dagger)^z\ket{0}=\Gamma(z+1)\oint_\gamma {ds\over 2\pi i}
s^{-z-1}e^{u^\dagger s}\ket{0}$, where $\gamma$ is the closed
contour encircling $s=0$ and $s=\infty$ with $\arg s=\pi+\arg
u^\dagger$ where $\arg u^\dagger$ is determined by the eigenvalue
of the dual coherent bra-state.}
\bea u (u^\dagger)^{m+1/2}\ket{0} &=& (m+1/2)
(u^{\dagger})^{m-1/2}\ket{0}\ ,\label{rule1}\\
\langle
0|u^{m+1/2}(u^{\dagger})^{n+1/2}\ket{0}&=&\d_{mn}\C(m+3/2)\langle
0|0\rangle\ .\label{rule2}\eea
The introduction of the non-unitary sectors is necessary in order
for the spaces to be an oscillator module. However, as we shall
see, the physical states will belong to the unitary subsector.

The $\msp(2)$ generators $K_{+-}=\frac12\{w^A,\bar w_A\}_\star$,
$K_{++}=(K_{--})^\dagger=w^Aw_A$ read
\bea K_{+-}&=& {1\over 2}(K_{11}+K_{22})\se a^\dagger_r \star
a_r-u^\dagger\star u - v^\dagger\star v+{D-3\over 2}\ ,
\label{K11}\\
K_{++}&=& {1\over 2}(K_{11}-K_{22}+2iK_{12})\se a_r a_r + 2u^\dagger v^\dagger \ ,\label{K12}\\
K_{--}&=&{1\over 2}(K_{11}-K_{22}-2iK_{12})\se a^\dagger_r
a^\dagger_r + 2u v \ .\label{K22}\eea
Therefore, a state $\ket{E,(s)}\in \cH$ with AdS energy $E$ and
spin $(s)$, which can be expanded as
\be \ket{E,(s)}\ =\ \Psi^{(s)}_{r_1\dots
r_s}~a_{r_1}^\dagger\cdots a_{r_s}^\dagger \Psi_{E,s}(
u^\dagger,v^\dagger,a^\dagger_r a^\dagger_r )\ket{0}\ ,\ee
with traceless $\Psi^{(s)}_{r_1\dots r_s}$, is governed by the
differential equations
\bea (\partial_\x-\partial_\eta-E)\Psi_{E,s}&=&0\ ,\label{omega1}\\
(2\partial_\zeta-\partial_\x-\partial_\eta+\epsilon_0+s)\Psi_{E,s}&=&0\ ,\label{omega3}\\
(2(\partial_\zeta^2+(\epsilon_0+s)\partial_\zeta)+e^{\x+\eta+\zeta})
\Psi_{E,s}&=&0\ ,\label{omega4}\\
(e^{\x+\eta+\zeta}+2\partial_\x\partial_\eta)\O_{E,s}&=&0\
,\label{omega5}\eea
where
\be \x\ =\ \log u^\dagger\ ,\quad \eta\ =\ \log v^\dagger\ ,\quad
\zeta\ =\ \log a^\dagger_r a^\dagger_r\ .\ee
Eqs.~\eq{omega1} and \eq{omega3} have the solution
\be \Psi_{E,s}\ =\ e^{\ft{E}2(\x-\eta)-\ft{\epsilon_0+s}2\zeta}
f_E(x)\ ,\qquad x\equiv e^{\x+\eta+\zeta}\ =\ u^\dagger v^\dagger a^\dagger_r a^\dagger_r\
.\label{PsiEs}\ee
The remaining conditions \eq{omega4} and \eq{omega5} imply
\be \left((x{d\over dx})^2+\ft12(x-\ft{E^2}2)\right)f_E(x)\ =\ 0\
,\qquad E^2\ =\ (\epsilon_0+s)^2\ ,\label{f0eq}\ee
\emph{i.e.} $f_E(x)$ obeys Bessel's differential equation in
$z=\sqrt{2x}$ with index $|E|$.

To describe the solutions one introduces the contour integral
\be F_{\nu,\gamma}(z)\ =\ \int_\gamma {ds\over 2\pi i}
(1-s^2)^{\nu-1/2}e^{sz}\ ,\ee
which solves the related differential equation
\be
\Big(z-(2\nu+1)\frac{d}{dz}-z\frac{d^2}{dz^2}\Big)F_{\nu,\gamma}(z)\
=\ 0\ ,\label{diffekv}\ee
provided that
\be \left[(1-s^2)^{\nu+1/2} e^{sz}\right]_{\partial\gamma}\ =\ 0\
,\ee
so that $F_{\nu,\gamma}(z)/z^\nu$ solves Bessel's differential
equation in $z$ with index $\nu$. The contours with
$\partial\gamma\in\{\pm 1,\pm\infty\}$ correspond to the Bessel
and Neumann functions. It is also possible to take
$\partial\gamma=\{-i\infty,i\infty\}$, by making use of the
prescription
\be \delta(z)\ =\ \int_{-i\infty}^{i\infty}{ds\over 2\pi i}
e^{sz+\epsilon s^2}\ ,\label{deltaZ}\ee
where $\epsilon\rightarrow 0^+$. This yields the distribution
\be F_\nu(z)\ =\ \sum_{k=0}^\infty{\nu-1/2\choose
k}(-1)^k\delta^{(2k)}(z)\ ,\label{dist}\ee
that has a well-defined action on test functions $h(z)$
($z\in\Real$) with $h^{(k)}(0)$ falling off with $k$ sufficiently
fast, as to justify the geometric series expansion. Indeed,
\eq{dist} solves \eq{diffekv} in the sense that
\be \int_{-\infty}^\infty dz~
h(z)~\left(z-(2\nu+1)\frac{d}{dz}-z\frac{d^2}{dz^2}\right)F_{\nu}(z)\
=\ 0\ .\ee
In summary, there are two solutions corresponding to states in the
generalized Fock spaces, namely the function
\be f^{(1)}_E(x)\ \equiv\  {\cal N}_{E}~J_{|E|}(\sqrt{2x})\
,\label{fE}\ee
which is analytic at the origin, and the distribution
\be f^{(2)}_E(x)\ =\ {\cal N}'_{E} (2x)^{|E|\over 2}
F_{|E|}(\sqrt{2x})\ .\label{distfE}\ee
The solution related to the Neumann function, which contains a
logarithm, will not be considered here. The existence of both
analytic and distributional wave-functions is analogous to the
existence of analytic and distributional phase-space and
spinor-space projectors found in \cite{Sagnotti:2005ns}.

The action of $L^{\pm}_r$ on the states $\ket{E,(s)}^{(q)}$ built
from $f^{(q)}_E(x)$ ($q=1,2$) is given by $
L^\pm_r\ket{E,(s);q}= c_{E,s}\ket{E\pm1,(s\pm {E\over
|E|});q}$. Since $s=|E|-\epsilon_0\geq 0$, it follows that if
$\epsilon_0\geq 1/2$, i.e. $D\geq 4$, then
\be D\geq 4~:\quad \ket{\O_\pm;q}\ \equiv\
\ket{\pm\epsilon_0,(0);q}\ ,
\label{groundstates}\ee
are singleton ground states, obeying
$L^\mp_r\ket{\O_\pm;q}=0$. Indeed,
$L^-_r\ket{\epsilon_0,(0);q}$ has a wave-function
proportional to
$(2\partial_\xi\partial_\zeta+e^{\x+\eta+\zeta})\Psi^{(q)}_{\epsilon_0,0}$,
that in turn vanishes due to \eqs{omega1}{omega5}. For $D=3$, the
states $L^\pm_r\ket{E,(s);q}$ with $|E|=s\geq 1$ are
non-vanishing, while $L^\pm_r\ket{0,(0);q}=0$. Thus $\ket{\pm
1,(1);q}$, each of which decompose under $\mso(2)_S$ into
helicity eigen-states $\ket{\pm 1,\l;q}$ with $\l=\pm 1$, are
the quasi-ground states for the spaces $D'(1,\l)$ and $\widetilde
D'(-1,\l)$ defined in \eq{3dsingleton}.

In summary,
\be \quad \cH\ =\ \sfD_0^{(1)}\oplus \sfD_0^{(2)}\
,\label{cHsummary}\ee
where $\sfD^{(q)}_0$ are extended singletons isomorphic to
\eq{singleton} and \eq{3dsingleton}. The phase-space reflection
\eq{pimap} acts as $\pi(\Psi(u^\dagger,v^\dagger,a^\dagger_r))=
\Psi(v^\dagger,u^\dagger,a^\dagger_r)$. Thus, declaring
$\pi\ket{0}=\ket{0}$, leads to that
$\pi(\ket{\O_\pm;q})=\ket{\O_\mp;q}$, so that
$D_0^{(q)}$ are irreducible under
$\{\mso(D-1,2),\pi\}$, in the sense of \eq{pidpm}.

The explicit expressions for the ground states read \cite{misha}
\be \ket{\O_\sigma;q}\ =\
\cN^{(q)}\left[(u^\dagger)^{(1+\s)/2}
(v^\dagger)^{(1-\sigma)/2}\right]^{\epsilon_0}\int_{\gamma^{(q)}}
ds~(1-s^2)^{\epsilon_0-1/2}~e^{s\sqrt{2 u^\dagger v^\dagger
a^\dagger_r a^\dagger_r}+\epsilon s^2}\ket{0}\ ,\ee
where $\gamma^{(1)}=[-1,1]$ and $\gamma^{(2)}=]-i\infty,i\infty[$.
The analytic ground states
\be \ket{\O_{\sigma};1}\ =\ \left[(u^\dagger)^{(1+\s)/2}
(v^\dagger)^{(1-\sigma)/2}\right]^{\epsilon_0}\sum_{n=0}^\infty
{(-1)^n\over 2^n n!} {\Gamma(\ft{D-1}{2})\over
\Gamma(n+\ft{D-1}2)}(u^\dagger v^\dagger a^\dagger_r
a^\dagger_r)^n\ket{0}\ ,\label{O0}\ee
are highly squeezed oscillator states, intersecting only the
positively normed subspaces of $\cF_{1/2,0}$ and $\cF_{0,1/2}$.
For $D=3$, the analytic quasi-ground states are given by,
\be \ket{\s,\l;1}\ =\ \l^r a^\dagger_r(u^\dagger)^{(1+\s)/2}
(v^\dagger)^{(1-\sigma)/2}\sum_{n=0}^\infty {(-1)^n\over 4^n
n!(n+1)!}(u^\dagger v^\dagger a^\dagger_r a^\dagger_r)^n\ket{0}\
.\label{3Dvacuum}\ee

The norms of the ground states have a divergent nature,
\emph{e.g.}
\be \|\ket{\O_\pm;1}\|^2\ =\ \Gamma(\ft{D-1}2) \sum_{n=0}^\infty
\langle 0|0\rangle\ .\label{normO0}\ee
Similarly, the inner products $\langle\Omega_\pm;2|\Omega_\pm;2\rangle$
and $\langle\Omega_\pm;2|\Omega_\pm;1\rangle$ have divergent
integral representations.

We \emph{conjecture that there exists a well-tempered linear
combination of the analytical and the distributive ground states
with finite norm},
\be \ket{\widehat\O_\pm}\ =\ c_1 \ket{\O_\pm;1}+c_2\ket{\O_\pm;2}\
,\label{conj}\ee
where $c_{1,2}$ are finite constants and
\be \|\ket{\O_{\pm}} \|^2\ =\ 1\ ,\ee
corresponding to a proper normalizable singleton subspace of
\eq{cHsummary},
\be \widehat\cH\ =\ {\rm diag} \left(c_1\sfD_0^{(1)}\oplus
c_2\sfD_0^{(2)}\right)\ \simeq \sfD_0\ . \label{widehatcH}\ee


\scsss{Remark on the $D$-Dimensional Scalar}\label{sec:Dsc}


Out of curiosity, let us examine the weakly $\msp(2)$-invariant
spaces
\be \cH_{\pm}\ =\ \Big\{\ket{\Psi_\pm}\ :\quad
K_{\pm\pm}\ket{\Psi_\pm}\ =\ K_{\pm\mp}\ket{\Psi_\pm}\ =\ 0\Big\}\
,\ee
which contain the singleton, $\cH_{\pm}\supset\cH$, that can be
projected out by imposing either $K_{\mp\mp}\ket{\Psi_\pm}=0$ or,
\emph{equivalently}, one of the lowest-weight conditions
$L^-_r\ket{\Psi_\pm}=0$ or $L^+_r\ket{\Psi_\pm}=0$. Hence, the
only lowest-weight module contained in $\cH_{\pm}$ is the
singleton. The spaces $\cH_\pm$ also contain a $D$-dimensional
scalar, residing in the manifestly Lorentz-covariant $\mso(D-1,2)$
module
\be V_\pm\ =\ \bigoplus_{n=0}^\infty P_{\{a_1}\star
P_{a_n\}}\ket{\O_\pm}\ ,\qquad
K_{\pm\pm}\ket{\O_\pm}=K_{\pm\mp}\ket{\O_\pm}=M_{ab}\ket{\O_\pm}=0\
,\ee
which is not of lowest-weight type.

Using the Lorentz-covariant oscillator vacuum ($a=0,\dots,D-1$)
\be \a_a\ket{\hat 0}\ =\ \beta\ket{\hat 0}\ =\ 0\ ,\ee
where $\a_a\equiv w_a$ and $\beta\equiv \bar w_{0'}$, the
conditions on
$\ket{\O_+}=\O_+(\a^{a\dagger}\a^\dagger_a,\beta^\dagger)\ket{\hat
0}$ become
\bea
(\a^a\a_a-(\b^\dagger)^2)~\O_+(\a^{a\dagger}\a^\dagger_a,\beta^\dagger)\ket{\hat
0 }\
&=&0\ ,\\
(\a^{a\dagger}\a_a-\b^\dagger\b+{D-1\over
2})~\O_+(\a^{a\dagger}\a^\dagger_a,\beta^\dagger)\ket{\hat 0}&=&0\
,\eea
which are equivalent to $\O_+=(\beta^\dagger)^{D-1\over 2}f_+(y)$,
$y^2=\a^{a\dagger}\a^\dagger_a(\beta^\dagger)^2$, with
\be \left(y{d^2\over dy^2}+(D-1){d\over dy}-y\right)f_+(y)\ =\ 0\
,\ee
\emph{i.e.} $y^{-\nu}f_+(y)$ obeys Bessel's differential equation
in $y$ with index $\nu={D-2\over 2}$ (that happens to coincide
with the index of the Bessel functions related to the singular
projector $M$ defined in \eq{singproj}). This implies that
\be P_a\ket{\O_+}\ =\ i(3-D)(\b^\dagger)^{D+1\over 2}\a^\dagger_a
(2y)^{-1}f'_L(y)\ket{0_+}\ .\ee
For $D=3$, $V_+$ is trivial, and indeed $K_{--}\ket{\O_+}=0$, so
that $V_+=\sfD(0,0)$. For $D>3$, $V_+$ is a non-trivial
representation space for $\mso(D-1,2)$, with
$C_2[\mso(D-1,2)|V_+]$ given by \eq{id4}, since
$C_2[\msp(2)|V_+]=0$.


\scss{$\msp(2N)$-Gauged Brane-Parton
Model}\label{sec:matrix}


In this section we examine the tensionless limit of the discretized
$p$-brane in anti-de Sitter spacetime at the classical level.
We are working covariantly in embedding space, \emph{i.e.} we do
not solve the gauge conditions on the embedding fields. As a result, we
end up with a model with an enhanced gauge symmetry,
restricting the classical partonic phase space substantially,
while being less restrictive at the quantum level, leaving
the expected partonic quantum states, as we shall discuss in Section \ref{sec:global}.

To discretize the Nambu-Goto action \eq{eq:ng} \cite{deWit:1988ig}\footnote{The discretization
procedure can also be performed fully covariantly using the einbein-formulation given in
\cite{lindstrom}.}, it is convenient to follow the approach
in which one introduces an auxiliary metric, $\gamma_{\alpha\beta}$, so that
\be S_p\ =\ -\frac{T_p}2 \int d^{p+1}\sigma
\sqrt{-\det{g}}\left(\gamma^{\alpha\beta}
g_{\alpha\beta}-(p-1)+\Lambda(X^2+L^2)\right)\ ,\ee
and then integrating out the spatial metric $\gamma_{rs}$ in the
gauge $\gamma_{00}=-\det \gamma_{rs}$ and $\gamma_{0r}=0$
($\alpha\rightarrow (0,r)$, $r=1,\dots,p$), imposed using the
$(p+1)$-dimensional diffeomorphisms. After re-scaling
$\sigma^r\rightarrow T_p^{-1/p}\sigma^r$, one finds
\be S_p\ =\ \frac12 \int d^{p+1}\sigma \left( \dot X^2-T_p^2\det'
g+\Lambda(X^2+L^2)\right)\ ,\ee
where $\det' g=\det g_{rs}$. The physical states obey the
mass-shell condition and Gauss' law,
\be \dot X^2+T_p^2\det' g\ =\ 0\ ,\qquad \dot X^A~\partial_r X_A\
=\ 0\ ,\ee
and the second-class embedding constraints
\be X^2+L^2\ =\ 0\ ,\qquad \dot X^A X_A\ =\ 0\ .\ee

To discretize one replaces
\be X^A(\tau,\sigma^r)\rightarrow
\left\{~X^A(\tau;\xi)~\right\}_{\xi=1}^N\ ,\ee
where $X^A(\tau;\xi)$ are degrees of freedom, referred to as partons
(see also \cite{thorn}), living on $N$ sites labelled by $\xi$.
This requires a prescription for replacing derivatives with
respect to $\sigma^r$ by difference operators,
\be {\partial X^A\over\partial \s^r} \rightarrow \mu^{-1/p}
\sum_\eta w_{\xi,\eta} X^A(\eta)\ ,\qquad \sum_\eta
w_{\xi,\eta}=0\ ,\ee
where $\mu$ is a fixed mass parameter, and $w_{\xi\eta}$ are some
weights, which give rise to interactions that are irrelevant in
the limit $T_p\ll \mu^{p+1}\rightarrow 0$. Gauss' law now takes
the form $\sum_\eta w_{\xi,\eta} \dot X^A(\xi) X_A(\eta)=0$.
Demanding closure of the constraint algebra in the limit $\mu L=0$
\cite{zeroradius} \emph{and} $\mu^{-p-1}T_p=0$
\cite{schild,lindstrom,Bonelli}, we find that for \emph{any}
choice of weights the only possibility is the $\msp(2N)$
constraint
\be \dot X^A(\xi) X_A(\eta)\ =\ 0\ ,\qquad \forall ~\xi,\eta\ .\ee
Thus we find an $\msp(2N)$-gauged sigma model, with the
following phase-space action
\be S_N\ =\ \frac14 \int Y^{AI}DY_{AI}\
,\label{eq:SN1}\ee
where $I$ labels an $2N$-plet, and ($\xi=1,\dots,N$; $i=1,2$)
\be Y^{AI}=\{Y^{Ai}(\xi)\}\ ,\qquad Y^{Ai}(\xi)\ =\
\sqrt{2}\left(X^{A}(\xi),P^{A}(\xi)\right)\ ,\ee
are the phase-space coordinates of the partons, and the $\msp(2N)$-covariant derivative is defined by
\be DY^{AI}\ =\ dY^{AI}+\Lambda^{IJ}Y^A_J\ ,\label{eq:2Pcd}\ee
where $\Lambda^{IJ}$ is a $(0+1)$-dimensional $\msp(2N)$
gauge field. The symplectic indices are raised and lowered as
$Y^{AI}=\Omega^{IJ}Y^{A}_{J}$ and $Y^A_I=Y^{AJ}\Omega_{JI}$ with
$\Omega_{IJ}=\epsilon_{ij}\delta_{\xi\eta}$.

The $\msp(2N)$ generators
\be T_{IJ}\ =\ \frac12 Y^A_IY_{AJ}\label{eq:KIJ}\ee
consist of $N$ $\msp(2)$ blocks $K_{ij}(\xi)$ along the
diagonal containing the mass-shell conditions and geometric
constraints of the separate partons, and off-diagonal generators
$T_{i,j}(\xi,\eta)=Y^A_i(\xi) Y_{Aj}(\eta)$, $\xi\neq\eta$,
comprising the discretized Gauss' law as well as additional ${\cal
W}$-constraints of a continuum limit to be discussed in more
detail below.

The large gauge transformations are generated by phase-space
\emph{reflections}
\be \rho_\xi:\ Y^{Ai}(\eta)\rightarrow (-1)^{\delta_{\xi\eta}}
Y^{Ai}(\eta)\ ,\label{refl}\ee
and \emph{permutations}
\be P_{\xi\eta}:\ Y^{Ai}(\xi)\rightarrow Y^{Ai}(P(\xi))\
.\label{perm}\ee
The latter are large discrete diffeomorphisms arising at the point
$T_p=0$, where thus the original $p$-dimensional nature of the
worldvolume is lost.

The classical equations of motion,
\be T_{IJ} \ =\ 0\ ,\qquad \dot Y^A_I+\Lambda_I{}^JY^A_J=0\
,\label{EoM}\ee
imply that the classical space-time angular momenta, defined by
\be M_{AB}=\frac12 Y^I_A Y_{IB}\ ,\ee
are light-like,
\be M_{A}{}^{C}M_{BC}\ =\ 0\ ,\ee
where we note that
\be M_A{}^CM_{BC}\ =\ T^{IJ}L_{IJ,AB}\ ,\qquad L_{IJ,}{}^{AB}\ =\
\frac 12 Y^A_{(I}Y^B_{J)}\ .\ee
This degeneracy is lifted in the quantum theory, which has a
discrete spectrum, essentially due to the underlying $\mso(D-1,2)$
covariance.

The naive dimension of the space ${\cal M}_N$ of
classical solutions is given by $2N(D+1)-2\times
\frac{2N(2N+1)}2=2N(D-2N)$, that becomes negative for large enough $N$. The
actual dimension is, however, positive for all $N$. To describe the solutions
one can fix the harmonic gauge
\be \Lambda^I{}_J\ =\ \mathbf 1_{N\times N}\otimes \epsilon\ ,\label{eq:sp2Ngauge}\ee
where $\epsilon=i\sigma^2$. The solutions to \eq{EoM} are now
given by
\be Y^A_1(\xi)\ =\ q^A(\xi)\cos \tau+p^A(\xi)\sin \tau\ ,\ee
where $q^A(\xi)$ and $p^A(\xi)$ are constants of motion obeying
\be q(\xi)\cdot q(\eta)=q(\xi)\cdot p(\eta)=p(\xi)\cdot p(\eta)=0\
,\qquad \forall \xi,\eta\ .\ee
The space of classical solutions is obtained by factoring out the
group $G$ of residual gauge transformations preserving \eq{eq:sp2Ngauge},
\be {\cal M}_{N}\ =\
\left\{\big(q^A(\xi),p^A(\xi)\big)\right\}/G\ , \qquad G=SO(2N)\cap
Sp({2N})\ .
\ee
A particularly simple class of solutions is thus given by
\be q(\xi)\ =\ q\ ,\qquad p(\xi)\ =\ p\ ,\qquad (q,p)\in {\cal
M}_1\ .\ee
For $q\neq p$ these solutions are rotating clusters of partons,
with
\be M_{AB}(N)\ =\ N M_{AB}(1)\ ,\ee
while $q=p$ leads to ``pulsating'' clusters, with
\be M_{AB}(N)\ =\ 0\ .\ee
We see that $\dim{\cal M}_{N}\geq \dim {\cal M}_1$ for all $N$,
while $\dim{\cal M}_{N}\ll N\dim {\cal M}_1$ as $N$ becomes large.
Thus, the space of classical solutions to the discretized
tensionless model is much smaller than the space of cusp solutions
of the tensile branes. This puzzle is resolved, however, at the
quantum level, where the gauge symmetry is much less restrictive,
as we shall see in Section \ref{sec:global}.

Before turning to this crucial issue, let us briefly address the
universality of the mechanism by which the constraints of the
singular geometry and the $p$-dimensional diffeomorphisms combine
into an $\msp(2N)$ gauge algebra. This suggests that different
$p$-branes result from different tensile deformations. However, as
described in Section \ref{sec:rotp}, the cusps on $p$-branes with
$p>2$ give rise to additional degrees of freedom that cannot
directly be related to singletons. Moreover, for $p=1$ the cusps
interact via a linear potential, suggesting that $S_N$-invariance
is actually broken down to $\integ_N$-invariance, as discussed in
Section \ref{sec:gas}. A natural resolution could be that tensile
deformations lead to the string in case they preserve
$\integ_N$-invariance and to the membrane in case they preserve
$S_N$-invariance. Let us demonstrate the latter in a particular
example.


\scss{A Tensile Membrany Deformation}\label{sec:M2def}


A small tensile perturbation of the $\msp(2N)$-gauged
action that preserves the $S_{N}$-invariance is given by
\be \la{defaction} S\ =\ {\mu\over 2}\int d\t\sum_{\xi=1}^{N}\sum_{\eta\neq \xi}\Big\{{\dot X}^2(\xi)
-k^2\big(X(\xi)-X(\eta)\big)^2+\L(\xi)\big(X^2(\xi)+L^2\big)\Big\}\
, \ee
with corresponding Virasoro-like constraints
\bea
&&{\dot X}^2(\xi)+k^2\sum_{\eta\neq\xi}\big(X(\xi)-X(\eta)\big)^2\;\;=\;\;0\ , \\
&&\sum_{\eta\neq\xi}\Big(X(\xi)\cdot {\dot X}(\eta)-X(\eta)\cdot
{\dot X}(\xi)\Big)\;\;=\;\;0\ , \eea
where $k\equiv T_p/\mu^{p+1}\ll1$, the long-range difference
operator corresponds to the behavior of the leading-order
interactions between the cusps, and, in a self-consistent fashion,
the $\msp(2N)$ gauge symmetry has been partly gauged
fixed and partly broken, leaving $N$ Lagrange multiplies
$\L(\xi)$ that will be determined by the equations of motion.

The classical solution describing co-planar and unison rotation
with mutual relative azimuthal angles $2\pi/N$ reads
\bea
X^{0'}(\xi)+iX^0(\xi)&=&L\cosh\frac{r_0}{L}e^{i\o(r_0)\t}\ , \\
X^{1}(\xi)+iX^2(\xi)&=&L\sinh\frac{r_0}{L}e^{i\tilde\o(r_0)\t}e^{2\pi(\xi-1)/N}\ , \\
X^A&=&0\ ,\quad A=3,\ldots,D-1\ , \eea
where the equations of motion and constraints determine the
angular velocities to be $\o^2(r_0)=\tilde\o^2(r_0)-2k^2N=4k^2N\sinh^2(r_0/L)$. The discrete symmetry of the
configuration implies that $M_{AB}(\xi)$ is independent of $\xi$,
resulting in total energy and spin given by
\bea
E_{\rm cl}&=&NL^2\o(r_0) \cosh^2\frac{r_0}{L}\ , \\
S_{\rm cl}&=&NL^2\tilde\o(r_0) \sinh^2\frac{r_0}{L}\ .
\eea
In the limit $r_0/L\ll1$, the dispersion relation becomes
$E^2_{\rm cl}\sim N^{3/2}L^2kS_{\rm cl}$, reminiscent
of that of a short string, \emph{i.e.} a string in flat spacetime.
On the other hand, for $r_0/L\gg1$, the classical charges scale
like $kL^2\exp(3r_0/L)$, and
\bea E_{\rm cl}-S_{\rm cl}&\sim &(k^2S_{\rm cl})^{1/3}\ ,\qquad
k\ll1\ , \eea
in agreement with the near tensionless behavior of the folded
rotating tensile membranes in AdS$_D$ found in \eq{m2anomaly}.
This result is consistent with the expected $S_N$-invariance of
the two-dimensional gas of cusps on the tensile membrane. Indirectly,
this also lends suppert to the idea that tensile deformations that break
$S_N$ to $\integ_N$ should lead to stringy dispersion relations.


\scss{Quantization and Global Wave-Function Anomalies}\label{sec:global}


As we found in Section \ref{sec:matrix}, the phase space of the
classical $\msp(2N)$-gauged model is too restricted to
match that of the cusps on the tensile brane.
However, at the quantum level, the expectation values of the
$\msp(2N)$ generators vanish provided the $N$-parton states $\ket{\Psi}$ obey\footnote{The
$N$-oscillator analog of \eq{id},
$$M_{AC}\star M_{B}{}^C\ =\ K^{IJ}\star L_{IJ,AB}-\frac{i}2
(D-1)M_{AB}-\frac{N}2(D-2N-1)\eta_{AB}\ ,$$
shows that the strong $\msp(2N)$-invariance condition
$K_{IJ}\ket{\Psi}=0$ is incompatible with the unitarity bounds,
except for certain ``sporadic'' values of $N>1$ and $D$,
that we shall not consider here. }
\be K_{ij}(\xi)\ket{\Psi}\ =\ 0\ ,\qquad \xi=1,\ldots,N\ .\label{strong}\ee
This condition can be solved by tensoring
$N$ normalizable single-parton states belonging to
$\widehat\cH \simeq \sfD_0$, \emph{i.e.} the scalar singleton, as given in \eq{widehatcH}.
Thus, the quantum state space is much richer than the classical phase space, \emph{i.e.} the correspondence principle is not valid.

Turning to the invariance under large $Sp(2N)$
transformations, the gauging of the permutations \eq{perm}, which
amounts to imposing
\be P_{\xi\eta}\ket{\Psi}\ =\ \ket{\Psi}\ ,\ee
implies that the multi-parton spectrum is fully Bose symmetrized,
\emph{viz}
\be {\cal S}_{\rm partons}\ =\ \bigoplus_{N=1}^\infty
\left[\sfD_0\right]^{\otimes N}_{\rm symm}\
.\label{Sbits}\ee
Thus, there is a correspondence between the quantized discretized tensionless model
and the partonic region of the phase space of the tensile brane. We also note that
the state space \eq{Sbits} is isomorphic to the the spectrum \eq{1elbrane} of the singular
conformal field theory, which, strictly speaking, also contains
singletons and anti-singletons in $\sfD_0$ realized as
second-quantized creation and annihilation operators.

However, in view of \eq{glan}, the gauging of the reflections
\eq{refl} is, however, possible in ${\cal S}_{\rm partons}$ only for
\be D\ =\ 3\ {\rm mod}\ 4\ ,\label{3mod4}\ee
while it leads to the unnatural truncation $N\e_0=0$ mod
$2$ for other values of $D$.

The global anomaly seems unnecessary in the massless sector, and
moreover, it would be desirable with a more transparent
formulation free from the subtleties associated with the
wave-functions.


\scss{Phase-Space Approach}\label{sec:phase}


The phase-space formulation of quantum mechanics is based on a
deformed product of phase-space functions, denoted by $\star$; a
set of functions with well-defined $\star$-composition, referred
to as operators; and, a trace operation $Tr$ analogous to
phase-space integration. There is no \emph{a priori} reference to
states, or any probabilistic interpretation, so the quantum theory
presents a smooth deviation away from the classical Liouville
equation
\cite{Weyl:1927vd,Wigner:1932eb,Moyal:1949sk,Bayen:1977ha}.

Field-theoretically, the approach is based on phase-space
correlators with boundary conditions of various complexity. The
basic formulation uses symmetric conditions at $\tau\rightarrow
\pm\infty$, \emph{viz.}
\be \left\langle{\cal O}\right\rangle_y\ \equiv\
\int_{Y(\pm\infty)=y}{\cal D} Y {\cal O}[Y] e^{{i\over
\hbar}S[Y]}\ ,\label{phsp}\ee
where $Y$ denote the phase-space coordinate, $y$ a point in phase
space, so that the standard trace becomes
\be Tr {\cal O}\ =\ \int [dy] \left\langle{\cal O}\right\rangle_y\
.\label{phsptrace}\ee
The sigma-model action $S$ is assumed to be diffeomorphism
invariant, which in general may require additional gauge fields,
so that expectation values of ultra-local operators ${\cal
O}_f=f(Y(\t))$, where $f$ are phase-space functions, are given by
$\left\langle {\cal O}_f \right\rangle_y=f(y)$, independently of
$\tau$. The $\star$-product is then induced from expectation
values
\be \left\langle {\cal O}_{f_1}{\cal O}_{ f_2}\right\rangle_y\ =\
\e(\t_1-\t_2)(f_1\star f_2)(y)+\e(\t_2-\t_1)(f_2\star f_1)(y)\
.\ee
We shall use the notation $Tr {\cal O}_f=Tr f$ and $Tr[{\cal
O}_{f_1}{\cal O}_{ f_2}]=Tr[f_1\star f_2]$.

In the phase-space formulation, external multi-singleton states $\Psi_{1\dots
P}=[\ket{\a_1}\otimes\cdots \otimes\ket{\a_{P}}]_{\rm symm}$ are
represented by vertices
\be {\cal V}\ \in\ \bigoplus_{\tiny\ba{c}P_1+P_2=P\\P_{1,2}>0\ea}
\left[(\sfD^{\otimes P_1}\otimes \widetilde \sfD^{\star\otimes
P_2})\oplus (\widetilde \sfD^{\otimes P_1}\otimes
\sfD^{\star\otimes P_2})\right]_{\rm symm}\ ,\label{calV}\ee
describing processes in which $P_2$ singletons are emitted while
$P_1$ singletons are absorbed, whereby the total number of
internal singletons is changed by $P_1-P_2$. The vertices act in
the space ${\cal S}_{\rm partons}$ given in \eq{Sbits}, which we may
view as a generalized Chan-Paton factor representing an internal
singleton cluster.

The two-singleton vertices act invariantly (and in fact
irreducibly) on each term in the generalized Chan-Paton factor.
Their action on the single singleton corresponds to
$\msp(4)$-gauge invariant and higher-spin invariant world-line
correlators related to topological open-string correlators with
consistent ``factorization'' properties manifesting themselves in
higher-spin gauge-field equations.

In what follows we shall discuss the world-line correlators, the
higher-spin structures and the topological closed string underlying
the generalized Chan-Paton factor, and then turn in Section
\ref{sec:tos} to the open-string reformulation of the singleton.


\scss{$\msp(4)$-invariant World-line Correlators}\label{sec:sp4}


Here we shall apply the phase-space approach to singleton quantum mechanics.
The steps leading from \eq{masslessst} to \eq{defvertex} via \eq{vertex} are
heuristic in nature, motivated by physical considerations, and
guided by algebraic structures lifted from higher-spin gauge theory.

The symmetric correlators of the free singleton theory based on
the $\msp(2)$-gauged action $S[Y,\L]=\frac14 \int Y^{Ai}DY_{Ai}$
are given by
\be \left\langle {\cal O}_f \right\rangle_y\ =\
\int_{Y^A_i(\pm\infty)\ =\ y^A_i}\left[{\cal D}Y{\cal
D}\L\right]~f(Y^A_i(\t))~e^{iS[Y,\L]}\ .\ee
Barring details of the BRST treatment, the gauge fixing amounts
to setting the gauge field $\L_{ij}$ equal to a fixed
value, which we shall take to be $\L_{ij}=0$, and imposing its classical equation
of motion, that is $Tr[K_{ij}{\cal O}]=0$. This is tantamount to decoupling
ideal elements belonging to
\be {\cal I}[\msp(2)]\ =\ \left\{ K_{ij}\star X^{ij}\in {\cal
P}[{\cal Z}]\right\}\ ,\ee
where $X^{ij}$ denote arbitrary triplets, and
${\cal P}[{\cal Z}]$ is the space of $\msp(2)$-invariant
functions
\be {\cal P}[{\cal Z}]\ =\ \left\{f\in{\cal W}[{\cal Z}]\ :\quad
[K_{ij},f]_\star=0\right\}\ .\label{PY}\ee
This is solved by
taking the observables to be elements of the weakly
$\msp(2)$-projected space
\be {\cal P}_0[{\cal Z}]\ =\ {{\cal P}[{\cal Z}]\over {\cal
I}[\msp(2)]}\ ,\ee
and inserting a projector -- the phase-space analog of an ordinary
space-time propagator -- into the trace of the ungauged theory.
The resulting gauged traces $Tr_\pm$ are given by
\be Tr_\pm f\ =\ tr_\pm [f\star M]\ ,\label{singTr}\ee
where $M=M(K^2)$ is the \emph{singular phase-space projector}
defined by \cite{Vasiliev:2003ev,misha,Sagnotti:2005ns} (see also
\cite{Sezgin:2001zs,Vasiliev:2001wa,Sezgin:2001yf,Sezgin:2001ij})
\be K_{ij}\star M\ =\ 0\ ,\qquad M(0)\ =\ 1\ ,\label{singproj}\ee
and $tr_\pm$ are the $(\pm)$-graded traces of ${\cal W}[{\cal Z}]$
defined by
\bea tr_+ f&=& \int_{{\cal Z}}
{d^{2(D+1)}y\over{(2\pi)^{D+1}}}\left\langle {\cal
O}_f\right\rangle_y\ =\ \int_{{\cal Z}}
{d^{2(D+1)}y\over{(2\pi)^{D+1}}} f(y)\
,\label{Trpl}\\[5pt] tr_-f &=& \left\langle {\cal O}_f \right\rangle_0\ =\ f(0)
\ . \label{Trmi}\eea
One may view $Tr_-$ and $Tr_+$ as singular-geometry counterparts
of the bulk-to-bulk and boundary-to-boundary Green functions in
anti-de Sitter spacetime, and the traces obey
\bea tr_\pm[f(y)\star g(y)]&=&tr_\pm[g(\pm y)\star f(y)]\ .\eea
Drawing on the properties of the spinor-oscillator realization of
the five-dimensional higher-spin gauge theory
\cite{Sezgin:2001zs,Vasiliev:2001wa}, it has been proposed
\cite{Sagnotti:2005ns} that there exists a \emph{normalizable
phase-space projector} $\D=\D(K^2)$ that is a distribution living on
the hypercone obeying
\be K_{ij}\star \D\ =\ 0\ ,\qquad \D\star\D\ =\ \D\ ,\qquad
\D\star M\ =\ M\ .\label{Delta}\ee
The functions $M(K^2)$ and $\D(K^2)$ are phase-space counterparts of
state-space projectors built from non-normalizable and
normalizable singleton states, respectively.

\begin{figure}[t]
\begin{center}
        \includegraphics[width=10cm,bb=0 0 700 500]{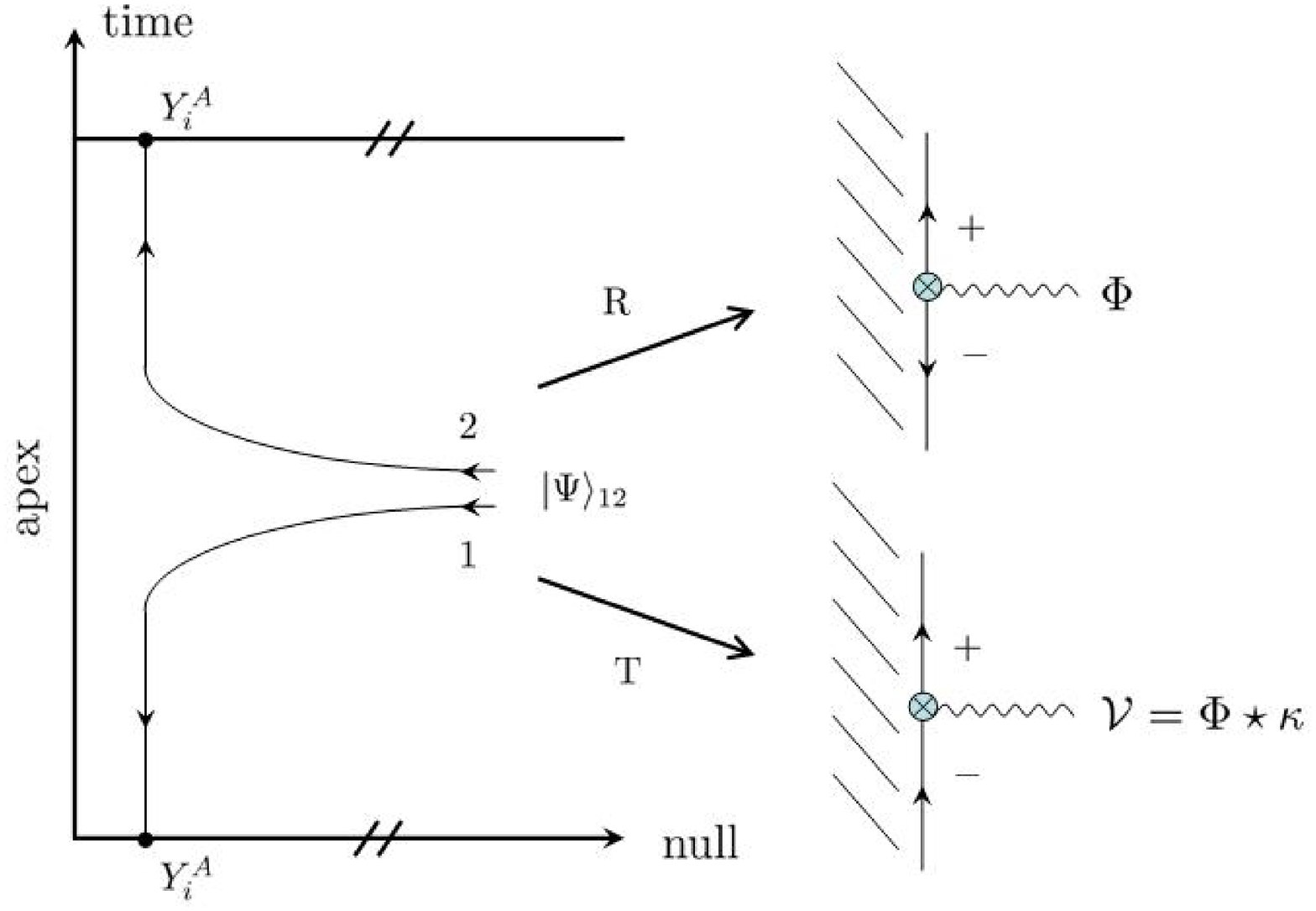}
\end{center}
\caption{\emph{Amputation and Role of Intertwiner}: An external
two-singleton composite
$\ket{\Psi}_{12}=\ket{\a}_1\ket{\b}_2+\ket{\widetilde\a}_1\ket{\widetilde\b}_2$,
is mapped to a twisted-adjoint vertex operator
$\Phi=\ket{\a}\bra{\b}+\ket{\widetilde\a}\bra{\widetilde\b}$
inserted into an $R$-ordered disc correlator, or, equivalently, to
a massless vertex operator ${\cal V}
=\Phi\star\kappa=\ket{\a}\bra{\widetilde\b}+\ket{\widetilde\a}\bra{\b}$
inserted into a $T$-ordered boundary correlator. The arrows
indicate the orientations of the corresponding worldlines, and the
amputation amounts to replacing $\Phi=C\star M$ by $C\star\D$
where $M$ and $\D$ are the singular and normalizable $\msp(2)$
projectors. }\label{fig:intw}
\end{figure}

An external massless particle is described by a $\pi$-invariant
symmetrized two-singleton state
\be \ket{\Psi}_{12}\ =\ \left[\ket{\a}_1\otimes
\ket{\b}_2+\ket{\widetilde\a}_1\otimes
\ket{\widetilde\b}_2\right]_{\rm symm}\ ,\label{masslessst}\ee
where $\ket{\a}$ and $\ket{\b}$ denote $\msp(2)$-invariant
singleton states. As drawn in Fig.~\ref{fig:intw}, this state
joins the internal worldline by
\begin{itemize}
\item[\emph{i)}] reversal of the
orientation of one of the incoming singletons;
\item[\emph{ii)}] amputation of external propagators.
\end{itemize}

Using \eq{tauket}, the
operation \emph{(i)} yields the \emph{un-amputated vertex
operator}
\be {\cal V}\ =\ \frac12\left[\ket{\a}\otimes
\bra{\widetilde\b}+\ket{\b}\otimes
\bra{\widetilde\a}+\ket{\widetilde\a}\otimes
\bra{\b}+\ket{\widetilde\b}\otimes \bra{\a}\right]\
,\label{vertex}\ee
with definite transformation properties under the discrete maps
$\pi$ and $\t$ given in \eq{piket} and \eq{tauket}. This vertex
corresponds to a phase-space function,
\be {\cal V}\ =\ \Phi\star \kappa\ ,\ee
where $\Phi$ belongs to the \emph{strongly projected
twisted-adjoint representation} \cite{Sagnotti:2005ns}
\be T[\mhso_0(D-1,2)]\ \equiv\ \big\{~K_{ij}\star \Phi\ =\ \Phi\star
K_{ij}\ =\ 0\ ,\quad \pi\tau(\Phi)\ =\ \pi(\Phi^\dagger)\ =\
\Phi~\big\}\ ,\label{C3}\ee
of the higher-spin algebra $\mhso_0(D-1,2)$ to be defined below (see eq. \eq{minalg}). Here $\pi$ and $\tau$ acting according to \eq{pimap} and
\eq{tauosc}, and $\kappa$ is an oscillator implementation of the
involution $\pi$ given in \eq{piket} and \eq{pimap}, that is
\be \pi(f(y))\ =\ \kappa\star f(y)\star \kappa\ ,\ee\be
\kappa\ket{\a}\ =\ \ket{\widetilde\a}\ ,\qquad \bra{\a}\kappa\ =\
\bra{\widetilde\a}\ .\ee
The element $\kappa$ can be realized on the internal worldline as
$\kappa=\exp_\star (i\pi v_A v_B w^A \star\bar w^B)$, where the
oscillators are defined in \eq{za1} and \eq{za2} and $\exp_\star
A\equiv 1+A+\frac12 A\star A+\cdots$. This element is, however,
not Weyl-ordered. As we shall see in Section \ref{sec:kappa}, this
subtlety is resolved in the two-dimensional theory, which admits a
well-defined Weyl-ordered implementation of $\kappa$.

The strongly projected elements $\Phi$ can be
represented as either normalizable or non-normalizable phase-space functions,
\be \Phi\ =\ \mx{\{}{ll}{C\star M&
\mbox{non-normalizable}\\C\star\D&\mbox{normalizable}}{.}\ ,\ee
where $C$ belongs to the \emph{weakly projected twisted-adjoint
representation} defined by \cite{Vasiliev:2003ev}
\be T_0[\mhso_0(D-1,2)]\ = \ \left\{C\in {\cal P}_0[{\cal Z}]\
:\quad \pi\tau(C)\ =\ \pi(C^\dagger)\ =\ C\right\}\ .\label{C2}\ee
Assuming that the amputation \emph{(ii)} leads to normalizable states,
we define the \emph{amputated vertex operator}
\be V\ =\ C\star \D\star\kappa\ .\qquad C\in T_0[\mhso_0(D-1,2)]\ .
\label{defvertex}\ee
Thus, given $N$ external massless states in a fixed cyclic order,
the resulting singleton correlator reads
\be {\cal A}^{(\pm)}_N\ =\ Tr_\pm\left[V_1\star\cdots\star
V_N\right]\ =\ tr_\pm\left[C_1\star\kappa\cdots\star
C_N\star\kappa\star M\right]\ .\label{corr1}\ee
For even $N$, say $N=2n$, the insertions of $\kappa$ cancel, and
the correlator can be expressed as a trace of Weyl-ordered
operators,
\be {\cal A}^{(\pm)}_{2n}\ =\ tr_\pm\left[C_1\star
\pi(C_2)\star\cdots \star\pi(C_{2n})\star M\right]\ .\ee

Turning to the $\msp(4)$-gauge transformations, the hyper-cone
reflections \eq{refl} -- which generate an anomalous
transformation of the external state \eq{masslessst}  -- leave the
twisted-adjoint field invariant, \emph{i.e.}
\be \rho(C)\ =\ C\ ,\ee
essentially due to the fact that the phase factor cancels in
\be \rho(\ket{\a}\bra{\b})\ =\
(-1)^{\e_0}\ket{\a}\bra{\b}[(-1)^{\e_0}]^\dagger\ =\
\ket{\a}\bra{\b}\ .\ee
Moreover, infinitesimal $\msp(4)$ transformations, given by
\be \d V_r \ =\ \varepsilon_{(1)r}^{ij}K_{ij}\star V_r
+\varepsilon^{ij}_{(2)r} V_r\star K_{ij}+
(\varepsilon^{ij}_r-\varepsilon'_r\e^{ij})y^{A}_{i}\star
\cV_r\star y_{Aj}\ ,\ee
with local, \emph{i.e.} $r$-dependent, parameters, leave the
correlators invariant. To show this, it suffices to use the weaker
$\msp(2)$-invariance condition \eq{PY}: one first observes that
the transformations induced by parameters that are triplets under
the $\msp(2)$ vanish, since they can be written as commutators
with $K_{ij}$ and all other operators are $\msp(2)$-invariant. The
closure of $\msp(4)$ then implies that the transformations induced
by the singlets $\varepsilon'_r$ must vanish as well.

More generally, the world-line correlators are invariant under
local transformations
\be \delta V_r=\varepsilon_{(1)r}\star V_r\star\varepsilon_{(2)r}\
,\ee
with parameters carrying general non-zero $\msp(2)$ charges, and
the element $\delta V_r$ defining a descendant of the ``primary''
operator $V_r$. Roughly speaking, one may view the above
transformations as a discrete analog of a continuous symmetry
group generated by the bilinear $\msp(2)$ current and
stress-energy tensor studied in Section \ref{sec:con}.

The space-time gauge symmetries, that will play an important role
in the master-field equations to be discussed below, arise as rigid
symmetries of the world-line correlators, \emph{i.e.} transformations that leave the
correlators invariant \emph{and} preserve the $\msp(2)$-projection
conditions on the vertices. These transformations are
\be \delta V_r\ =\ [\e,V_r]_\star\ ,\qquad \delta \Phi_r\ =\
\e\star\Phi_r-\Phi_r\star\pi(\e)\ ,\ee
with rigid, \emph{i.e.} $r$-independent, parameters $\e$ belonging
to the \emph{off-shell bosonic higher-spin algebra}
\be \mhso(D-1,2)\ \equiv\ \Big\{P\in{\cal P}[{\cal Z}]\ :\quad
\t(P)\ =\ P^\dagger\ =\ -P\Big\}\ ,\label{offhsa}\ee
containing as a subalgebra the vector-oscillator realization of
the \emph{minimal bosonic higher-spin algebra}
\be \mhso_0(D-1,2)\ =\ \left\{P\in {\cal P}_0[{\cal Z}]\ :\quad
\t(P)\ =\ P^\dagger\ =\ -P\right\}\ ,\label{minalg}\ee
where the ideal elements correspond to trivial transformations of
the vertices.

Let us examine the algebraic properties of the adjoint and twisted-adjoint
representations in more detail.

\scss{Adjoint and Twisted-Adjoint Fields}\label{sec:alg}

Here we high-light properties of the higher-spin algebra that
arise directly from the singleton. The extent to which these
suffice by themselves for constructing an interacting theory is
clearly a challenging problem to address in its generality,
linking the open string to be discussed in Section \ref{sec:tos}
to purely algebraic constructions based on spin-chains and
Yangians \cite{Yangian1,Yangian2}.

\scsss{The Minimal Bosonic Higher-Spin Algebra}\label{sec:hsa}

The linear action of $M_{AB}$ in $\sfD$ is not transitive,
\emph{i.e.} the singleton is not a fundamental representation of
$\mso(D-1,2)$. The minimal \emph{Lie-algebra} extension of
$\mso(D-1,2)$ with this property is the bosonic higher-spin
algebra
\cite{Eastwood:2002su,Vasiliev:2003ev,misha,Sagnotti:2005ns}
\be \mhso_0(D-1,2)\ =\ \left\{Q\in{{\cal U}(\mso(D-1,2))\over
{\cal A}[\sfD]}\ :\quad \t(Q)\ =\ Q^\dagger\ =\ -Q \right\}\
,\label{ho}\ee
where the singleton annihilator is defined by
\be {\cal A}[\sfD]\ =\ \left\{ X\in {\cal U}(\mso(D-1,2))\ :\quad
X\ket{\Psi}\ =\ 0\ ,\quad \forall \ket{\Psi}\in \sfD\right\}\ ,\ee
and $\t$ is the involutive anti-automorphism defined in \eq{tau}. The
Lie bracket of $\mhso_0(D-1,2)$ is given by
\be [Q,Q']\ =\ Q\star Q'-Q'\star Q\ ,\ee
where $\star$ is the product on ${\cal U}(\mso(D-1,2))$.

The algebra has the following \emph{level decomposition}:
\be \mhso_0(D-1,2)\ =\ \bigoplus_{\ell=0}^\infty {\mathfrak
L}_\ell\ ,\ee
where the $\ell$'th level is an irreducible $\mso(D-1,2)$ tensor
of highest weight $(2\ell+1,2\ell+1)$, comprising the generators
\be Q_\ell\ =\ Q_{A_1\dots A_{2\ell+1},B_1\dots
B_{2\ell+1}}^{(2\ell+1,2\ell+1)} M^{A_1B_1}\cdots
M^{A_{2\ell+1}B_{2\ell+1}}\ .\label{Qell}\ee
This can be shown using the triple-grading \eq{grading} of
$\mso(D-1,2)$, and the decomposition of the singleton into energy
levels defined in \eq{singleline}:

We first decompose $Q_\ell$ into irreducible $\mso(D-1,2)$
tensors, $Q_\ell\ =\ \sum_{i=1}^{r_\ell} Q_{\ell\,[i]}$. By
hypothesis, the spaces $Q_{\ell\,[i]}|\epsilon_0\rangle$ are
invariant under $\mg^{(-)}\oplus \mg^{(0)}$. On the other hand,
the states in $Q_\ell|\epsilon_0\rangle$ have energy eigenvalues
$\leq \epsilon_0+2\ell+1$, and the lower energies can be reached
from the maximal energy via repeated action with $\mg^{(-)}$, so
that $Q_{\ell}|\epsilon_0\rangle$ is irreducible under
$\mg^{(-)}\oplus \mg^{(0)}$. Hence, by a choice of the label $i$,
$Q_{\ell\,[i]}|\epsilon_0\rangle=0$ for $i>1$. Acting on these
identities with $\mg^{(+)}$, and using the irreducibility of
$Q_{\ell\,[i]}$, yields $Q_{\ell\,[i]}\in {\cal A}[\sfD]$ for
$i>1$, which shows the assertion that $Q_\ell$ is irreducible. As
a byproduct, we have also found that the highest $\mso(D-1,2)$
weights of the $\ell$'th level are given by $(2\ell+1,2\ell+1)$,
since this projection contains the generator $E^{2\ell+1}$, that
acts non-trivially in $\sfD$.~$\Box$

We note that the off-shell algebra \eq{offhsa} decomposes into levels given by
irreducible $\msl(D+1)$ tensors with highest
weights $(2\ell+1,2\ell+1)$ for $\ell=0,1,2,\dots$,
with $0$'th level given by $\mso(D-1,2)$.

The twisted-adjoint vertices are arranged into a massless
multiplet of $\mhso_0(D-1,2)$.

The rule for multiplying $\mso(D-1,2)$ tensors imply
\be [Q_\ell,Q_{\ell'}]_\star\ =\
\sum_{\ell''=|\ell-\ell'|}^{\ell+\ell'}
f_{\ell\ell'}{}^{\ell''}~Q_{\ell''}\ ,\label{alg1}\ee
where the structure coefficients $f_{\ell\ell'}{}^{\ell''}$ are
invariant $\mso(D-1,2)$ tensors times non-vanishing constants. It
is now easy to see that $\mhso_0(D-1,2)$ is minimal: Suppose that
$\mathfrak L$ is a Lie subalgebra of $\mhso_0(D-1,2)$ containing
$\mso(D-1,2)$. Then $\mathfrak L=\bigoplus_\ell c_\ell {\mathfrak
L}_\ell$ with multiplicities $c_\ell\in\{0,1\}$. Clearly, one
possibility is $\mathfrak L=\mso(D-1,2)$, for
$c_\ell=\delta_{\ell,0}$. Adding one extra generator, say at level
$\ell>0$, closure with level $0$ requires adding all of
${\mathfrak L}_\ell$, and in turn, via \eq{alg1}, all levels up to
$2\ell$, and so on.

The precise strengths of $f_{\ell\ell'}{}^{\ell''}$ can be
evaluated using oscillator representations, relying on the
\emph{uniqueness of the minimal extension}:

To show uniqueness, we assume $\widetilde{ \mh}$ to be a
non-trivial extension of $\mso(D-1,2)$ that is minimal and admits
$\sfD$. Minimality implies that $\widetilde {\mh}$ does not
contain any other ideals than $\mso(D-1,2)$, so that in particular
$\widetilde{\mh}\cap {\cal A}[\sfD]=0$. Let us decompose the
generators into irreducible $\mso(D-1,2)$ tensors, $\widetilde
Q_{\tilde l}$, where $\tilde l=(\tilde l_1,\tilde l_2,\dots)$
($\tilde l_1\geq \tilde l_2\geq\cdots)$) are highest weights, and
consider the set of generators with fixed $\tilde l_1$, say
$\widetilde Q_{[i]}$, $i=1,\dots,r$. The spaces $\widetilde
Q_{[i]}|\epsilon_0\rangle$ are invariant under $\mg^{(-)}\oplus
\mg^{(0)}$, and consist of states with energies ranging from
$\epsilon_0$ up to $\epsilon_0+\tilde l_1$. Thus, by the argument
given above, $\widetilde Q_{[i]}|\epsilon_0\rangle=0$ for $i>1$
(by a choice of label $i$), that in turn implies $\widetilde
Q_{[i]}\in {\cal A}[\sfD]$ for $i>1$, \emph{i.e.}
$\widetilde{\mh}$ contains at most one irrep for each $\tilde
l_1$, that we can denote by $Q_{\tilde l_1}$. Since $\widetilde
Q_{\tilde l_1}|\epsilon_0\rangle$ is irreducible under
$\mg^{(-)}\oplus \mg^{(0)}$ it follows that $\widetilde Q_{\tilde
l_1 }|\epsilon_0\rangle= c_\ell Q_\ell|\epsilon_0\rangle$, where
$\tilde l_1=2\ell+1$, the constants $c_\ell\in\{0,1\}$ (without
loss of generality), and $Q_\ell$ ($\ell=0,\frac12,1,\frac32$) are
the generators of the minimal associative higher-spin algebra
${\cal U}(\mso(D-1,2))/{\cal A}[\sfD]$. Acting with ${\mg}^{(+)}$
yields $\widetilde Q_{\tilde l_1}-c_\ell Q_\ell\in {\cal
A}[\sfD]$. Hence, by minimality, $\widetilde Q_{\tilde
l_1}=Q_\ell$ for integer $\ell$, and $\widetilde Q_{\tilde l_1}=0$
for half-integer $\ell$, which completes the proof.~$\Box$

Having defined the algebra, we turn to its physical
representations.

\scsss{Multipletons}\label{sec:multipletons}

The Young projections $Y[\sfD^{\otimes N}]$, referred to as
multipletons\footnote{The term ``doubleton'' also refers to the
singletons of $\msu(2,2)\simeq \mso(4,2)$, which arise most
naturally from the $\msu(2,2)$-spinor oscillator splitting into
$(2,1)\oplus (1,2)$ of $\mso(4)\simeq \msu(2)\oplus \msu(2)$.
Historically, Flato and Fronsdal referred to $\sfD(1,\frac12)$ and
$\sfD(\frac 12,0)$ as ``singletons'', presumably referring to
their single-oscillator realization. They also named them
``$\mathfrak{Di}$'' and ``$\mathfrak{Rac}$'', to honor
$\mathfrak{Di}\otimes \mathfrak{Rac}=\Psi_{\rm Dirac}+$
higher-spin friends.} \cite{Bianchi:2003wx,Beisert:2004di}, are
irreducible $\mhso_0(D-1,2)$ representations. To show this it
suffices to show that there does not exist any non-trivial
$\mhso_0(D-1,2)$-invariant tensor $I_N:\ \sfD^{\otimes
N}\rightarrow \Comp$. By definition an invariant $I_N$ obeys
\be I_N\sum_{\xi=1}^N Q(\xi)\ =\ 0\quad \forall ~Q\in
\mhso_0(D-1,2)\ ,\label{I_M}\ee
and we need to show that this implies that $I_N=0$.

For $N=2$ and $D>3$, the $\mso(D-1)$ invariance implies
\be I_2\ =\ \sum_{n=0}^\infty c_n\Big[\bra{E_n,(n)}\otimes
\bra{E_n,(n)}\Big]_{(0)}\ ,\ee
where $c_n$ are constants, and the subscript $(0)$ denotes the
$\mso(D-1)$-singlet projection of $d(n)\otimes d(n)$, where $d(n)$
is defined in eq.~\eq{singleline}. We note that the singlets
belong to the symmetrized tensor product, so that already at this
level it is clear that the anti-symmetric doubleton is
irreducible. Taking $Q=E$, one finds that $\sum_{n=0}^\infty 2E_n
c_n[\bra{n}\otimes \bra{n}]_{(0)}=0$ implying $c_n=0$ since all
$E_n\neq 0$. Hence, $I_2=0$, and also the symmetrized doubleton is
irreducible.

To show \eq{I_M} for a general finite rank $N$, we use the basic properties of
the $\pi$ and $\tau$ maps given in \eq{piket} and \eq{tauket},
which imply that an element $Q\in \mhso_0(D-1,2)$ can be expanded
as
\be Q\ =\ \sum_{\a,\b}
Q_{\a\b}\Big(\ket{\a}\bra{\b}-\ket{\widetilde\a}\bra{\widetilde\b}\Big)\
,\qquad (Q_{\a\b})^\star\ =\ -Q_{\b\a}\ .\ee
Thus, by various level truncations, it can be seen that $\mhso_0(D-1,2)$
contains $\mathfrak u(k)$ subalgebras (not containing
$\mso(D-1,2)$) for arbitrarily large values of $k$,
that rule out $\mhso_0(D-1,2)$-invariant tensors of finite rank. ~$\Box$

Let us proceed by taking a closer look at the doubletons.


\scsss{Doubletons and The Flato-Fronsdal
Theorem}\label{sec:flatofronsdal}


The fact that the singleton is the square root of infinitely many
spins was originally found by Flato and Fronsdal in $D=4$
\cite{Flato:1978qz}, and later generalized to higher dimensions in
\cite{Sezgin:2001zs,Sezgin:2001ij,misha}.

A state $|n\rangle \in \sfD(\e_0,(0))^{\otimes 2}$ with energy
$E=n+D-3$, $n=0,1,2,\dots$, can be expanded as
\be |n\rangle\ =\ \sum_{s_1+s_2=n}\Psi^{(s_1,s_2)}_{q_1\dots
q_{s_1},r_1\dots r_{s_2}}(L^+_{q_1}\cdots
L^+_{q_{s_1}})(1)(L^+_{r_1}\cdots L^+_{r_{s_2}})(2)|\Omega\rangle\
,\ee
where $|\Omega\rangle=|\epsilon_0\rangle(1)\otimes
|\epsilon_0\rangle(2)$, and the tensors $\Psi^{(s_1,s_2)}$ belong
to $(s_1)\otimes (s_2)$ of the diagonal $\mso(D-1)$,
\be \Psi^{(s_1,s_2)}_{pp q_3\dots q_{s_1},r_1\dots
r_{s_2}}=0=\Psi^{(s_1,s_2)}_{q_1\dots q_{s_1},pp r_3\dots
r_{s_2}}\ .\ee
The lowest weight condition
\be L^-_r|n\rangle\ =\ (L^-_r(1)+L^-_r(2))|n\rangle\ =\ 0\ ,\ee
amounts to
\bea c_{n}\Psi^{(n,0)}_{pq_1\dots
q_{n-1}}+c_{1}\Psi^{(n-1,1)}_{q_1\dots q_{n-1},p}&=&0\
,\label{eq1}\\[5pt]
c_{n-1}\Psi^{(n-1,1)}_{pq_1\dots
q_{n-2},r_1}+c_2\Psi^{(n-2,2)}_{q_1\dots q_{n-2},pr_1}&=&0\ ,\label{eq2}\\
&\vdots&\nn\\ c_1 \Psi^{(1,n-1)}_{p,r_1\cdots
r_{n-1}}+c_n\Psi^{(0,n)}_{pr_1\dots r_{n -1}}&=&0\
,\label{eq3}\eea
where $c_{k}=2k(k+\epsilon_0-1)$. Eq.~(\ref{eq1}) forces all
$\mso(D-1)$ irreps in $\Psi^{(s-1,1)}$ to vanish except $(n)$,
which is set equal to a constant times $\Psi^{(n,0)}$.
Eq.~(\ref{eq2}) eliminates all irreps in $\Psi^{(n-2,1)}$ in a
similar fashion, and so on. Thus, $|n\rangle$ contains precisely
one lowest-weight state, given by
\be |2\epsilon_0+n,(n)\rangle\ =\ \Psi^{(n)}_{r_1\dots r_n}
\sum_{k=0}^n{(-1)^k {n\choose k}{n+\epsilon_0-1\choose k}\over
{k+\epsilon_0-1\choose k}}(L^+_{r_1}\cdots
L^+_{r_k})(1)(L^+_{r_{k+1}}\cdots L^+_{r_n})(2)|\Omega\rangle\
,\label{lws}\ee
which belongs to the symmetric tensor product for even $n$ and the
anti-symmetric part for odd $n$. Hence the product of two
singletons decompose into two doubletons
\bea [\sfD\otimes \sfD]_{\rm symm}&=&\bigoplus_{s=0,2,4,\dots}
\sfD(2\epsilon_0+s,(s))\
,\label{masslessspectrum}\\[5pt]
[\sfD\otimes \sfD]_{\rm anti-symm}&=&\bigoplus_{s=1,3,5,\dots}
\sfD(2\epsilon_0+s,(s))\ .\label{addspectrum}\eea
Group-theoretically, the masslessness of the doubletons expresses
itself in the presence of the singular vector \eq{lowestnull},
corresponding to a gauge condition on a $D$-dimensional symmetric
rank-$s$ tensor field, or, equivalently, a $(D-1)$-dimensional
conservation law. Hence $|2\epsilon_0+s-1,(s-1)\rangle$ must
vanish identically in the singleton-composite realization
(\ref{lws}) as a consequence of $L^+_rL^+_r|\epsilon_0\rangle= 0$,
as one can check explicitly.

The doubletons have a dual enveloping-algebra presentation,
referred to as the \emph{twisted-adjoint representation}, defined
by
\be T_0[\mhso_0(D-1,2)]\ =\ \left\{C\in{{\cal
U}(\mso(D-1,2))\over{\cal A}[\sfD]}\ :\quad \pi\tau(C)\ =\
\pi(C^\dagger)\ =\ C\right\}\ .\label{C3b}\ee
Its elements are equivalence classes, as opposed to the strongly projected
twisted-adjoint representation \eq{C3} whose elements are fixed by the strong
$\msp(2)$ projection. The equivalence classes can be represented by elements with
the following level decomposition
\be C\ =\ \sum_{\ell=-1}^\infty C_s\ ,\qquad s\ =\ 2\ell+2\
,\label{levelexp}\ee \be C_s\ =\ \sum_{k=0}^\infty
C^{(s+k,s)}_{a_1\dots a_{s+k},b_1\dots b_s} M^{a_1b_1}\cdots
M^{a_sb_s}P^{a_{s+1}}\cdots P^{a_{s+k}}\ ,\ee
where the Lorentz traces belong to the annihilating ideal, and the
$\mhso_0(D-1,2)$-transformation rule reads
\be \d_\e C\ =\ \e\star C-C\star \pi(\e)\ .\ee
In the real basis obeying \eq{finorm}, \eq{piket} and \eq{tauket},
the expansion of $C$ reads
\be C\ =\ \sum_{\a,\b}\left(C_{\a\b}\ket{\a}\bra{\b}+(
C_{\a\b})^\star\ket{\widetilde\a}\bra{\widetilde\b}\right)\
,\qquad C_{\a\b}\ =\ C_{\b\a}\ , \label{C1}\ee
which establishes the isomorphism between $T_0[\mhso_0(D-1,2)]$ and
$[\sfD^{\otimes 2}]_{\rm symm}$.


\scss{Continuum Limits}\label{sec:con}


Here we examine the continuum limit $N\rightarrow
\infty$ of the discretized $p$-brane action \eq{eq:SN1}, and find
a topological closed string inside the sigma model on the phase
space of the 7D Dirac hypercone. This model is related to those of
\cite{lindstrom,Witten:2003nn,Berkovits:2004hg,Berkovits:2004tx,Siegel:2004dj,Bars:2004dg}
although our results and interpretations are slightly different.
First, in the way the limits are taken, \emph{tensile strings do
not have any a priori privileged status}. Second, the singling out
of $D=7$ depends crucially on the normal-ordering prescriptions
following the usage of the standard Neveu-Schwarz vacuum
$\ket{0}$, in turn obeying $E_0\ket{0}=0$ corresponding to an
$AdS$-covariant regularization of the zero-point energy of the
discrete system which differs from the regularization used in
\cite{lindstrom,Hwang:1998gs,Hwang:1999yh}.


\scsss{The 7D Closed Singleton String}\label{sec:7D}


The discretized phase-space action \eq{eq:SN1} can be written as
\be S\ =\ \frac12 \oint d\t \left\{\sum_\x ( \dot X^A(\x)
P_A(\x)+\L^{ij}K_{ij}(\x))+\sum_{\x\neq\eta}\L^{i,j}(\x,\eta)
T_{i,j}(\x,\eta)\right\}\ ,\ee
where $K_{ij}(\x)=\frac12 Y^A_i(\x) Y_{Aj}(\x)$ and
$T_{i,j}(\x,\y)=\frac12 Y^A_i(\x)Y_{Aj}(\y)$. In Section
\ref{sec:global} we found that effectively the physical-state
conditions arise from the $\msp(2)$-blocks along the diagonal and
the large gauge transformations. Thus, there are two ways of
taking the continuum limit: \emph{i)} truncate $\msp(2N)\rightarrow \bigoplus_\x \msp(2)_\x$ and send $N\rightarrow \infty$ while imposing invariance under large
gauge transformations by hand; \emph{ii)} send $N\rightarrow\infty$ in $\msp(2N)$.

We shall interpret the result of \emph{(i)} as the
$\msp(2)$-gauged chiral (closed) string
\be S\ =\ \frac12 \int_\S d\t d\s (P^A \dot X_A +\L^{ij}K_{ij})\ ,\ee
where the worldsheet $\S$ is periodic both $\s$ and $\t$, in
accordance with the prescriptions in \eq{phsp} and \eq{phsptrace}.
The weights of $(X^A,P_A;\L^{ij};K_{ij})$ under
$\s$-reparameterizations are given by
\be {\rm
wt}(X^A,P_A;\L^{11},\L^{12},\L^{22};K_{11},K_{12},K_{22})\ =\
(0,1;0,1,2;2,1,0)\ .\ee
Going to the complex coordinate $z=\sigma+i\tau$, and twisting the
fields using the $\mathfrak{u}(1)$ current $K_{12}$, one obtains a
one-parameter family of models described by the chiral action
($\l\in \Real$)
\be S_{\l}\ =\  \frac12 \int_\S dz d\bar z \phi^{Ai}\bar D
\phi_{Ai}\ ,\ee
where the conformal weights of $(\phi^{i A} ;K_{ij})$ are given by
\be {\rm wt}(\phi^{1 A},\phi^{2 A};K_{11},K_{12},K_{22})\ =\
(\l,1-\l;2(1-\l),1,2\l)\ .\ee
The levels of the $\msp(2)$ currents formed out of matter and
ghost fields are given by\footnote{In this section we use the
conventions of \cite{DiFrancesco:1997nk}. For a recent, related work on
the equivalence between free field theories and topological WZW models,
see \cite{Schiappa:2005mk}.}
\be k^{(\phi)}_{\msp(2)}\ =\ -{D+1\over 2}\ ,\qquad
k_{\msp(2)}^{(gh)}\ =\ 4\ ,\ee
which implies that the $\msp(2)$ BRST operator is nilpotent in
\be D_{\rm crit}\ =\ 7\ .\ee
Indeed, the central charge of the twisted systems, given by
\be c_{\rm tot}\ =\ c_{(\phi)}+c_{(gh)}\ ,\qquad c_{(\phi)}\ =\
(12\l^2-1)(D+1)\ ,\qquad c_{(gh)}\ =\ -2(48\l^2+3)\ ,\ee
add up to a total $\l$-independent total central charge only for
$D=7$, where it is given by the critical value
\be c_{\rm crit}\ =\ -14\ .\ee
The critical system gives rise to a BRST-invariant scalar
singleton with conformal weight
\be h_{\rm crit}(\sfD)\ =\ -1\ .\ee
For example, at the manifestly $\msp(2)$-invariant point $\l=1/2$
the matter fields are $2(D+1)$ symplectic bosons with weight
$1/2$, and the extended scalar singleton \eq{singleton} arises in
the R-sector where the zero-modes of $\phi^{Ai}$ act on an
oscillator module created by a spin field with weight
\footnote{Bosonization yields the integer as well as half-integer
Fock spaces \eq{Fockspace} and \eq{genFockspace}.}
\be \l\ =\ \frac 12\ :\quad h(\sfD)\ =\ -{D+1\over 8}\ .\ee

The critical $\msp(2)$-invariant system of matter and ghosts can
equivalently be presented as an $\widehat{\mso}(6,2)_{k_{\rm
crit}}$ WZW model with critical level identified with the
free-field value, that is\footnote{In $D=7$, the stress tensor
built from the total $\msp(2)$ current is trivial, while the pure
matter current with $k^{(\phi)}_{\msp(2)}=-1=-g(\msp(2))$ yields a
contracted Virasoro algebra \cite{nocritdim,Lindstrom:2003mg}.}
\be k_{\rm crit}\ =\ -2\ .\ee
Indeed, the Sugawara charge for $\widehat\mso(D-1,2)_k$, given by
\be c_{\rm sug}\ =\ {k ~{\rm dim} \,\mso(D-1,2)\over k+g}\ ,\qquad
g\ =\ D-1\ ,\label{sugawara}\ee
coincides with $c_{\rm crit}$ for the critical values of $k$ and
$D$, where also the weight of the primary singleton field, given
by
\be h_{\rm sug}({\sfD})\ =\ {C_2[\mso(D-1,2)|\sfD]\over
2(k+g)}\ ,\ee
agrees with the free-field value.

The BRST-invariant sector of the above critical system is,
however, too large, and needs to be projected further by gauging
the maximal compact subalgebra $(\widehat{\mso}(6)\oplus
\widehat{\mso}(2))_{-2}$ \cite{Dixon:1989cg,Bars:1989ph}.
Remarkably, in the critical dimension\footnote{In $D=7$, the
central charge vanishes also for the coset
$\widehat\mso(6,2)_{-2}/\widehat\mso(6,1)_{-2}$, whose
relation to the $7$-dimensional scalar found in Section
\ref{sec:Dsc} would be interesting to spell out in more detail. }
\be c_{\rm sug}[\widehat{\mso}(6)_{-2}]\ =\ -15\ ,\qquad c_{\rm
sug}[\widehat{\mso}(2)_{-2}]\ =\ 1\ , \ee
so that the GKO coset construction has vanishing central charge,
\be c_{\rm gko}\ =\ -14-(-15+1)\ =\ 0\ .\ee
Moreover, in the critical dimension
\be h_{\rm sug}[\widehat{\mso}(6)_{-2}|\sfD]\ =\ 0\ ,\qquad
h_{\rm sug}[\widehat{\mso}(2)_{-2}|\sfD]\ =\ -1\ ,\ee
resulting in gauged primary singletons with vanishing conformal
weight
\be h_{\rm gko}(\sfD)\ =\ -1-(0-1)\ =\ 0\ .\ee
We conjecture that the corresponding spin field has a local
operator product algebra so that its normal-ordered products
generate a chiral ring ${\cal S}_{\rm gko}$ consisting
group-theoretically of symmetrized singletons,
\be {\cal S}_{\rm gko}\ \simeq \ \bigoplus_{N=1}^\infty
[\sfD^{\otimes N}]_{\rm symm}\ ,\ee
to be identified as the generalized Chan-Paton factor. We expect
the corresponding generalized operators \eq{calV} to be free of
anomalies under the remaining large $\msp(2\infty)$-gauge
transformations since the 7D singleton obeys \eq{3mod4}.

In $D=7$ both vector and spinor oscillators consist of $8$ complex
components. The spinor oscillators give rise to the higher-spin
algebra $\mathfrak{hs}(8^*)$, defined as \cite{Sezgin:2001ij}
\be \mathfrak{hs}(8*)\ =\ \{~Q\in {{\cal P}[{\cal Z}_s]\over
I[\msu(2)]}\ :\quad Q^\dagger\ =\ \t(Q)\ =\ -Q~\}\ ,\ee
where ${\cal Z}_s$ denotes the non-commutative spinor space and
$\msu(2)$ is generated by spinor-oscillator bilinears. One can
show that $\mathfrak{hs}(8^*)$ is minimal and admits the scalar
singleton, so that by uniqueness
\be \mathfrak{hs}(8^*)\ \simeq\ \mhso_0(6,2)\ .\label{iso}\ee
Hence the 7D singleton string has a dual free-field formulation
based on an $\msu(2)$-gauged sigma model in spinor space.

The simple tensile deformation studied in Section \ref{sec:M2def}
is membrany in nature, and it is tantalizing to speculate about a
natural embedding of the closed singleton string into String
Theory would be via a discretization of the tensile supermembrane
in AdS$_7\times S^4$. However, adding free $USp(4)$ fermions to
the world-sheet theory modifies the central charges.


Clearly, the above analysis leaves several issues open, in particular
the details of the free-field construction of the vertex operators \eq{calV}
and the trace operation. We plan to return to this in a more conclusive report \cite{wip}.


The limit \emph{(i)} leads to a critical system
admitting a natural further projection down to the physical
topological string. This raises the question whether the
projection can be derived from the original $\msp(2N)$
gauging by taking the limit \emph{(ii)}.

The off-diagonal $\msp(2N)$ generators with $|\x-\y|=n$ may be
interpreted as discretizations of $\mso(D-1,2)$-invariant
bilinears in $\phi^{Ai}$ containing $n$ derivatives. At $\l=1/2$,
these form a higher-spin current algebra, that we denote by
\be {\cal W}_{\msp(2)+\infty}\ \equiv\ \bigoplus_{s=1,3,5,\dots} {\cal
W}^{(s)}_{ij}~\oplus~\bigoplus_{s=2,4,\dots}{\cal W}^{(s)}\ ,\ee
where the currents with odd and even conformal weight $s$ are
$\msp(2)$ triplets and singlets, respectively, with ${\cal
W}^{(1)}_{ij}$ and ${\cal W}^{(2)}$ being the $\msp(2)$ current
and the free-field stress tensor, respectively. The algebra is
reducible, and contains a number of subalgebras: ${\cal
W}^{(1)}_{ij}\oplus {\cal W}^{(2)}$; ${\cal W}_{\infty}^{(2)}$
consisting of all even-weight currents; ${\cal W}_{1+\infty}$
consisting of ${\cal W}_{\infty}^{(2)}$ plus the odd-weight
currents ${\cal W}_{12}^{(s)}$. Gauging ${\cal
W}_{\msp(2)+\infty}$, the resulting ghost contribution to the
$\widehat\msp(2)$ and Virasoro central charges become
\bea k_{\msp(2)}^{(gh_{\msp(2)+\infty})}&=& \sum_{s=1,3,\dots} 1\ =\ 0\
,\\ c_{\rm vir}^{(gh_{\msp(2)+\infty})}&=&\left(3
\sum_{s=1,3,\dots}+\sum_{s=2,4,\dots}\right)(-2)(6s^2-6s+1)\ =\ 2\
,\qquad\eea
that cannot be cancelled simultaneously against the matter
contributions, $k_{\msp(2)}^{(\phi)}=-(D+1)/8$ and $c_{\rm
vir}^{(\phi)}=-(D+1)$. Hence the continuum limit \emph{(ii)} does
not appear to lead to an interesting model.

One might instead consider gauging ${\cal W}_{1+\infty}$, which
yields separately vanishing central charges in the matter and
ghost sectors in $D=-1$. However, as we shall see, this type of
critical gauging does make sense in supersymmetric setups where
the central charges cancel separately in the matter sector.


\scsss{Remarks on Twistor Superstrings}


The tensionless limits of discretized standard Green-Schwarz
actions will yield bosonic vector oscillators and fermionic spinor
oscillators. The vectors can be converted to bosonic spinor
oscillators in $D\leq 7$ as shown in \cite{misha} and demonstrated
in Appendix \ref{sec:ver} in the case $D=7$, which suggests the
existence of analogous dualizations of the fermionic spinors
resulting in super-oscillator analogs of \eq{eq:SN1} in turn
leading in the continuum limit to chiral sigma models containing
topological closed-string realizations of generalized
super-singleton Chan-Paton factors.

It is interesting that under the vector-spinor dualities, the
$\msp(2)$ is converted into $\msu(2)$ and $\mathfrak{u}(1)$ in
$D=7$ and $D=5$, respectively, in turn leading to master-field
equations involving phase-space projectors, while the 4D spinors
have trivial internal gauge group. On the other hand, in all
spinor-oscillator realizations the lowest-weight state
$\ket{\e_0,(0)}$ is given by the Fock-space vacuum and hence has a
manifestly normalizable analytical wave-function in spinor space.

An alternative approach is to view the chiral phase-space strings
as fundamental. Remarkably, in this spirit, the $\psu$-covariant
free-field model \cite{Bars:2004dg} can be viewed as a critical
model based on gauging the ${\cal W}_{1+\infty}$-extension of the
free-field stress tensor and the $\mathfrak u(1)_Z$-current of the
$\widehat{\mathfrak{su}}(2,2|4)_0$ current algebra\footnote{We
note that the ${\cal W}_{1+\infty}$-gauging is critical in itself
without any need for internal ``matter'', unlike the model based
on gauging the $U(1)$ and Virasoro currents, which requires an
internal sector with $c_{\rm vir}=24$
\cite{Berkovits:2004hg,Berkovits:2004jj}.}, corresponding to an
$N\rightarrow \infty $ limit of a discrete $U(N)$-gauged sigma model \cite{johanphd}. Essentially this is due
to the fact that the central charges vanish separately in the
matter and ghost sectors for $(1+\infty)$-gauging. Moreover, the
central charges vanish also in the compact subalgebra
$(\widehat\msu (2|2)\oplus \widehat\msu(2|2))_0$, whose gauging
gives rise to a critical GKO model providing the framework for
constructing the generalized Chan-Paton factor.

The above construction contains in itself no dynamics other than
linearized on-shell conditions inherited from the singletons. In
the case of the single singleton truncation, to be considered in
the next section, the dynamics follows from the open-string
reformulation of phase-space quantization, in which a crucial role
is played by bi-local operators describing interacting
two-singleton composites arising either as external states or
internal states resulting from factorization\footnote{Dynamical
phase-space Lagrangians have been proposed in \cite{Bars:2001ma}
although their relation to the Vasiliev equations remains to be
worked out.}. The analogous treatment of the complete phase-space,
or spinor, closed strings presumably requires an extension based
on open topological membranes refined with multi-local operators.


\scs{THE OPEN SINGLETON STRING}\label{sec:tos}


In the phase-space approach, the dynamical equations are the
conditions on deformations of the $\star$-product algebra
preserving the gauge symmetries. Indeed, this is how the internal
part of the Vasiliev equations arises from the phase-space
quantization of a single singleton. One may think of the full
master fields as analogs of exactly marginal operators on tensile
closed-string worldsheet (or other physical deformations of
tensile $p$-branes), and the Vasiliev equations as a classically
consistent truncation of the string-field theory based on the
chiral sigma model mentioned in Section \ref{sec:con}.

The deformations of the singleton worldline are parameterized by
$\msp(2)$-projected twisted-adjoint master fields built from
$(X^A,P^A)$, which one may view as analogs of the
Virasoro-projected vertices on the first Regge trajectory built
from $(X^\mu,\dot X^\mu)$. The latter arise as a sub-sector of the
open-string phase space $(X^\mu,\dot X^\mu,\ddot X^\mu,\dots)$
with non-local world-line Green function $\left\langle X^\mu(\t)
X^\nu(\t')\right\rangle\sim i\eta^{\m\n}\alpha'\log|\t-\t'|$,
while the former arise from a topological action with local
Green function $\left\langle X^A(\t) P^B(\t')\right\rangle \sim
i\eta^{AB}(\theta(\t-\t')-\th(\t'-\t))$ that extends \'a la Cattaneo-Felder into a
topological open string on a worldsheet $\S$ as we shall discuss
in Sections \ref{sec:sp2} and \ref{sec:bv}.

Thus, by analogy, the \emph{interactions giving rise to non-linear
deformations require a notion of breaking up the worldline and
inserting a composite realized as a topological open-string
excitation} (see Fig.~\ref{fig:intw}). As we shall discuss in
Sections \ref{sec:kappa} and \ref{sec:osc}, this can be
implemented by generalizing the local boundary condition in
\eq{phsp} to \emph{$2$-punctured boundary conditions}
corresponding to adding branch-points $p\in\partial\S$ where the
embedding fields behave inside correlators as
\be Y^{Ai}(u,\bar u)\ \sim \ y^{Ai}+{1\over2\pi i}y^{\prime
Ai}\log{u\over\bar u}+\cdots\ ,\label{asymptp}\ee
where $u$ is a local coordinate vanishing at $p$, the omitted
terms either vanish at $\partial \S$ or are $q$-exact, and $q$ is
the shift-symmetry BRST operator. Such branch-points are generated
by insertions of \emph{bi-local operators} with a non-trivial
dependence on the $q$-closed zero-mode $y^{Ai}$ as well as the
$q$-non-closed shift-mode $y^{\prime Ai}$. Indeed, the above
geometric structures combined with ordinary canonical quantization
yield Vasiliev's bi-local algebraic structures -- the oscillator
algebra \eq{intro:osc}, the intertwiner \eq{Jprime} and the
operator $d'$ --  in turn facilitating the construction of the
observable \eq{observables} giving rise to the internal part
\eq{vas1pr} and \eq{vas2pr} of the weakly projected Vasiliev
equations, which we shall analyze in Section \ref{sec:vas}.

The phase-space approach converts the standard
notions of scattering amplitudes and an effective action into that
of a BRST-cohomological master-field
equation in non-commutative phase space ${\cal Z}$, which one may
also think of as a non-linear phase-space counterpart of the
singleton Schr\"odinger equation. In this formalism, the \emph{true} space-time
geometry arises as a result of Vasiliev's unfolding
procedure, whose main features are high-lighted in Section
\ref{sec:unf}.


\scss{Two-dimensional Action and $\msp(2)$-Gauging}\label{sec:sp2}


A phase space with non-degenerate symplectic structure $\O=d\o$
admits Kontsevich's covariant $\star$-product, in turn
identifiable with the algebra of boundary operators in the
two-dimensional Cattaneo-Felder-Kontsevich model
\cite{Kontsevich:1997vb,Cattaneo:1999fm} based on the Poisson
sigma model $S=\frac12\int_{\S}(d Y^M+\frac12\O^{MN}\eta_N)\wedge
\eta_M$ on a disc $\S$ with $\eta_M|_{\partial\S}=0$, which is a
parent of the point-particle action $S=\frac12 \oint_{\partial \S}
\omega$.

The index contraction in the kinetic $dY^M\eta_M$-term is
background independent, so that standard perturbative quantization
methods give rise to a manifestly \emph{background covariant}
$\star$-product. Hence, in the case of the singleton, the
subsequent master-field equations can be derived without the need
to single out any specific component fields, such as metric or
Lorentz connection, and this applies as well to the tentative
open-membrane reformulations mentioned above.

The two-dimensional parent action of the phase-space action
\eq{eq:psa} is given by
\be S\ =\ \frac12 \int_{\Sigma}\left( D Y^{M}\wedge \eta_{M}
+\frac12 \eta^{M}\wedge \eta_{M}+\xi^{MN} F_{MN}\right)\
,\label{eq:cfSP}\ee
where $\partial\S$ is the singleton worldline; $\eta_{M}$ and
$\xi^{MN}\equiv\eta^{AB}\xi^{ij}$ are Lagrange multipliers;
$F^{MN}\equiv d \Lambda^{MN}+\L^{MP}\wedge \L_P{}^N$ with $\Lambda^{MN}=\eta^{AB}\Lambda^{ij}$;
and the symplectic index $M\equiv Ai$ is raised and lowered using
\be \Omega_{MN}\ =\ -\e_{ij}\eta_{AB}\ .\ee
The stationary configurations obey
\be  D Y^{Ai}+\eta^{Ai}\ =\ 0\ ,\quad  D\eta^{Ai}\ =\ 0\ ,\quad
 D\xi^{ij}+\eta^{A(i}  Y^{j)}_A\ =\ 0\ ,\quad
 F^{ij}\ =\ 0\ ,\label{eq:cfCIS}\ee
where $ D\x^{ij}=d\x^{ij}+2 \L^{k(i}\x_{k}{}^{j)}$, and the
boundary conditions
\be \oint_{\partial\S} \delta  Y^{Ai}\eta_{Ai}\ =\ 0\ ,\qquad
\oint_{\partial\S} \x^{ij}\delta  \L_{ij}\ =\ 0\
.\label{varprin}\ee
A unique configuration is singled out by imposing Dirichlet
conditions on the one-forms,
\be \eta_{Ai}|\ =\ 0\ ,\qquad \L_{ij}|\ =\ {\rm fixed}\
.\label{Dirbc}\ee

Let us derive the relation between \eq{eq:cfSP} and \eq{eq:psa}.
From \eq{eq:cfCIS} it follows that
\be \x_{ij}-K_{ij}\ =\ U_i{}^r U_j{}^s k_{rs}\ ,\ee
where $dk_{rs}=0$ and $U_{ir}$ is a coset element defined by
$DU_{ir}\equiv dU_{ir}+\L_i{}^j U_{jr}=0$ and $U_{ir}|_p=\e_{ir}$, with
$p\in\S$ being a given fixed point. Moreover, from \eq{eq:cfCIS}
and \eq{Dirbc} it follows that $D\x_{ij}|=0$, which implies that
\be \x_{ij}|\ =\ U_i{}^r| U_j{}^s| k'_{rs}\ ,\ee
where $k'_{rs}$ is another constant. The boundary field equations
are therefore equivalent to the equations of motion of \eq{eq:psa}
with $V_{ij}$ identified as
\be V_{ij}\ =\ U_i{}^r| U_j{}^s|(k_{rs}-k'_{rs})\ .\ee
Thus, from the two-dimensional point of view the singleton limit
corresponds to taking $k_{rs}=k'_{rs}$, \emph{i.e.}
\bea K_{ij}|&=&0\ ,\label{kkprime}\eea
which we shall assume henceforth.


\scss{Brst Formulation and Truncation of Triplets}\label{sec:bv}


The systematic gauge-fixing procedure is provided by the BV
formulation, for which we use the conventions and notations
collected in Appendix \ref{app:BV}.

The BRST transformations leaving the field equations \eq{eq:cfCIS}
invariant, are
\bea \d \eta^M&=&D C^M+C^{MN}\eta_N\ ,\qquad \d\L^{MN}\ =\ -D
C^{MN}\ ,\\\d Y^M&=& -C^M+C^{MN}Y_N\ ,\qquad \d\x^{MN}\ =\
2C^{K(M}\x_K{}^{N)}-Y^{(M}C^{N)}\ ,\label{BRST1}\eea
where the vector indices are suppressed, and $C^{MN}$ and $C^M$
are ghost fields obeying the boundary conditions
\be DC^M|\ =\ 0\ ,\qquad DC^{MN}|\ =\ 0\ .\ee
Demanding that $\d^2$ vanishes on-shell fixes
\be \d C^M\ =\ C^{MN}C_N\ ,\qquad \d C^{MN}\ =\ C^{MP}C^{N}{}_P\
.\label{BRST2}\ee
The resulting BRST transformations leave the classical action
invariant
\be \d S\ =\ 0\ .\ee
The anti-field dependence of the classical BV action is expanded
as
\be {\cal S}\ =\ \sum_n S_n+{\cal O}(\hbar)\ ,\qquad S_0\ =\ S\
,\qquad S_1\ =\ \sum_{\phi\in \aleph} \int_{\S} \phi^+\wedge
\d\phi\ ,\ee
with
$\aleph=\{Y^M,\eta_M,\x_{MN},\L^{MN};C_M,C^{MN};B^M,B_{MN};\l^M,\l_{MN}\}$,
where the $B$ and $\l$ fields, which are required to write
gauge-fixing terms, are assigned the BRST transformations
\be \d B^M\ =\ -\l^M\ ,\qquad \d\l^M\ =\ 0\ ,\qquad \d B^{MN}\ =\
\l^{MN}\ ,\qquad \d\l^{MN}\ =\ 0\ .\label{BRST3}\ee
The BV-bracket can then be expanded as $({\cal S},{\cal
S})=(S_1,S_1)+2(S_0,S_2)+\cdots$, where
\be (S_1,S_1)\ =\ 2\int_\S \sum_{\phi\in\aleph} \phi^+\wedge
\d^2\phi\ ,\ee
vanishes identically, \emph{i.e.} $\d\d=0$ off-shell. Furthermore,
the action of the BV Laplacian $\D$ on $S_1$ yields terms proportional
to $C^M{}_M=0$, implying $\D S_1=0$. The BV master
action is thus given by
\be {\cal S}\ =\ S+\sum_{\phi\in \aleph} \int_{\S}
\phi^\star\wedge \d\phi\ ,\ee
with $\d\phi$ given by \eq{BRST1}, \eq{BRST2} and \eq{BRST3}.

Using the gauge-fixing fermion
\be\Psi\ =\ \int_\S\left(\eta_M\wedge \star DB^M-\L^{MN}\wedge
\star dB_{MN}\right)\ ,\ee
where the Hodge $\star$ is defined using an auxiliary Euclidean
metric, the anti-fields become
\bea B^+_M&=& -D\star\eta_M\ ,\qquad \eta^{+M}\ =\ \star DB^M\ ,\\
\L^+_{MN}&=& -\star (dB_{MN}-\eta_{(M}B_{N)})\ ,\qquad B^+_{MN}\
=\ d\star\L_{MN}\ ,\eea
resulting in the following gauge-fixed action
\bea S_{\rm gf}&=&\frac12\int_\S\left\{DY^I\wedge\eta_I+\frac12
\eta^M\wedge \eta_M+\xi_{MN}F^{MN}+\star DB^M\wedge
(DC_M+C_M{}^N\eta_N)\right.\nn\\&&\qquad\quad\left.+\star(dB^{MN}-\eta^M
B^N)\wedge DC_{MN}+\l^M D\star\eta_M+\l^{MN}d\star\L_{MN}\right\}.\eea
The gauge-fixed classical field equations read
\be D\eta^M\ =\ 0\ ,\qquad D\star\eta^M\ =\ 0\ ,\ee  \be F^{MN}\
=\ 0\ ,\qquad d\star\L^{MN}\ =\ 0\ ,\ee\be D\star DB^M\ =\ 0\
,\qquad D\star DC^M\ =\ 0\ ,\ee\be D\star dB^{MN}\ =\ 0\ ,\qquad
d\star DC^{MN}\ =\ 0\ ,\ee \be DY^M+\eta^M+\star D\l^M+\star D(B^N
C_N{}^M)\ =\ 0\ ,\ee \\[-40pt]\bea
D\x_{MN}+\eta_{M}Y_{N}-\star\eta_{M}\l_{N}+\star DB_{M} C_N
-B_M\star DC_N+(M\leftrightarrow N)&&\nn\\=\ \star d\l_{MN}-2\star
dB_{M}{}^P C_{PN}+\star\eta_M B^P C_{NP}+(M\leftrightarrow N)&&\
.\eea
Taking further curls yields simpler harmonic equations, of which
it is worth noting $D\star d\l_{MN}+2\star dB_{(M}{}^P\wedge
DC_{N)P}=0$.

The full BRST current reads
\bea \star J&=&-C^M\eta_M-Y^M DC_M+\l^M\star
DC_M\nn\\&&+C^{MN}\star (d\l_{MN}+B_M DC_N)-\star
DC^{MN}\l_{MN}-\star dB_{MN}C^{MP}C_P{}^N\ .\eea
The BRST charge is given by integrals of $\star J$ along open
contours $L$ with endpoints at the boundary of $\S$, \emph{viz.}
\be Q\ =\ \int_L\star J\ ,\qquad L\subset\S\ ,\qquad \partial
L\subset
\partial\S\ .\ee
The charge is conserved if $\star J|=0$, which can be achieved
(with some loss of generality) by imposing the homogenous
Dirichlet conditions
\be \eta_M|\ =\ 0\ ,\qquad DC^M|\ =\ 0\ ,\qquad C_{MN}|\ =\ 0\
,\ee
\be  dB^M|\ =\ 0\ ,\qquad dB_{MN}|\ =\ 0\ ,\qquad \l^M|\ =\ 0\
,\qquad \l_{MN}|\ =\ 0\ ,\ee
where we note that the condition on $C_{MN}$ sharpens that
required by the BV master equation.

The ghost field $C^M$ contains a zero-mode obeying
\be DC^M_{(0)}\ =\ 0\ ,\qquad F_{MN}\ =\ 0\ ,\ee
which drops out of the gauge-fixed action. This mode must be
deleted from the path-integral measure, which we define
schematically as
\bea \int_{\L^{MN}|={\rm fixed}} {\cal D}\L_{MN}\int_{DC^M|=0}
{\cal D}C^M\d(C^M_{(0)}) \int {\cal D}\x^{MN}\cdots\
,\label{measure}\eea
where the integration over $\x^{MN}$ ensures $F_{MN}=0$. The ghost
zero-mode corresponds to a physical zero-mode in $Y^M$, as can be
seen from the BRST transformations given in \eq{BRST1}. Moreover,
the non-zero modes in $Y^M$ are either paired with non-zero modes
in $C^M$, in which case they are unphysical, or unpaired, in which
case they in fact are BRST exact (see eq.~\eq{ymboundary}).

The $\mso(D-1,2)$ current
\be J_{AB}\ =\
Y^i_{[A}\star\eta_{B]i}+\l^i_{[A}\eta_{B]i}+B^i_{[A}(DC_{B]i}+C_i{}^j\eta_{B]j})
-DB^i_{[A}C_{B]i}\ ,\ee
in general has an anomaly, given by
\be \star J_{AB}|\ =\-\star DB^i_{[A}C_{B]i}|\ .\label{anomaly}\ee
However, for certain deformations of the theory, such as those
giving rise to the Vasiliev equations, it is possible to redefine
the Lorentz generators, so that the resulting master-field
equations are manifestly Lorentz invariant.

To simplify the gauge-fixed Lagrangian it is convenient to specify the
Dirichlet condition \eq{Dirbc} to the homogeneous condition
\be \L^{MN}|\ =\ 0\ .\ee
At the level of field equations, the remaining $\msp(2)$ triplets
(recall that $\L^{MN}=\eta^{AB}\L^{ij}$ \emph{idem} $B^{MN}$ and $C^{MN}$)
are then given by
\be \L_{MN}\ =\ 0\ ,\qquad C_{MN}\ =\ 0\ ,\qquad dB_{MN}\ =\ 0\
,\ee
and
\be d\x_{MN}\ =\ -\eta_{(M}Y_{N)}+\star\eta_{(M}\l_{N)}-\star
dB_{(M} C_{N)} +B_{(M}\star dC_{N)}\ ,\label{xi}\ee
where the last equation is part of the $\msp(2)$-invariance
condition discussed in Section \ref{sec:sp2}.

At the quantum level, and considering correlators of operators
independent of the triplets, the above equations constitute a
quasi-consistent truncation to the free action
\be s_{\rm gf}\ =\ \frac12\int_\S\left(dY^M\wedge\eta_M+\frac12
\eta^M\wedge \eta_M+\l^Md\star\eta_M+\star dB^M\wedge dC_M
\right)\ ,\label{gftr}\ee
the boundary conditions $\eta_M|=0$, $\l^M|=0$, $dC^M|=0$ and
$dB^m|=0$; the $\msp(2)$ constraint following from \eq{xi} and
\eq{kkprime}; and, the shift-symmetry BRST charge
\be q\ =\ \int_L\star j\ ,\qquad \star j= -C^M\eta_M-Y^M
dC_M+\l^M\star dC_M\ ,\ee
generating the transformations
\be \d_q Y^M\ =\ -C^M\ ,\qquad \d_qB^M\ =\ -\l^M\ .\label{q}\ee

Before turning to the canonical quantization, we shall discuss in
more detail the boundary conditions and ordering prescriptions for
boundary and bulk correlators.


\scss{On Open-String Vertex Operators}\label{sec:kappa}


In this section we give a heuristic discussion of boundary correlators of
bi-local operators, \emph{i.e.} correlators of operators that depend on
both end points of the string. The two main proposals are that these
correlators contain the $\star$ product \eq{intro:osc}, and
that radial and time-ordered boundary correlators are related by
$\star$-multiplication by the intertwining operator $\k$ defined in
\eq{Jprime}.

The quantization of the model gives rise to \emph{$N$-punctured
correlators} of the form
\be \left\langle {\cal O}\right\rangle_{\{y(\x)\}_{\x=1}^N}\ =\
\left\langle T_\pm [{\cal O}_{\partial\S}] R[{\cal
O}_\S]\right\rangle_{\{y(\x)\}_{\x=1}^N}\ ,\label{RTcorr}\ee
where ${\cal O}_{\partial\S}$ and ${\cal O}_\S$ are operators
inserted on $\partial\S$ and $\S$, respectively; $T_\pm$ denote
path ordering with respect to the two orientations of
$\partial\S$; $R$ denotes ordering in terms of increasing radius
of concentric circles defined using an auxiliary Euclidean metric;
and $y(\x)$ are boundary conditions at points $p_\x\in\partial\S$,
$\x=1,\dots,N$, that we shall consider in the cases of $N=1$ and
$N=2$.

The path orderings are related via the outer anti-automorphism
$\tau$ corresponding to reversal of the orientation of
$\partial\S$, defined by
\be \left\langle\t\Big( T_+ [{\cal O}_{\partial\S}]R[{\cal
O}_\S]\Big)\right\rangle\ =\ \left\langle T_-[\t{\cal
O}_{\partial\S}]R[\tau{\cal O}_\S]\right\rangle\
,\label{tauTpl}\ee
which act on the zero-mode in the embedding field as in
\eq{tauosc}. Moreover, letting $R_p$ denote the ordering of the
radial evolution emanating from a point $p\in\partial\S$, we shall
assume that ${\cal O}_\S$ is built from operators
whose mutual products are local in the sense of analytical continuation, so that
\be R_p[{\cal
O}_\S]\ =\ R[{\cal O}_\S]\ ,\label{rt}\ee
is independent of $p$.

More general correlators, \emph{e.g.} the variations of a local
correlator, may require a specific choice of $p$ and a
prescription for combining the $T$ and $R$ orderings, which we
shall refer to as an intertwiner. To formulate this algebraically,
we consider the ``amputation'' of an external massless string
state, drawn in Fig.~\ref{fig:intw}, which shows that the
insertions of an operator ${\cal O}(p)$ into the radial and
path-ordered parts of $T_+[{\cal O}_{\partial\S}]
R[{\cal O}_\S]$, respectively, are related as
\be \left\langle T_+\left[{\cal
O}_{\partial\S}\right]R_p\left[{\cal O}_\S{\cal
O}(p)\right]\right\rangle\ =\ \left\langle T_+\left[{\cal
O}_{\partial\S}({\cal O}{\cal O}_\kappa)(p)\right]R\left[{\cal
O}_\S\right]\right\rangle\ ,\label{vfik}\ee
where ${\cal O}_\kappa$ generates the inner automorphism
corresponding to reversal of the orientation of $\partial\S$, as
drawn in Fig.~\ref{fig1}, that is, ${\cal O}_\kappa$ implements
the world-sheet extension of the automorphism $\pi$ defined in
\eq{piket} and \eq{pimap}.

Turning to the boundary data, the case $N=1$ is implemented as the
symmetric boundary condition
\be \lim_{\tau\rightarrow\pm\infty} \left\langle {\cal
O}Y^A_i(\tau)\right\rangle_{y}\ =\ y^A_i \left\langle {\cal
O}\right\rangle_{y}\ .\ee
As a result, $Y^{Ai}|$ is given by a zero-mode identified as
$y^A_i$ plus $q$-exact non-zero modes, so that symmetric boundary
correlators of $q$-invariant \emph{local} operators ${\cal O}_f$,
where $f$ are functions on ${\cal Z}$, are equivalent to the
algebra ${\cal W}[{\cal Z}]$ based on the Moyal $\star$-product
\eq{star} \cite{Vasiliev:1986qx}.

The insertion of more general operators ${\cal O}(p)$
requires a specification of the behavior of the embedding fields
at $p$ in the form of the \emph{$2$-punctured boundary
condition}\footnote{There is also a branch-point at infinity, with
asymptotic behavior determined by $y$ and $y'$. Alternatively,
\eq{bilocBC} can be imposed as in \eq{RTcorr} with $p_1$ and $p_2$
chosen as two points on the two regular branches of $\partial\S$,
respectively, in which case $y(p_1)=y$ and $y(p_2)=y+y'$.}
\be \lim_{\tiny\ba{c}\tau\rightarrow \tau(p)^\pm\\\s\rightarrow
0\ea} \left\langle R_p\left[{\cal O}Y^A_i(u,\bar
u)\right]\right\rangle_{y,y^{\prime}} \ = \ (y^A_i+{\arg u|_\pm\over \pi}
y^{\prime A}_i)\left\langle{\cal O} \right\rangle_{y,y'}\
,\label{bilocBC}\ee
modulo $q$-exact terms, and where $u=\t+i\s$ is a local coordinate
vanishing at $p$. The determination of the phase-factor in
\eq{bilocBC} is chosen as
\be \arg u|\ =\ {\pi\over 2}\left(1-\e\big(\t-\t(p)\big)\right)\
.\label{determination}\ee
Thus one may identify
\be y^M\ ,\qquad \widetilde z^M\ \equiv\ y^M+y^{\prime M}\
,\label{boundarycoord}\ee
with the phase-space coordinates of $T_+$ and $T_-$-ordered
portions of the boundary to the right and left of $p$. The
$T_-$-ordered coordinate $\widetilde z^M$ is mapped to a
$T_+$-ordered counterpart $z^M$ by the anti-automorphism $\t$
acting as in \eq{tauosc}, and we fix conventions by choosing
\be \tau(\widetilde z^M)\ =\ iz^M\ .\label{tauz}\ee
As we shall demonstrate in the next section, the
resulting $\star$-product algebra for the $(y,z)$-oscillators is given by
\eq{intro:osc}, and
\be Y^{Ai}|\ =\ y^{Ai}+{1\over 2\pi i} \log{u\over \bar u}
y^{\prime Ai}+\{q,\b^{Ai}\}\ ,\ee
where the shift-mode $y^{\prime Ai}$ is paired with a non-zero
mode $c^{\prime Ai}$ in $C^{Ai}$ and hence unphysical. Thus,
$q$-closed local operators ${\cal O}_f$ are independent of the shift-mode,
and $2$-punctured boundary correlators of such operators are
equivalent to ${\cal W}[{\cal Z}]$.

Non-trivial deformations of the $q$-cohomology are generated by
operators that are \emph{covariantly $q$-closed}, i.e. closed up
to covariantisations that drop out of the trace. The relevant
operators are constructed from functions $\widehat f$ on ${\cal
Z}\times {\cal Z}$, depending on two insertion points, \emph{viz.}
the \emph{bi-local operators}
\be {\cal O}_{{\widehat f}}\ =\
\widehat f(Y^{Ai}(p_1),Y^{Ai}(p_2))\ ,
\qquad p_{1,2}\in\S\ .\ee
We shall assume that in $R[{\cal O}_\S\widehat
f(Y^{Ai}(p_1),\widetilde Y^{Ai}(p_2))]$ with $p_{1,2}\in\partial
\S$, the boundary coordinates $Y^{Ai}(p_1)$ and $\widetilde
Y^{Ai}(p_2)$ evolve under $T_+$ and $T_-$ order, respectively.
Hence, assuming that $q$-exact contributions cancel,
\be \left\langle T_+[{\cal O}_{\partial\S}]R[{\cal O}_\S{\cal
O}_{\widehat f}|]\right\rangle_{y,y'}\ =\ \left\langle T_+[{\cal
O}_{\partial\S}]R[{\cal O}_\S]\right\rangle_{y,y'}\star
\widehat f(y,\widetilde z)\ ,\label{whfR}\ee
and analogously
\be \left\langle T_+[{\cal O}_{\partial\S}{\cal O}_{\widehat
f}|]R[{\cal O}_\S]\right\rangle_{y,y'}\ =\ \left\langle T_+[{\cal
O}_{\partial\S}]R[{\cal O}_\S]\right\rangle_{y,y'}\star \widehat f(y,z)\
.\label{whfT}\ee
Under $R$-ordering, the worldlines emanating from ${\cal
O}_{\widehat f}$ connect to $T_+$-ordered worldlines to the left
and $T_-$-ordered dittos to the right. Transporting ${\cal
O}_{\widehat f}$ from the $R$-ordered sector to the $T_+$-ordered
sector requires a reversal of the orientation of the worldlines
emanating on the right side. Thus, in \eq{vfik} the role of ${\cal
O}_\kappa$ is to exchange the orientation of the worldlines in
accordance with the automorphism $\pi$ drawn in Fig. \ref{fig1}.
Thus, for $p_{1,2}\in\partial\S$ we take
\be \widehat f(Y^{ai}(p_1),Y^i(p_1);\widetilde
Y^{ai}(p_2),\widetilde Y^i(p_2))~{\cal O}_\kappa\ =\
:\widehat f(Y^{ai}(p_1),\widetilde Y^i(p_2);\widetilde
Y^{ai}(p_2),Y^i(p_1)){\cal O}_\k:\ .\ee
Alternatively, applying \eq{vfik} to ${\cal O}(p)={\cal
O}_{L+}{\cal O}_{L_-}$, where $L_\pm$ denote the portions of
$\partial\S$ on which $R_p$ is equivalent to $T_\pm$,
\be T_+[{\cal O}_{\partial\S}]R[{\cal O}_\S{\cal O}(p)]\ =\ T_+[{\cal O}_{\partial\S}\left(T_+[{\cal
O}_{L_+}]T_-[{\cal O}_{L_-}]\right){\cal O}_\kappa]R\left[{\cal O}_\S \right]\ ,\ee
which yields a well-defined gluing of worldlines provided
\be T_-[{\cal O}_{L_-}]{\cal O}_\kappa\ =\ {\cal O}_\kappa
T_+[{\cal O}_{L_-}]\ .\ee
Hence, in accordance with \eq{whfT}, ${\cal O}_\kappa$ can be
implemented taking $\kappa$ to be a bi-local phase-space function
with the property\footnote{The classical multiplication by
$\kappa$ may follow from boundary terms in turn required as
counter-terms in order for the variational principle \eq{varprin}
to hold also at the branch points. }
\bea \widehat f(y^{ai},y^i; z^{ai}, z^i)\star \kappa&=&
\kappa\widehat f (y^{ai},i z^i ;z^{ai},-i y^i)\
,\label{kappaproperty}\\\kappa\star\widehat f(y^{ai},y^i; z^{ai},
z^i)&=& \kappa\widehat f (y^{ai},-i z^i ;z^{ai},i y^i)\ ,\\
\kappa\star \widehat f(y^{ai},y^i; z^{ai}, z^i)\star \kappa&=&
\widehat f (y^{ai},-y^i ;z^{ai},- z^i)\ ,\label{piyz}\eea
which we identify with the action of the intertwiner in
\eq{Jprime}.

With these preliminary considerations in mind, we turn to the
canonical quantization with the primary aim of deriving the
$\star$-product algebra and $q$-transformations of the
$(y,z)$-oscillators.


\scss{$\star$-products and $q$-transformations}\label{sec:osc}


In this section we analyze the oscillator algebra
and BRST transformations of the free field theory defined by the
truncated Lagrangian \eq{gftr}.
To begin, we choose the auxiliary metric to be the flat metric on the
upper half-plane
\be \S\ =\ \left\{u\ =\ \t+i\s\,:\ \s\ >\ 0\right\}\ ,\qquad ds^2\
=\ |du|^2\ ,\qquad \star du\ =\ -idu\ ,\ee
and consider the radial evolution around the point $p\in
\partial \S$ with $u(p)=0$, governed by
the field equations
\be d\star\eta^M\ =\ d\eta^M\ =\ 0\ ,\quad dY^M+\eta^M+\star
d\l^M\ =\ 0\ ,\ee
subject to the boundary conditions
\be \eta^M|\ =\ 0\ ,\quad \l^M|\ =\ 0\ ,\ee \be d\star dB^M\ =\
d\star dC^M\ =\ 0\ ,\qquad dB^M|\ =\ dC^M|\ =\ 0\ ,\ee
imposed on
\be \partial \S\setminus\{p\}\ =\ L_+\cup L_-\ ,\qquad L_{\pm}\ =\
\{\s\ =\ 0\ ,\pm \t>0\}\ .\ee
The fields are also subject to the Euclidean-signature reality
conditions
\be (Y^M(u,\bar u))^\dagger\ =\ Y^M(u,\bar u)\ ,\quad (\l^M(u,\bar
u))^\dagger\ =\ \l^M(u,\bar u)\ ,\ee\be(B^M(u,\bar u))^\dagger\ =\
B^M(u,\bar u)\ ,\qquad (C^M(u,\bar u))^\dagger\ =\ C^M(u,\bar u)\
.\ee
The resulting mode expansions read
\bea Y^M&=& y^M+{1\over 2\pi i} y^{\prime M}\log {u\over \bar
u}+i\sum_{n\neq
0}{1\over n}(y^M_n u^{-n}-\bar y_n^M\bar u^{-n})\ ,\label{modeexp1}\\
\l^M&=&i\sum_{n\neq
0}{1\over n}{\rm Im} y^M_n( u^{-n}-\bar u^{-n}) \ ,\label{modeexp2}\\
B^M&=& b^M+{1\over 2\pi i} b^{\prime M}\log {u\over \bar
u}+i\sum_{n\neq
0}{1\over n}b^M_n( u^{-n}-\bar u^{-n})\ ,\label{modeexp3}\\
C^M&=& c^M+{1\over 2\pi i} c^{\prime M}\log {u\over \bar
u}+i\sum_{n\neq 0}{1\over n}c^M_n( u^{-n}-\bar u^{-n})\
,\label{modeexp4}\eea
where $\bar y^M_n=(y^M_n)^\dagger$ and the remaining operators are
real, and we have also given the results for the ghosts for later
reference.

To determine the $(y,z)$-oscillator algebra it suffices to compute
the radial-ordered two-point functions subject to the homogeneous
boundary condition
\be \lim_{|u|\rightarrow\infty} \left\langle R\left[Y^M(u,\bar
u){\cal O}\right]\right\rangle_{0} \ =\ 0\ .\ee
The $\eta Y$ and $\eta\l$ contractions are the inverses of the
kinetic terms in
\be s_{\rm kin}[Y,\eta,\l]\ =\
\frac12\int_\S\left(dY^M\wedge\eta_M+\l^Md\star\eta_M\right)\ .\ee
Using conventions in which $\int f(u,\bar u) d\star d{1\over
4\pi}\log |u|^2= \int f(u,\bar u) d d{1\over 4\pi i}\log
{u\over\bar u}=f(0,0)$, one finds
\be \left\langle R\left[\eta_M(1) Y^N(2)\right]\right\rangle_0 \
=\ d_1\Psi(1,2)\d_M^N\ ,\qquad \left\langle R\left[\eta_M(1)
\l^N(2)\right]\right\rangle_0 \ =\ -id_1\Phi(1,2)\d_M^N\ ,\ee
 \be \Psi(1,2)\ =\ {1\over 2\pi}\log {u_1-u_2\over \bar u_1-\bar
u_2}{ \bar u_1- u_2\over u_1-\bar u_2}\ ,\qquad \Phi(1,2)\ =\
{i\over 2\pi}\log {|u_1-u_2|^2\over |u_1-\bar u_2|^2}\ .\ee

The $YY$ contractions arise via the vertex $\frac i4 \int_\S
\eta^M\wedge \eta_M$, leading to
\be \left\langle R\left[Y^M(1) Y^N(2)\right]\right\rangle_0\ =\
2\times \frac i4 \int_\S \left\langle Y^M(1) \eta^P(u,\bar
u)\right\rangle_0 \left\langle \eta_P(u,\bar u)
Y^N(2)\right\rangle_0\ ,\ee
that can be evaluated by dissecting $\S$ into three parts: two
discs, $D_k$ ($k=1,2$), with evanescent radii centered on $u_k$;
and the remaining part of the upper half-plane, $\S'$, with branch
cuts $L_{k}$ from $\bar u_k$ to $u_k$. The contributions from
$D_{k}$ vanish, while that from $\S'$ can be rewritten using
Stokes' theorem as
\be \int_{\S'} d\Psi(1)\wedge d\Psi(2)\ =\
\oint_{\partial\S'}\Psi(1) d\Psi(2)-\int_{\S'} \Psi(1)d^2\Psi(2)\
,\ee
where $d=du\partial_u+d\bar u\partial_{\bar u}$, $\Psi(k)\equiv
\Psi(u,\bar u;k)$ and the last term vanishes identically, since
$\Psi(u,\bar u;2)$ is regular in $\S'$. In the first term
\be \partial\S'\ =\ \{{\rm Re} ~u=0\}~\cup~ \bigcup_k\,(-\partial
D_k)\cup L_{k,+}\cup (-L_{k,-})\ ,\ee
where $L_{k,\pm}$ denote contours drawn on each side of the two
branch cuts. The contributions from $\{{\rm Re}~u=0\}\cup
(-\partial D_1)\cup L_{2,+}\cup (-L_{2,-})$ vanish, and
\be \oint_{-\partial D_2}\Psi(1) d\Psi(2)\ =\ -2i\Psi(2,1)\ ,\ee
\be \int_{L_{1,+}\cup (-L_{1,-})}\Psi(1) d\Psi(2)\ =\ 2i\int_q^1
d\Psi(2)\ =\ 2i \Psi(1,2)\ ,\ee
where it has been used that $\Psi(1)|_+-\Psi(1)|_-=2i$ and
$\Psi(u|,\bar u|;2)=0$.

Hence, we have arrived at
\be \left\langle R\left[Y^M(1) Y^N(2)\right]\right\rangle_0\ =\
-\O^{MN}\left\{\Psi(1,2)-\Psi(2,1)\right\}\ = {1\over \pi}
\O^{MN}\log{u_1-\bar u_2\over \bar u_1-u_2}\ .\label{gfyy1}\ee
In the limit ${\rm Im}\,u_k\rightarrow 0$, this two-point function
yields
\be\left\langle R\left[Y^M(\t_1) Y^N(\t_2)\right]\right\rangle_0\
=\ -i \O^{MN}\e(\t_1-\t_2)\ ,\label{gfyy2}\ee
that in turn produces the $T_\pm$-ordered
two-point functions
\be \left\langle T_\pm\left[Y^M(\t_1)
Y^N(\t_2)\right]\right\rangle_0\ =\ - i \O^{MN}\e(\t_1-\t_2)\
,\qquad \pm \t_{1,2}>0\ ,\ee
from which it follows that
\be \left\langle y^M\star y^N \right\rangle_0\ =\ -i \O^{MN}\
,\qquad \left\langle \widetilde z^M\star \widetilde z^N
\right\rangle_0\ =\ i \O^{MN}\ .\label{yyzz}\ee

To determine their mutual contractions it suffices to consider the
following holomorphic moments of the $R$-ordered two-point
function,
\bea \left\langle y^{\prime M}\star Y^N(v,\bar v)
\right\rangle_0&=&\oint_{|u|>|v|} du{\partial\over\partial u}
\left\langle R\left[Y^M(u,\bar u) Y^N(v,\bar v)\right]
\right\rangle_0\ =\ 2i \O^{MN}\ ,\\
\left\langle Y^M(v,\bar v) \star y^{\prime N}
\right\rangle_0&=&\oint_{|u|<|v|} du{\partial\over\partial u}
\left\langle R\left[Y^M(v,\bar v) Y^N(u,\bar u)\right]
\right\rangle_0\ =\ 0\ ,\\
\left\langle y^{\prime M}\star y^{\prime N}
\right\rangle_0&=&\oint_{|u|>|v|} du \oint_0
dv{\partial\over\partial u}{\partial\over\partial v}\left\langle
R\left[Y^M(u,\bar u) Y^N(v,\bar v)\right] \right\rangle_0\ =\ 0\
.\qquad\qquad\eea
Setting ${\rm Im}~v=0$ then gives
\be \left\langle y^{\prime M}\star y^N \right\rangle_0\ =\
2i\O^{MN}\ ,\qquad \left\langle y^M\star y^{\prime
N}\right\rangle_0\ =\ \left\langle y^{\prime M}\star y^{\prime N}
\right\rangle_0\ =\ 0\ ,\ee
which are consistent with \eq{yyzz}, and we note that $y^{\prime
M}$ is a commuting element.

The resulting contraction rules read
\be \left\langle y^M\star y^N \right\rangle_0\ =\ \left\langle
y^M\star \widetilde z^N \right\rangle_0\ =\ -\left\langle
\widetilde z^M\star y^N \right\rangle_0\ =\ -\left\langle
\widetilde z^M\star \widetilde z^N \right\rangle_0\ =\ -i \O^{MN}\
,\label{sum1}\ee
or equivalently, using \eq{tauz}, as
\be \left\langle y^{M}\star y^{N} \right\rangle_0\ =\
i\left\langle y^{M}\star z^{N} \right\rangle_0\ =\ -i\left\langle
z^{M}\star y^{N} \right\rangle_0\ =\ \left\langle z^{M}\star z^{N}
\right\rangle_0\ =\ -i\O^{MN}\ ,\ee
which we identify as the contraction rules following from
Vasiliev's $\star$-product rule \eq{intro:osc}. In \cite{Cattaneo:2001bp}
this ``non-commutativity'' of the string end points was analyzed classically.

Since $y$ and $z$
arise under a common $T_+$-ordered evolution, it follows from
\eq{tauTpl} that their contractions are invariant under the
anti-involution $\t$, and for the same reason also under
$\dagger$, whose transformation properties we identify as
\bea \t(\widehat f(y^M,z^M))&=&\widehat f(iy^M,-iz^M)\ ,\qquad
\t(\widehat f\star\widehat g)\ =\ \t(\widehat g)\star \t(\widehat
f)\ ,\label{tauyz}\\ (\widehat f(y^M,z^M))^\dagger&=&\widehat
f^{\,\dagger}(y^M,z^M)\ ,\qquad \,\quad(\widehat f\star\widehat
g)^\dagger\ =\ \widehat g^{\,\dagger}\star \widehat f^{\,\dagger}\
.\label{daggeryz}\eea

The linearized $q$-transformations read
\be [q,{\rm Re}~y^M_n]\ =\ -c^M_n\ ,\qquad \{q,b^M_n\}\ =\ -{\rm
Im}~y^M_n\ ,\label{BRS1}\ee \be [q,y^{\prime M}]\ =\ -c^{\prime
M}\ ,\qquad [q,y^M]\ =\ 0\ ,\ee\be \{q,b^M\}\ =\ \{q,b^{\prime
M}\}\ =\ 0\ , \label{BRS2}\ee
where the contribution from the zero-mode $c^M$ drops out due to
the $\d$ function inserted into the measure \eq{measure}. The
resulting cohomology is thus generated by $y^M$ and the two
$B^M$-ghost modes $b^M$ and $b^{\prime M}$. Thus, the
two-singleton system in Fig.~\ref{fig:intw} does not depend on the
relative distance between the two singletons\footnote{In case
$\partial\S$ has $N-1$ punctures, $p_{\x}$, $\x=1,\dots,N-1$,
there are $N-1$ shift-modes $y^{\prime M}_{\x}$ paired with
corresponding ghost modes $c^{\prime M}_{\x}$, so there still
remains a single physical bosonic zero-mode.}. Moreover, on the
boundary the embedding field is given by
\be Y^M|\ =\ y^M+{1\over 2\pi i} y^{\prime M}\log {u\over \bar
u}|-2\left\{q~,~\sum_{n\neq 0}{b^M_n\over n} \t^{-n}\right\}\
,\label{ymboundary}\ee
justifying \eq{whfR} and \eq{whfT}.

The form of the $q$-transformations \eq{BRS1} and \eq{BRS2}
implies that $q$ acts as the exterior derivative \eq{dprime} in
the ghost-extended system of master fields of the form
\be \widehat \O\ =\ \widehat \O(y^M,z^M,dz^M)\ ,\qquad dz^{M}\ =\
c^{\prime M}\ .\label{whO}\ee
In what follows we shall, however, truncate the ghost sector, and
build observables using only matter fields, using classical
Grassmann-odd parameters in the $q$-transformations.


\scss{The Vasiliev Observables}\label{sec:vas}


So far we have exhibited a number of features of the
two-dimensional topological field theory based on the
undeformed action \eq{eq:cfSP}. This model describes not only the free propagation
of topological open strings on the singleton phase space,
but also their interactions. In principle, it should be possible to
assemble the data contained in
the BRST operator and a suitable set of fundamental
string vertices into an interacting open-string field action.
Here we shall instead exploit an alternative
approach, whereby open-string field equations are
derived by requiring that deformations of \eq{eq:cfSP}
preserve the BRST symmetry. This is analogous to marginal deformations
of ordinary tensile string theory, such
as those used to derive the effective field equations in the massless sector.
Clearly, the drawback of this approach is the somewhat \emph{ad hoc}
introduction of the deformations, as opposed to the unified treatment
offered by a proper action approach.

Let us consider the observables
\be \left.\widehat{Tr}_\pm\left\{ T_+\big[\exp\oint_{p_2\in\partial\S} dY^M
\widehat A'_M\big]~ R\big[\exp {i\over 2} \int_{p_2\in\S} dY^i\wedge dY_i
\widehat \Phi'\,\big]\right\}\right|_{p_1\in\partial\S}\ ,\label{obs}\ee
with traces to be specified below, and where the operators are
bi-local with $Y^{Ai}(p_1)$ identified
with the zero-mode $y^{Ai}$ in accordance with \eq{whfR} and \eq{whfT}
(justified by \eq{ymboundary}), \emph{i.e.}
\be \widehat A'_M\ =\ \widehat A'_M(y^N,Y^N(\t(p_2)))\ ,\qquad \widehat
\Phi'\ =\ \widehat \Phi'(y^N,Y^N(u(p_2),\bar u(p_2)))\ .\ee
The second argument is integrated over $\partial\S$ or $\S$, and
we shall assume that the surface deformation is local in the sense
of \eq{rt}.

Using \eq{q}, the variations under shift-symmetry BRST
transformations with local parameter $\varepsilon^M$ are given by
\bea \d_{\varepsilon q} T[e^{\oint \widehat A'}]&\!\!=\!\!&T[e^{\oint
\widehat A'}~\oint dY^M
\varepsilon^N \widehat F'_{MN}]\ ,\\
\d_{\varepsilon q} R[e^{ \frac i2\int d^2Y \widehat
\Phi'}]&\!\!=\!\!&-iR\left[e^{ \frac i2\int d^2Y \widehat \Phi' }(\oint
\varepsilon^i dY_i \widehat \Phi'+{1\over 2}\int dY^i(dY_i
\varepsilon^{aj}-2\varepsilon_i dY^{aj})\partial_{aj}\widehat \Phi')\right]\quad\qquad
\label{var2}\eea
where $\widehat F'_{MN}$ are the components of the field strength
\be \widehat F'_{MN}\ = \ \partial_{M}\widehat A'_N+\widehat
A'_M\star \widehat A'_N-(M\leftrightarrow N)\ ,\label{FMN}\ee
with $\partial_M=\partial/\partial Y^M(\t)$, and $\star$ referring
to the operator product on $\partial\S$. The non-abelian extension
arises from the expansion of $T[e^{\oint \widehat A'}]$ in terms
of $T$-ordered products of integrals $\int_L \widehat A'$ over
open intervals $L\subset\partial\S$, with $\d_{\varepsilon
q}\int_L\widehat A'$ given by one contribution on $L$ which yields
the abelian part and one contribution on $\partial L$ which
covariantises lower orders in the expansion of the exponential.

In order to cancel the variations on $\partial\S$ we use the
intertwining relation \eq{vfik} and the replacement \eq{whfT}
(again relying on \eq{ymboundary}), that is
\be \left.\widehat A'_M(y^N,Y^N(\t))\right|\rightarrow \widehat
A'_M(y^N,z^N)\ ,\qquad \left.\widehat \Phi'(y^M,Y^N(\t)){\cal
O}_\kappa\right|\rightarrow \widehat \Phi'(y^M,z^N)\star\kappa\
,\ee
resulting in the component form\footnote{Inclusion of ghosts
should lead to direct contact with \eq{vas1pr} via the
identification \eq{whO}. } of the internal two-form curvature
constraint \eq{vas1pr}. The one-form curvature constraint
\eq{vas2pr} then follows from integrability
\cite{Vasiliev:2003ev}, in turn implying that the bulk variation
in \eq{var2} can be rewritten as $[\widehat A'_{ai},\widehat
\Phi']_\pi$ and then cancelled by means of \eq{vfik}, the graded
cyclicity and assuming the $\msp(2)$-invariance that assures that
$\widehat \Phi'$ is an even element.

Turning to the traces, the truncated $\msp(2)$-triplet sector must
be replaced by a singular projector, following essentially the
same reasoning as in Section \ref{sec:sp4} using full
$\msp(2)$-generators inducing canonical rotations of all doublet
indices $\widehat A'_{Ai}(y^{Bj},z^{Bj})$ and $\widehat
\Phi'(y^{Ai},z^{Ai})$ \cite{Vasiliev:2003ev}. In the oscillator
formulation, these can be shown to be
\be \widehat K'_{ij}\ =\ K^{(tot)}_{ij}-\frac 14 \{\widehat
S^{\prime A}_i,\widehat S'_{Aj}\}_\star\ =\
K_{ij}-\partial^A_{(i}\widehat A'_{j)A}+2\widehat A^{\prime A}_{(i}\star
\widehat A'_{j)A}\ ,\ee 
where
\be \widehat S'_{Ai}\ =\ z_{Ai}-2i\widehat A'_{Ai}\
.\label{widehatS}\ee
Consequently, the observables \eq{obs} require the deformed trace
\be \widehat{Tr}_\pm{\cal O}\ =\ \widehat{tr}_\pm[{\cal O}\star
\widehat M]\ ,\ee
with $\widehat M$ defined by \eq{widehatMintro}, and
$\widehat{tr}_\pm$ defined by
\bea \widehat{tr}_+ \widehat {\cal O}&=& \int {d^{2(D+1)}y
d^{2(D+1)}y'\over (2\pi)^{2(D+1)}} ~\langle\widehat{{\cal
O}}\rangle_{y,y'}\ ,\\
\widehat{tr}_- \widehat {\cal O}&=& \widehat{tr}_+[\widehat {\cal
O}\star\rho]\ ,\eea
with $\rho$ an implementation of the large gauge transformation
\eq{rho}, determined by the oscillator algebra \eq{intro:osc} to
be
\bea  \rho&=&\exp \left({y^{M}z_{M}}\right)\ ,\eea
with the properties
\be  \rho\star \widehat f(y,z)\ =\ \rho\widehat f(-i z,i y)\
,\qquad \widehat f(y,z)\star \rho\ =\ \rho\widehat f(i z,-i y)\
.\ee

Concerning the Lorentz invariance, eq.~\eq{anomaly} implies that
$\mso(D-1,2)$-anomalies are induced by non-vanishing
$C^{Ai}$-ghosts on the boundary corresponding to $dY^{Ai}|$
insertions in the observable \eq{obs}. The Lorentz generators can
be replaced, however, by improved full dittos generating canonical
Lorentz transformations of all Lorentz indices in the master
fields \cite{Vasiliev:2003ev,Sagnotti:2005ns},
\be \widehat M'_{ab}\ =\ M^{(tot)}_{ab}-\frac12\{\widehat
S^{\prime i}_a,\widehat S'_{bi}\}_\star\ =\ M_{ab}-\partial^i_{[a}\widehat
A'_{b]i}+2\widehat A^{\prime i}_{[a}\star \widehat A'_{b]i}\ ,\ee
where $M^{(tot)}_{ab}=M_{ab}+\frac12 z^i_a z_{bi}$. We note that $\widehat
M'_{ab}$ reduces to $M_{ab}$ in the gauge where $\widehat A'_{ai}=0$
and the remaining master-field components are taken to be
independent of $z^{ai}$.

At the level of the master-field equations, the minimal bosonic
model based on the adjoint and twisted-adjoint representations
given in \eq{offhsa} and \eq{C2}, respectively, results from
imposing the discrete symmetries on the full master fields using
$\pi$, $\t$ and $\dagger$ given by \eq{piyz}, \eq{tauyz} and
\eq{daggeryz}. This corresponds to unoriented open strings,
leading to factorization of \eq{obs} using symmetrized phase-space
propagators along the lines stressed in Section \ref{sec:sp4}, and
we hope to report on this elsewhere.

This concludes our analysis of the Vasiliev observables \eq{obs},
which clearly leaves a number of important details to be worked
out, as discussed in Section \ref{sec:hsgt} below eq.
\eq{observables}.

We next turn to some remarks on the 4D spinor string, and then end
our considerations by discussing some aspects of the unfolding
procedure.


\scss{Remark: 4D Spinor String}\label{sec:4d}


The 4D bosonic spinor string is defined by the action
\be S\ =\ \frac 12 \int_\S (dY^{\a}\wedge \eta_\a+\frac12
\eta^\a\wedge \eta_\a+d\bar Y^{\ad}\wedge \bar\eta_{\ad}+\frac12
\bar\eta^{\ad}\wedge \bar\eta_{\ad})\ ,\ee
where $Y^\a$ and $\bar Y^{\ad}=(Y^\a)^\dagger$ embed the
worldsheet into $SL(2,\Comp)$-spinor space ${\cal Z}$. There is no
internal gauge group, the world-line correlators are invariant
under the spinor-realization $\mhs(4)\simeq \mhso_0(3,2)$
\cite{hssymmetries1,Sezginhs4,Sezgin:2002ru}, and Vasiliev's $\star$-product
algebra of functions $\widehat f(y^\a,\yb^{\ad};z^\a,\zb^{\ad})$
\cite{Vaseqs} follows from the boundary correlators of bi-local
operators in a fashion completely analogous to that of the vector
model.

However, unlike the vector model, the $\pi$-map, which is
originally defined by its action on the higher-spin algebra
induced via \eq{pi}, can be lifted to two inequivalent maps $\pi$
and $\bar\pi$ reflecting the holomorphic and anti-holomorphic
coordinates of ${\cal Z}$, respectively, generated by
\be \kappa\ =\ e^{iy^\a z_\a}\ ,\qquad \bar\kappa\ =\
e^{-i\yb^{\ad}\zb_{\ad}}\ .\ee
This suggests a complexification in which $\kappa$ and
$\bar\kappa$ intertwine boundary correlators involving explicit
holomorphic and anti-holomorphic free-field insertions,
respectively. We propose that with this prescription the
observable
\be \widehat{tr}_\pm\left\{ T\big[\exp\oint_{\partial\S} (dY^\a
\widehat A_\a+{\rm h.c.})\big]~ R\big[\exp {i\over 4} (\int_\S
dY^\a\wedge dY_\a b\widehat \Phi+{\rm h.c.})\,\big]\right\}\
,\label{obs4d}\ee
where $\widehat{tr}_-[{\cal O}]=\widehat{tr}_+[{\cal
O}\star(\k\bar\k)]$, has $q$-transformations proportional to the
4D spinor-oscillator Vasiliev equations, \emph{viz.}
\be \widehat F\ =\ \frac {ib}4 dz^\a\wedge dz_\a
\widehat\Phi\star\kappa-{\rm h.c.}\ ,\qquad \widehat
D\widehat\Phi\ =\ 0\ ,\label{vas4d}\ee
with curvatures given by direct analogs of those in \eq{vas1}. The
truncation to the minimal bosonic Type A and Type B models
\cite{Sezgin:2003pt} for $b=1$ and $b=i$, respectively, then
follows the steps out-lined above for the vector model, with
analogous definitions of $\t$ and $\dagger$ and either $\pi$ or
$\bar\pi$ can be used to define the twisted-adjoint master field.
The observables \eq{obs4d} are zero-form invariants on ${\cal M}$,
and we expect the expansion of the Wilson-loop to produce the
invariants examined recently in \cite{solutions}.

In both \eq{obs} and \eq{obs4d}, the two-component nature of $Y^i$
and $Y^\a$, respectively, is crucial for the $q$-variation of the
surface deformation to simplify, hindering a straightforward
extension to spinor strings in $D>4$. This is the basic reason
that for the supersymmetric algebras
\cite{Sezgin:1998gg,Sezgin:2001yf,Sezgin:2001ij}
\be \mhs(8|4)\supset \mathfrak{osp}(8|4)\ ,\qquad
\mhs(2,2|4)\supset\mathfrak{psu}(2,2|4)\ ,\qquad
\mhs(8^*|4)\supset\mathfrak{osp}(8^*|4)\ ,\label{susyalg}\ee
only the corresponding 4D theory is known fully, based on an
extension of \eq{vas4d} by a single fermionic oscillator in a
vector representation of $SO(8)$ (see \cite{Sezgin:1998gg} for
details). The open-string realization leads, however, to
interesting modifications, such as a fermionic superpartner of
$z_\a$, and also suggests natural refinements, such as a
fundamental $2$-form potential.


\scss{Space-time Unfold and The Doubling Proposal}\label{sec:unf}


In the doubling proposal, the background-covariant quantization in
the fiber ${\cal Z}$ leads to weakly projected bi-local master
fields $\widehat\Phi'$ and $\widehat A'$. This data is in turn
unfolded \emph{\'a la} Vasiliev into a classical geometry
\be E[{\cal M};\widehat \Phi,\widehat A]\ ,\ee
which is a bundle with base manifold ${\cal M}$, and sections
given by bi-local master fields obtained by extending $\widehat
\Phi'$, $\widehat A'$ and $d'$ from rank-$r$ differential forms on
${\cal Z}$ to rank-$r$ dittos on ${\cal M}\times {\cal Z}$,
\emph{viz.}
\bea \widehat d&=&dx^\mu\partial_\mu+dz^M\partial_M\ ,\\
\widehat A&=& dx^\mu\widehat A_\m(x^\n;y^N,z^N)+dz^M\widehat
A_M(x^\n;y^N,z^N)\ ,\\ \widehat\Phi&=&\widehat \Phi(x^\m;y^M,z^M)\
,\eea
obeying \emph{i)} identical kinematical constraints with
$\tau(x^\m)=(x^\m)^\dagger=x^\m$ and
\be  \t(\widehat A_\m)\ =\ (\widehat A_\m)^\dagger\ =\ - \widehat
A_\m\ ,\ \ee
so that $\widehat A_\m\in \mhso(D-1,2)$ defined in \eq{offhsa};
and \emph{ii)} the extended strongly projected Vasiliev equations
\eq{vas1} and \eq{vas2}.

Since \emph{all curvatures tangent to ${\cal M}$ vanish}, the
bi-local master fields can be expressed in coordinate balls
$U\subset {\cal M}$ using a gauge function,
\be \widehat A_\mu\ =\ \widehat g^{-1}\star\partial_\mu\widehat g\
,\quad \widehat A_M \ =\ \widehat g^{-1}\star(\widehat
A'_M+\partial_M)\star\widehat g\ ,\quad \widehat \Phi\ =\ \widehat
g^{-1}\star \widehat \Phi'\star \pi(\widehat g)\ ,\label{g}\ee
in turn determining local master fields
\be A_\m\ =\ \widehat A_\m|_{z=0}\ ,\qquad \Phi\ =\
\widehat\Phi|_{z=0}\ ,\ee
containing the physical fields given by the scalar
$\phi=\Phi|_{y=0}$ and symmetric-tensor gauge fields given by in
general $\Phi$-dependent directions in the $y$-expansion of
$A_\mu$ \cite{Sagnotti:2005ns}. Due to these field redefinitions,
the metric and symmetric-tensor gauge fields are non-trivial even
for simple choices of gauge function \cite{solutions}.

The global geometry is in general non-trivial. This state of
affairs -- that a given Weyl zero-form determines a geometry -- is
completely analogous to that in lower-spin systems involving
gravity, only that in higher-spin theory the vielbein is extended
by an infinite tower of higher-spin gauge fields and a scalar
packed away into local master fields $\Phi$ and $A_\m$ in turn
incorporated into bi-local master fields $\widehat\Phi$ and
$\widehat A$ for which the local gauge fixing assumes the simple
form \eq{g}. In this respect, the higher-spin extension presents a
tremendous simplification, facilitating a purely algebraic
calculation of the metric and its higher-spin counterparts
starting from $\widehat\Phi'$ \cite{solutions}.

Alternatively, determining the $z$-dependence in a
$\Phi$-expansion, the remaining constraints are equivalent to
\be \widehat F_{\m\n}|_{z=0}\ = 0\ ,\qquad \widehat
D_\m\widehat\Phi|_{z=0}\ =\ 0\ ,\label{fmn}\ee
constituting an integrable set of differential constraints on the
local master fields. Formally, the $y$-expansion of \eq{fmn}
yields an infinite set of fundamental $r_i$-forms $\a^i$ and
composite interactions $f^i(\a^j)$ obeying
\be d\a^i+f^i\ =\ 0\ ,\qquad f^j\wedge {\partial
f^i\over\partial\a^j}\ =\ 0\ ,\ee
defining a non-linear cohomology, or graded homotopy Lie algebra,
which can also be expressed as a nilpotent $(d+Q)$-structure
acting on arbitrary composites $W(\a^i)$ \cite{Vasiliev:2005zu}
\be (d+Q)W\ =\ 0\ ;\qquad Q\ \equiv -f^i{\partial\over
\partial\a^i}\ ,\ee\be d^2 W\ =\ \{d,Q\}W\ =\ Q^2W\ =\ 0\ .\ee
By construction, the system is invariant under the gauge
transformations
\be \d \a^i\ =\ d\e^i+\e^j\wedge{\partial f^i\over\partial\a^j}\
,\ee
with $\e^i\equiv 0$ if $r_i=0$, so that locally each positive-rank
form can be gauged away, which is essentially the non-linear
version of the Poincar\'e's lemma.

Starting from $E[{\cal M};\widehat\Phi,\widehat A]$ one may view
the open-string field $\widehat\Phi'=\widehat\Phi|_{x=0}$ as an
ultra-local holographic dual of the higher-spin gauge theory in
the simply connected coordinate ball $U$. More generally, the homotopy invariance -- which does require
the vielbein to be invertible -- facilitates many other
holographic duals.

Referring to a $D$-dimensional base manifold; taking the vielbein
$e^a$ to be invertible; assuming the solution to be asymptotically
Weyl flat \cite{solutions}; requiring regularity in the center of
spacetime; and assuming the existence of an \emph{on-shell action}
$S=\int_{\cal M}L_D$ where $L_D$ is a suitable $Q$-closed $D$-form
\cite{Vasiliev:2005zu}, the perturbative expansion around the
anti-de Sitter metric yields holographic $n$-point correlation
functions ${\cal C}_n(\widehat X_1,\dots,\widehat X_n)$ where
$\widehat X$ are boundary coordinates.

Assuming instead the vielbein to admit a null direction given by a
globally well-defined vector field $R$ obeying $i_R e^a=0$,
$a=0,\dots,D-1$, the Vasiliev equations imply flow equations along
$R$. In particular, the perturbative expansion around Dirac's
$D$-dimensional hypercone defines an evolution from an initial
value at $R=\infty$ to a final value at $R=0$. For the zero-form,
the constraint in \eq{fmn} reads
\be \nabla_{(0)}\Phi+{1\over 2i}e^a_{(0)}\{P_a,\Phi\}+P\ =\ 0\
,\label{nablaphi}\ee
where $\nabla_{(0)}$ contains the hyper-cone Lorentz connection
and $P$ is linear in the $e^a_{(0)}$ and given by an expansion in
fluctuation fields starting in the second order. Hitting
\eq{nablaphi} with $i_R$ yields the flow equations
\be {\nabla}_{(0)R}\Phi+i_RP\ =\ 0\ .\label{flow}\ee
Focusing on the scalar-field sector, the independent component
fields are given by
\be \phi\ =\ \Phi|_{y=0}\ ;\qquad \widetilde \phi\ =\ ({\rm ker}~
e^a)\cap \phi_a\ ,\quad \phi_a\ =\ i{\partial^2\Phi\over\partial
Y^{ai}\partial Y_i}{\Big |}_{y=0}\ ,\ee
where $e_\m{}^a$ is viewed as a linear map from Lorentz vectors to
tangent vectors, and they obey
\be {\cal L}_R \phi\ =\ -i_R P|_{y=0}\ ,\qquad {\cal
L}_R\widetilde\phi\ =\ -i_R \widetilde P\ .\label{calLn}\ee

Based on the $\mhso_0(D-1,2)$ symmetry and the properties of
\eq{calLn} -- whose right-hand side vanishes in the leading order
-- we propose that ${\cal C}_n(\widehat X_1,\dots,\widehat X_n)$
are the correlators of bilinear operators built from a singleton
$\varphi$ on the conifold, and that $P[\phi,\widetilde\phi,\dots]$
is related to the $\beta$-functional
$\beta[\phi,\widetilde\phi,\dots]$ governing the anomalous scale
dependence of renormalized finite local coupling constants.

Under this hypothesis, the scalars $\phi$ and $\widetilde\phi$ are
related to the source of ${\cal O}=\varphi^2$ and a Lagrange
multiplier used to switch on an ${\cal O}^2$-interaction giving
rise to the scale \cite{On}. It is interesting that in $D=7$ the
above model interpolates between a free theory in the IR with
$\D({\cal O})=\D_+=4$ and a UV fixed point with $\D({\cal
O})=\D_-=6-4=2$, that is, ${\cal O}$ and the interaction ${\cal
O}^2$ become a fundamental free field and its mass-term,
respectively, serving as a form of UV ``safety net''.


\scs{Conclusions}


A new angle on string quantization in curved backgrounds is
provided by the doubling proposal, whereby the quantum strings
live in fibers over classical space-time ''unfolds''. The crucial
input is the singleton determining the structure group of the
fiber -- the classical geometry is then an output. The structure
group contains a higher-spin algebra acting ``diagonally'' on the
generalized Chan-Paton factor -- and as gauge group determining
the self-interactions for massless doubletons. The full structure
group is some Yangian-style extension by ``off-diagonal''
generators, giving rise to a presently unknown massive gauge
theory.

The massless sector lives on the boundary of an open string --
corresponding to a single singleton worldline -- and the massive
sector lives on the boundary of an open membrane -- corresponding
to a topological closed string containing a chiral ring generated
by a singleton-valued weight-$0$ spin field. The purely bosonic
closed string has anomalies that cancel in $D=7$, while the
analogous singleton worldline has only a global anomaly that
cancels in $D=3$ mod $4$, raising a tricky question of whether the
open-string extension should be anomaly free as well.

Historically, higher-spin gauge theory was developed using
remarkably simple algebraic techniques, and we end by speculating
over the massive theory in the same spirit. The fact that the
massless gauge theory consists of total-rank $0$ and $1$ master
fields, suggests a massive theory consisting of master fields of
unlimited rank, say $\widehat A$ and $\widehat B$ with ghost
numbers $1,3,\dots$ and $2,4,\dots$, respectively, subject to
\be \widehat d\widehat A+\widehat A\star\widehat A+\widehat B\ =\
0\ ,\qquad \widehat d B+ \widehat A\star\widehat B-\widehat
B\star\widehat A\ =\ 0\ .\ee
This system becomes dynamical upon truncating $\widehat
B\rightarrow \widehat\Phi\star\widehat J'$, where $\widehat\Phi$
is a zero-form and $\widehat J'$ a fixed intertwiner. An
(associative) $\star$-product emulation of the multi-parton system
can be realized using multi-flavored oscillators, with $\widehat
f_M(\{y(\x),z(\x)\}_{\x=1}^M)\star \widehat
f_N(\{y(\x),z(\x)\}_{\x=1}^N)$ given by ``cluster'' expansion,
$\widehat f_M\star\widehat f_N=\sum_{k=1}^{{\min}(M,N)} \widehat
f_{M+N-k}$, wherein $k$ denotes the numbers of pairs of
oscillators that are identified and Weyl ordered.

{\bf Acknowledgements:}

We are grateful to P. Howe, U. Lindstr\"om, P. Rajan, L. Tamassia
and in particular D. Francia, A. Sagnotti, E. Sezgin and M.
Vasiliev for many discussions and remarks. We have also enjoyed
conversations with D. Anselmi, X. Bekaert, M. Bianchi, G. Bonelli,
A. Bredthauer, M. Cederwall, U. Danielsson, C. Iazeolla, R. Leigh,
S. Lyakhovich, J. Minahan, B. Nilsson, M. Nishimura, M. Olsson, J.
Persson, A. Petkou, K. Skenderis, D. Sorokin, B. Sundborg, M.
Tsulaia, M. Vonk and K. Zarembo. We wish to thank the
string-theory group at Universit\`a di Roma ``Tor Vergata'' and
the George P. and Cynthia W. Mitchell Institute for Fundamental
Physics at Texas A\&M University for their hospitality. Part of the
work was supported by INTAS grant 99-1-590. Finally we wish to thank the
referee for carefully reading the manuscript, and suggesting several
changes for this revised version of the paper.

\newpage

\begin{appendix}


\scs{Conventions for $\mso(D-1,2)$ Representation
Theory}\label{sec:rep}


The $\mso(D-1,2)$ generators $M_{AB}$ obey the commutation rules
\bea [M_{AB},M_{CD}]&=& i\eta_{BC}M_{AD}-
i\eta_{AC}M_{BD}-i\eta_{BD}M_{AC}+i\eta_{AD}M_{BC}\ ,\eea
where $\eta_{AB}={\rm diag}(--+\cdots+)$ and the $\mso(D-1,2)$
vector index $A=0',0,1,\dots,D-1$. The decomposition into
translations and Lorentz rotations is given by
\be P_a\ =\ v_a{}^Av^B M_{AB}\ ,\qquad M_{ab}\ =\ v_a{}^A
v_b{}^BM_{AB}\ , \label{pa}\ee
where $(v_a^A,v^A)\in SO(D-1,2)/SO(D-1,1)$, and $\eta_{ab}={\rm
diag}(-+\cdots +)$. The enveloping algebra ${\cal U}(\mso(D-1,2))$
has the outer anti-involution $\t$, defined by
\be \t(M_{AB})\ =\ -M_{AB}\ .\label{tau}\ee
The combined time reversal and parity operation $\pi$ is the outer
involution defined by
\be \pi(P_a)\ =\ -P_a\ ,\quad \pi(M_{ab})\ =\ M_{ab}\ .
\label{pi}\ee
The real form o$\mso(D-1,2)$ has maximal compact subalgebra
$\mg^{(0)}\simeq \mso(2)\oplus \mso(D-1)$. The eigenvalues of the
adjoint $\mso(2)$ action define a triple grading
\be \mso(D-1,2)\ =\ \mg^{(-)}\oplus \mg^{(0)}\oplus \mg^{(+)}\
,\qquad [E,\mg^{(p)}]\ =\ p\mg^{(p)}\ ,\label{grading}\ee
corresponding to the following decomposition of the generators
\bea \mso(2)_E&:& E\ =\ P_0\ =\ M_{0,0'}\ ,\label{e0}\\
\mso(D-1)_S&:& M_{rs}\ ,\label{mrs}\\
\mg^{(\pm)}&:& L^{\pm}_r \ \equiv\ M_{0r}\mp i P_r\ =\ M_{0r}\mp i M_{r,0'}\ ,\label{boosts}\\
&&{}[E,L^\pm_r]\ =\ \pm L^\pm_r\ ,\quad {}[L^-_r,L^+_s]\ =\
2(\d_{rs} E+i M_{rs})\ .\label{lpllmi}\eea
where the energy operator $E$ is the generator of time
translations; the spin operators $M_{rs}$ are the generators of
spatial rotations; and $L^\pm_r$ are the generators of
energy-raising and energy-lowering boosts.

The $D$-dimensional one-particle and one-anti-particle states form
unitary irreducible representations of $\mso(D-1,2)$ of
lowest-weight type with energy bounded from below and above by
eigenvalues $E_0>0$ and $-E_0<0$, respectively. These
representations are denoted by $\sfD(E_0,S_0)$ and $\widetilde
\sfD(-E_0,S_0)$, where $S_0=(s_1,\dots,s_\n)_{D-1}$
($\n=[(D-1)/2]$) is the spin, \emph{i.e.} highest $\mso(D-1)$
weights, carried by the states with extremal energy. We shall let
$(s_1,\dots,s_\mu)$ denote $(s_1,\dots,s_\mu,0,\dots,0)$ for
$\mu<\nu$.

The lowest-weight spaces are obtained starting from generalized
Verma modules\footnote{The compact subalgebra generators are
suppressed in the generalized modules.} $V(E_0,S_0)$ obtained
boosting the ground state $\ket{E_0,S_0}$ obeying the generalized
lowest-weight condition
\be L^-_r\ket{E_0,S_0}\ =\ 0\ ,\label{lwc1}\ee
and then factoring out the maximal ideal ${\cal N}(E_0,S_0)$
consisting of all submodules generated from singular vectors in
$V(E_0,S_0)$, \emph{i.e.} excited states in $V(E_0,S_0)$ obeying
the lowest-weight condition. In other words,
\be \sfD(E_0,S_0)\ =\ V(E_0,S_0)/\cN(E_0,S_0)\ ,\label{de0s0}\ee
where
\be V(E_0,S_0)\ =\ \bigoplus_{n=0}^\infty\bigoplus_{S}
\big\{\ket{E_n,S}=\underbrace{L^+_{r_1}\cdots
L^+_{r_n}\ket{E_0,S_0}}_{\mbox{$S$ projected}}\big\}\ ,\ee
and, using the natural inner product induced via
\be \bra{E_n,S}\ =\ \left(L^+_{r_1}\cdots
L^+_{r_n}\ket{E_0,S_0}\right)^\dagger\ =\
\bra{E_0,S_0}L^-_{r_1}\cdots L^-_{r_n}\ ,\ee
the maximal ideal can be said to consist of all states that vanish
in the inner-product sense, that is
\be \cN(E_0,S_0)\ =\ \big\{\ket{E_n,S}\in V(E_0,S_0):\
\left\langle E_n,S|E_n,S\right\rangle\ =\ 0\big\}\ .\ee
The non-degenerate inner-product space $\sfD(E_0,S_0)$ is positive
definite only for certain values of the lowest weights falling
into three categories \cite{unitarity}: massive, massless and
singleton-like. The latter two yield non-trivial singular vectors,
notably the longitudinal modes of the massless symmetric rank-$s$
tensor, given by
\be \ket{s+1+2\e_0,(s-1)}_{r_1\dots r_{s-1}}\ =\
L^{+}_r\ket{s+2\e_0,(s)}_{rr_1\dots r_{s-1}}\
,\label{lowestnull}\ee
and the scalar-singleton mass-shell condition
\be \ket{2+\e_0,(0)}\ =\ L^{+}_{r}L^+_r\ket{\e_0,(0)}\
.\label{singv}\ee
The singleton comprises a single line in weight space,
\be \sfD\ =\ \bigoplus_{n=0}^\infty d(n)\ ,\label{singleline}\ee
where each $d(n)$ is an $\mso(2)_E\oplus \mso(D-1)_S$ irrep with
energy eigenvalue $\e_n=\e_0+n$ and spin $(n)$.

One-particle and one-anti-particle spaces can be paired in a
natural fashion by extending the $\pi$-map \eq{pi} to an
involutive map acting also on states, defined by
\be \pi\ket{E_0,S_0}\ =\ \ket{-E_0,S_0}\ .\label{pilws}\ee
From $\pi L^-_r=L^+_r$, $\pi E=-E$ and $\pi M_{rs}=M_{rs}$ it then
follows that
\be \widetilde \sfD(-E_0,S_0)\ =\ \pi(\sfD(E_0,S_0))\
,\label{pidpm}\ee
so that $\sfD(E_0,S_0)\oplus\widetilde \sfD(-E_0,S_0)$ becomes
irreducible under $\{\mso(D-1,2),\pi\}$.


\scs{Verification of $\mhso_0(6,2)\ \simeq
\mathfrak{hs}(8^*)$}\label{sec:ver}


Let us verify the isomorphism \eq{iso} by showing that
$\mathfrak{hs}(8^*)$ and $\mhso_0(6,2)$ are isomorphic real forms
of a complex algebra based on the $GL(8;{\Comp})$-invariant
$\star$-product algebra ${\cal W}[{\cal X}]$ generated by
\be [X_I,\bar X^J]_\star\  =\ 2\d_I^J\ ,\label{XIXJ}\ee
where $X_I$ and $\bar X^J$ are not subject to any reality
condition. The complex higher-spin algebra is given by
\be \mhso_0(S;{\Comp})\ =\ \{~Q\in {{\cal W}[{\cal X}]\over
I[H_S]}\ :\quad  \t(Q)\ =\ -Q~\}\ ,\ee
where $\t(f(X,\bar X))=f(iX,i\bar X)$, and $I[H_S]$ is the ideal
generated by left and right $\star$-multiplication with the
generators of $H_S\simeq A_1$ given by\footnote{We use conventions
in which $[L_+,L_-]_\star=2L_3$ and $[L_3,L_{\pm}]_\star =\pm
L_{\pm}$. The two inequivalent real forms are $(L_3)^\dagger= L_3$
and $(L_{\pm})^\dagger =\e L_{\mp}$, giving $\msu(2)$ and
$\msp(2)$ for $\e=+1$ and $\e=-1$, respectively.}
\be L_+\ =\ \ft14 \bar X^I \bar X^J S_{IJ}\ ,\qquad L_-\ =\ -\ft14
X_I X_JS^{IJ}\ ,\qquad L_3\ =\ \ft14 \bar X^I X_I\
,\label{HsubS}\ee
where $S_{IJ}=S_{JI}$ and $S^{IJ}S_{KJ}=\d^I_K$. The $GL(8;{\Comp})$
 transformations $X\rightarrow gX$, $\bar X\rightarrow \bar X
(g^{-1})^T$ and $S\rightarrow gSg^T$ can be used to change the
signature $S$. For fixed $S$, the maximal finite-dimensional
subalgebra of $\mhso_0(S;{\Comp})$ is $\mso(S;{\Comp})$ generated
by $i\bar X^I \S_I{}^J X_J$ where
$S^{IJ}\S_J{}^K+(I\leftrightarrow K)=0$ and
$\S_I{}^JS_{JK}+(I\leftrightarrow K)=0$. The algebra
$\mhso_0(S;{\Comp})$ decomposes under $\mso(S;{\Comp})$ into
levels with highest weights $(2\ell+1,2\ell+1)$,
$\ell=0,1,2,\dots$. Using the basis
\be Q_{I(\ell),J(\ell)}\equiv M_{I_1 J_1}\cdots
M_{I_{2\ell+1}J_{2\ell+1}}\ ,\label{basisI}\ee
where $M_{IJ}=iX_I\bar X^K S_{KJ}$ and $I(\ell)\equiv I_1\dots
I_{2\ell+1}$, the resulting structure coefficients of
$\mhso_0(S;{\Comp})$ are
\be [Q_{I(\ell),J(\ell)},Q_{K(\ell'),L(\ell')}]=\sum_{k=0}^{2{\rm
min}(\ell,\ell')}
c_{I(\ell),J(\ell);K(\ell'),L(\ell')}^{M(\ell+\ell'-k),N(\ell+\ell'-k)}
Q_{M(\ell+\ell'-k),N(\ell+\ell'-k)}\ ,\label{alg2}\ee
where the coefficients are $\mso(S;{\Comp})$-invariant tensors
built from products of $S_{IJ}$ and $\d_I^J$.

Real forms $\mhso_0(S_\e)$ arise from reality conditions on the
oscillators,
\be \bar X^I\ =\ (X_J)^\dagger M_J{}^I\ ,\qquad \det M\neq 0\
,\qquad M^\star\ =\ M\ ,\ee
as
\be \mhso_0(S_\e)\ =\ \{~Q\in {{\cal W}[{\cal X}]\over I[H_\e]}\
:\quad \t(Q)\ =\ Q^\dagger\ =\ -Q~\}\ ,\ee
where $H_\e$ is defined by \eq{HsubS} with $S\equiv S_\e$ obeying
\be M (S_\e)^\star M^T\ =\ -\e S^{-1}\ ,\qquad H_{\e}\ =\
\mx{\{}{ll}{\msu(2)&\e=+1\\\msp(2)&\e=-1}{.}\ee

The algebras $\mathfrak{hs}(8^*)$ and $\mhso_0(6,2)$ arise from
\bea \mathfrak{hs}(8^*)&:& M\ =\ i\C^{08}\ ,\qquad S\ =\ C\ ,\qquad (\e=-1)\ ,\\
 \mhso_0(6,2)&:& M\ =\ \eta\ ,\qquad S\ =\ \eta\ ,\qquad (\e=+1)\ ,\eea
and their structure coefficients are given by \eq{alg2} with
$S_{IJ}$ replaced by $C_{\a\b}$ and $\eta_{AB}$, respectively.
These are numerically the same when expressed in spinor indices
$\a$ and vector indices $A$ indices.

Finally, to convert between vector and spinor indices we use a
triality-rotation matrix $E_A^\a$ obeying
\be E_A^\a E_B^\b C_{\a\b}\ =\ \eta_{AB}\ ,\qquad \C^9 E_A\ =\
E_A\ .\label{trail2} \ee
This matrix is invariant under $SO(6,2)'\otimes SO(6,2)''$
rotations, and hence the symbols
\be \C^{AB}_{CD}\ =\ E_C^\a E_D^\d (\C^{AB})_{\c\d}\ ,\ee
are invariant under the diagonal $SO(6,2)$. From
$\C^{(AB)}_{CD}=0=\C^{AB}_{(CD)}$ it then follows that
$\C^{AB}_{CD}=-\ft14 \d^{AB}_{[CD]}$, in turn implying
\be E_A^\a E_B^\b M_{\a\b} = M_{AB}\ ,\label{trial1}\ee
which together with \eq{trail2} implies that conversion between
spinor and vector indices in the explicit commutators descending
from \eq{alg2} does not affect the normalization of the structure
coefficients.

\scs{BV Formalism}\label{app:BV}

The BV expectation value is given by \cite{bv}

\be \left\langle {\cal O}\right\rangle_{\Psi}\ =\ \int d\phi
d\phi^+\d\left(\phi-{\partial \Psi\over
\partial\phi^+}\right){\cal O}\exp{\frac i\hbar{\cal S}}\ ,\ee
where $\phi^\a$ and $\phi^+_\a$ are the fields and anti-fields,
respectively; $\Psi$ is the gauge fermion; ${\cal S} =\int_{{\cal
M}} {\cal L}[\phi,\phi^+]$ is the BV action; and ${\cal O}$ are
the observables. The set of fields constitutes all the integration
variables, \emph{i.e.} the classical fields, ghosts and Lagrange
multipliers. The set of anti-fields constitutes the basis of the
dual of the infinite-dimensional space of differential forms on
field space. The fields and anti-fields are further characterized
by Grassmann parity $\e(\phi^\a)+\e(\phi^+_\a)=1$ mod $2$; ghost
number ${\rm gh}(\phi^\a)+{\rm gh}(\phi^+_\a)=-1$;
differential-form degree on ${\cal M}$, ${\rm deg}(\phi^\a)+{\rm
deg}(\phi^+_\a)=\dim {\cal M}$; and that they have dual boundary
conditions on $\partial{\cal M}$. Finally, the gauge fermion
$\Psi=\Psi[\phi,\phi^+;\gamma]$ is a functional with $\e(\Psi)=1$,
${\rm gh}(\Psi)=-1$ and ${\rm deg}(\Psi)=0$, depending on an
auxiliary metric $\gamma$ on ${\cal M}$.

The integrand is a formally well-defined measure, \emph{i.e.} if
$\Psi$ and $\Psi'$ are two different gauge fermions then
\be \left\langle {\cal O}\right\rangle_{\Psi}\ =\ \left\langle
{\cal O}\right\rangle_{\Psi'}\ ,\ee
provided that
\be \D({\cal O}\exp{\frac i\hbar{\cal S}})=0\ ,\qquad \D\ \equiv \
(-1)^{\e^a+1}{\partial_r\over\partial\phi^\a}{\partial_r\over\partial\phi^+_\a}\
,\label{mfe}\ee
where $\partial_r$ denotes derivatives taken from the right. This
is an infinite-dimensional analog of the fact that $\int f$ is
well-defined if $df=0$, with $\D\leftrightarrow d$,
$d\phi^\a\leftrightarrow{\partial_r\over\partial\phi^+_\a}$,
$\phi^+_\a\leftrightarrow i_\a$, so that the dual basis of forms
$(1,\phi^+_\a,\phi^+_\a\phi^+_\b,\dots)\leftrightarrow (V,i_\a
V,i_\a i_\b V,\dots)$, where $V$ is the volume form. The condition
\eq{mfe} amounts to the master equations
\be ({\cal S},{\cal S})-2i\hbar\D {\cal S}\ =\ 0\ ,\ee\be{\cal
Q}{\cal O}\ \equiv\ ({\cal O},{\cal S})-i\hbar\D {\cal O}\ =\ 0\
,\qquad {\cal Q}^2\ =\ 0\ ,\ee
where the anti-bracket, defined via
\be \D(XY)\ =\ X\D Y+(-1)^{\e(Y)}(\D X)Y+(-1)^{\e(Y)}(X,Y)\ ,\ee
is given explicitly by
\be (X,Y)\ =\ {\partial_r X \over\partial\phi^\a}{\partial_l Y
\over\partial\phi^+_\a}-{\partial_r
X\over\partial\phi^+_\a}{\partial_l Y\over\partial\phi^\a}\ =\
X\left({\overleftarrow\partial
\over\partial\phi^\a}{\overrightarrow\partial
\over\partial\phi^+_\a}-{\overleftarrow\partial
\over\partial\phi^+_\a}{\overrightarrow\partial\over\partial\phi^\a}\right)Y
\ .\ee
The bracket $(\cdot,\cdot)$ has a grading that can be derived by
assigning the comma ghost number $1$, implying
$\e((X,Y))=\e(X)+\e(Y)+1$, ${\rm gh}((X,Y))={\rm gh}(X)+{\rm
gh}(Y)+1$ and
\bea 0&=&(X,Y)+(-1)^{(\e(X)+1)(\e(Y)+1)}(Y,X)\ ,\\
0&=&((X,Y),Z)+(-1)^{(\e(X)+1)(\e(Y)+\e(Z))}((Y,Z),X)\nn\\ &&
+(-1)^{(\e(Z)+1)(\e(X)+\e(Y))}((Z,X),Y)\ . \eea
The master equation for ${\cal S}$ is solved in a double
$\phi^+$-expansion and $\hbar$-expansion subject to the boundary
condition and regularity conditions
\be {\cal S}|_{\phi^+=\hbar=0}\ =\ S[\phi_{\rm cl}]\ ,\qquad
\det~{\partial_l\partial_r {\cal S}\over
\partial\phi^\a\partial\phi^+_\b}\ \neq\ 0\ ,\ee
and the solution is assumed to be unique up to canonical
transformations generated by ${\cal S}$. The $\hbar$-corrections
may be thought of as Green-Schwarz anomaly cancellation terms. The
formalism extends to the effective master action $\C$ for which
$(\C,\C)$ measures the true quantum anomalies. The operators are
found likewise up to generalized BRST transformations $\d{\cal
O}=({\cal S},X)-i\hbar \D X$.

\end{appendix}


\end{document}